\documentclass[twocolumn, twocolappendix]{aastex701}

\newcommand{\colt}{\textsc{colt}}

\shorttitle{Column Density Estimator}

\begin{document}

\title{Machine-Learning Reconstruction of Intergalactic Medium \\ Column-Density Maps from Observed Ly$\alpha$ Emission Cubes}

\author{Xihan Deng}
\affiliation{California Institute of Technology, Pasadena, CA 91125, USA}
\email{xdeng@caltech.edu}

\author{Christopher Martin}
\affiliation{California Institute of Technology, Pasadena, CA 91125, USA}
\email{martinc@caltech.edu}

\author{Aaron Smith}
\affiliation{University of Texas at Dallas, Richardson, TX 75080, USA}
\email{asmith@utdallas.edu}

\author{Sultan Hassan}
\affiliation{STScI, Baltimore, MD 21218, USA}
\email{sultanier@gmail.com}

\author{Drew M. Miles}
\affiliation{California Institute of Technology, Pasadena, CA 91125, USA}
\email{drewmiles@caltech.edu}

\author{Nikolaus Z.\ Prusinski}
\affiliation{California Institute of Technology, Pasadena, CA 91125, USA}
\email{nik@astro.caltech.edu}

\author{Vincent Picouet}
\affiliation{California Institute of Technology, Pasadena, CA 91125, USA}
\email{vpicouet@caltech.edu}

\author{Mateusz Matuszewski}
\affiliation{California Institute of Technology, Pasadena, CA 91125, USA}
\email{matmat@caltech.edu}

\author{Zeren Lin}
\affiliation{California Institute of Technology, Pasadena, CA 91125, USA}
\email{zlin@caltech.edu}

\author{Jake S.\ Bennett}
\affiliation{School of Physics \& Astronomy, University of Nottingham, University Park, Nottingham NG7 2RD, UK}
\email{jake.bennett@nottingham.ac.uk}

\begin{abstract}
Understanding the large-scale structure of the intergalactic medium (IGM) is essential for studying galaxy evolution, particularly at $z \sim 2$--$6$, when cold filaments in the cosmic web are considered a primary mechanism for fueling galaxy growth. At these redshifts, Ly$\alpha$ emission shifts into the optical regime and serves as the primary tracer of diffuse IGM for ground-based integral field units (IFUs). However, detections remain limited by low signal-to-noise ratios (S/N) and the complexities of resonant scattering. We present a proof-of-concept machine-learning framework to reconstruct IGM gas column-density maps from noisy Ly$\alpha$ IFU emission cubes. Using Keck Cosmic Web Imager (KCWI) observations of $z \sim 2.3$ protoclusters as a case study, we train a custom convolutional neural network (CNN), the Column Density Estimator (CDE), on a simulated dataset combining IllustrisTNG100 with simplified radiative-transfer calculations and an approximate treatment of KCWI noise. The CDE successfully reconstructs broad filamentary structures in the simulated test set under these adopted assumptions. Applied to observed KCWI data, the CDE predicts gas column-density maps containing filamentary structures extending over $\gtrsim 40''$ ($\sim 1$~comoving Mpc), consistent with an IGM interpretation. The observed reconstructions remain conditional on the adopted models. These results illustrate the potential of the CDE framework for connecting diffuse Ly$\alpha$ observations with simulations of the IGM.
\end{abstract}


\section{Introduction} 
\label{sec:intro}

The intergalactic medium (IGM) is believed to contain the majority of baryons across most of cosmic time. In the present-day ($z \approx 0$) universe, \citet{2012ApJ...759...23S} estimated that roughly 60\% of the baryonic mass resides in the IGM, while about 30\% remained unaccounted for. More recent studies, based on X-ray absorption-line observations toward background quasars \citep{2018Natur.558..406N} and dispersion measures of localized fast radio bursts \citep{2020Natur.581..391M, 2022ApJ...940L..29Y,2025NatAs...9.1226C}, indicate that most of the previously ``missing'' baryons are indeed contained in the IGM, raising its total baryonic mass fraction to $\sim 80\%$ at $z \approx 0$. At higher redshifts, constraints from Ly$\alpha$ forest absorption provide indirect but robust evidence that the IGM contains the vast majority of baryons \citep{1998ARA&A..36..267R}.

As the dominant baryonic component of the universe, the IGM plays a central role in fueling galaxy formation and evolution. At redshifts $z \sim 2$--6, cosmological simulations consistently suggest that the filamentary structure of the IGM, also known as the cosmic web, is especially influential: halos are not yet too hot and their potential wells are not too deep during this epoch, allowing cold gas filaments to penetrate directly into halo centers and efficiently supply gas for galaxy growth \citep{2006MNRAS.368....2D,2009Natur.457..451D,2009Natur.460..213C,2020MNRAS.497..279C}. Understanding and characterizing these filaments is therefore essential for explaining how galaxies sustain their relatively high star formation rates (SFRs) in the early universe, particularly during cosmic noon ($z \sim 2$--3), when the cosmic SFR density reached its peak \citep{2014ARA&A..52..415M}.

Much of what we know about the large-scale filamentary structure of the cosmic web is based on Ly$\alpha$ absorption tomography, which probes the IGM along sparse sightlines toward bright background sources and reconstructs large-scale structure by interpolating many one-dimensional absorption spectra \citep{2001MNRAS.326..597P,2008MNRAS.386..211C,2014ApJ...788...49L,2019ApJ...887...61H,2022ApJ...930..109Q}. However, the effectiveness of absorption tomography rapidly degrades at high redshift, where bright background sources become increasingly rare, motivating alternative observational probes of the diffuse IGM.

In contrast to absorption-based approaches, Ly$\alpha$ emission in principle enables direct mapping of diffuse hydrogen in the IGM over spatially extended regions \citep[for a comprehensive review, see][]{2020ARA&A..58..617O}. However, the expected surface brightness of IGM Ly$\alpha$ emission can be as low as $\lesssim 10^{-20}\,\mathrm{erg\,s^{-1}\,cm^{-2}\,arcsec^{-2}}$ \citep{2018MNRAS.475.3854G}, yielding extremely low signal-to-noise ratios (S/N). In the past, only very long exposures could reveal such faint emission, and the few early detections of extended Ly$\alpha$ structures, including the quasar-illuminated Slug Nebula \citep{2014Natur.506...63C} and Cosmic Web Imager (CWI) observations \citep{2014ApJ...786..106M}, were subject to uncertain or debated interpretations and did not provide unambiguous evidence for a filamentary cosmic web.

The advent of powerful integral field units (IFUs), such as the Multi Unit Spectroscopic Explorer (MUSE; \citet{2010SPIE.7735E..08B}) and the Keck Cosmic Web Imager (KCWI; \citet{2018ApJ...864...93M}), has dramatically improved sensitivity to low-surface-brightness Ly$\alpha$ emission, enabling systematic studies of diffuse structures in the intergalactic environment. Deep MUSE observations have revealed extended filamentary Ly$\alpha$ emission \citep{2019Sci...366...97U,2021A&A...647A.107B,2025NatAs...9..577T}. KCWI observations have also revealed extended Ly$\alpha$ emission on $\sim$1\,comoving megaparsec (cMpc) scales in two massive protoclusters at $z \sim 2.3$ \citep{2023NatAs...7.1390M}, which will be used as the case study for this work.

These developments highlight a timely opportunity to connect observations of Ly$\alpha$ emission with theoretical predictions for the IGM. On the simulation side, state-of-the-art cosmological models such as IllustrisTNG \citep{2018MNRAS.473.4077P,2018MNRAS.475..624N,2018MNRAS.475..676S,2018MNRAS.477.1206N,2018MNRAS.480.5113M,2019ComAC...6....2N} provide detailed predictions for the density, temperature, ionization state, and velocity structure of the cosmic web on scales of hundreds of cMpc. These quantities form the essential inputs for radiative-transfer calculations of Ly$\alpha$ emission. Coupled with recent advances in Ly$\alpha$ radiative-transfer modeling \citep{2015MNRAS.449.4336S,2020ascl.soft02002M,2025arXiv250711603B}, it is now possible to generate realistic simulated Ly$\alpha$ emission cubes that enable meaningful, quantitative comparisons with observations. For example, \citet{2025NatAs...9..577T} used forward modeling to predict Ly$\alpha$ surface-brightness profiles from gas distributions in IllustrisTNG and compare them with MUSE observations.

Building on these developments, we aim to establish a framework for addressing the inverse problem: inferring the underlying gas column-density maps from observed Ly$\alpha$ emission subcubes. Doing so requires addressing several practical and conceptual challenges.

The first set of challenges involves observational and astrophysical systematics. The intrinsically low S/N of Ly$\alpha$ data mean that noise can easily blend with genuine Ly$\alpha$ emission, making realistic noise modeling essential for assessing the fidelity of any comparison between simulations and observations. In addition, despite the substantial progress in Ly$\alpha$ radiative–transfer modeling, predicted surface-brightness levels of IGM Ly$\alpha$ emission at $z\sim2$ can still vary by an order of magnitude depending on assumptions about escape fractions, resonant scattering, dust geometry, and continuum pumping \citep{2010ApJ...725..633F,2012MNRAS.419..780B,2012MNRAS.423..344R,2023MNRAS.523.5248B}. In this work we adopt a cosmological simulation (IllustrisTNG100) together with a physically motivated radiative-transfer model, while emphasizing that our framework is designed to be modular, enabling straightforward retraining as improved cosmological simulations and Ly$\alpha$ radiative-transfer models continue to develop.

Even with realistic forward modeling of both noise and astrophysical processes, a deeper conceptual difficulty remains: how to meaningfully compare simulated and observed data cubes in order to extract robust physical information. This challenge involves two primary components: identifying the physical quantities realistically recoverable from Ly$\alpha$ emission and establishing a methodology that reliably bridges the gap between simulations and observations.

To address the first component, we focus on physical quantities that are most directly linked to Ly$\alpha$ emission. The gas column density exhibits a strong and physically-motivated correlation with Ly$\alpha$ surface brightness as the Ly$\alpha$ emission is thought to trace the filamentary accretion of baryons from the surrounding cosmic web \citep{2010ApJ...725..633F, 2012MNRAS.419..780B, 2012MNRAS.423..344R, 2023MNRAS.523.5248B}, making gas column density an astrophysically informative target for inference. By contrast, quantities such as the dark matter distribution, gas kinematics, and ionization state, although of significant astrophysical interest, are far more difficult to infer, particularly given the limited sensitivity and spectral resolution of current observations like \cite{2023NatAs...7.1390M}. While recent investigations, such as those by \citet{2026arXiv260122677L}, have begun connecting dark-matter properties directly to Ly$\alpha$ emission by comparing simulated intensity maps under different paradigms (e.g., cold vs. warm dark matter), the predicted observational signatures remain subtle for current 10m class telescopes.

Moreover, inferring the full 3D gas distribution from Ly$\alpha$ IFU emission data (2D spatial maps with an additional spectral dimension) constitutes a highly degenerate inverse problem at the resolutions and S/N currently accessible, whereas mapping the emission cube to a line-of-sight integrated gas column-density map represents a more tractable inverse problem. We therefore adopt the gas column density as our starting point. If even this inverse problem, which is most directly constrained by the observed Ly$\alpha$ emission and substantially less degenerate than the full 3D reconstruction, cannot be solved reliably, then attempts to infer more weakly constrained IGM properties would be infeasible.

Having established this goal, the next step is to determine how simulated and observed cubes can be connected in a way that enables reliable physical inference. Several methodological pathways exist, each with different advantages and limitations, and we outline the most relevant approaches below.

Developing sufficiently realistic analytical models for Ly$\alpha$ emission from the diffuse IGM remains a significant challenge, as such models typically require strong approximations to account for the complexities of resonant scattering \citep[e.g.,][]{2025MNRAS.544.4246P}. Historically, predictions generally rely on cosmological hydrodynamical simulations followed by Monte Carlo radiative transfer (MCRT) calculations \citep[e.g.,][]{2021MNRAS.506.5129B,2023MNRAS.523.5248B}. While these simulations provide the most realistic modeling of IGM Ly$\alpha$ emission to date, the relationship between gas structure and emission is extremely complex, nonlinear, and high-dimensional \citep{2014PASA...31...40D}, making it nearly impossible to directly infer the underlying gas distribution from the emergent Ly$\alpha$ signal. This difficulty is further compounded by the intrinsically low S/N of the observations.

Machine learning (ML) offers a flexible and powerful pathway forward, and has already been applied successfully to a wide range of inverse problems in astronomy (see \citealt{2023RSOS...1021454S} for a review). ML models can learn complex, nonlinear mappings directly from data, enabling the recovery of physical quantities such as gas column density from noisy Ly$\alpha$ emission cubes even when no tractable analytic model exists. A key strength of ML is its modularity: the model can be retrained on improved or alternative Ly$\alpha$ simulations as they become available, ensuring compatibility with future theoretical developments.

Among the various ML architectures considered, convolutional approaches, including convolutional neural networks (CNNs) and CNN-based generative models, are particularly well suited to our goals. Such models have demonstrated strong performance in tasks such as deconvolution, denoising, and data reconstruction, all of which are essential components of the present problem \citep{2017MNRAS.467L.110S,2021MNRAS.503.3204V,2021ApJ...916...42W}. Motivated by these strengths, we develop a framework centered on a custom 2D CNN, the Column Density Estimator (CDE), designed to predict gas column-density maps from Ly$\alpha$ emission subcubes extracted from KCWI data.

A key feature of the CDE framework is its approximate treatment of KCWI noise, designed to reproduce selected noise properties of the observations. This enables the trained CDE to be applied directly to the processed KCWI observations, although the predictions remain conditional on the adopted physical and observational assumptions. We use IllustrisTNG100 data post-processed with simplified Ly$\alpha$ radiative-transfer calculations, together with this noise treatment, to construct a physically motivated training set designed to be observationally realistic.

Finally, we emphasize that the CDE framework is not limited to Ly$\alpha$ emission or to studies of the IGM. Instead, it represents a general inverse-modeling strategy that can be extended to a wide range of datasets and astrophysical applications. Throughout this work, we use ``CDE'' to refer to the neural network model itself, and ``CDE framework'' to denote the full framework including data generation, training, and application to simulated and observed data.

The paper is organized as follows. In Section~2, we describe the CDE model and its baseline validation. Section~3 presents an overview of the data generation and reduction procedures, as well as the training strategy. In Section~4, we apply the CDE to simulated data to evaluate its performance. Section~5 applies the trained CDE to observed data and presents the resulting column-density maps. In Section~6, we discuss the implications of our results, and Section~7 summarizes our conclusions.

\section{CNN Model Overview}

\subsection{High-level CNN Setup}

\begin{figure*}[ht]
  \centering
  \includegraphics[width=0.75\linewidth]{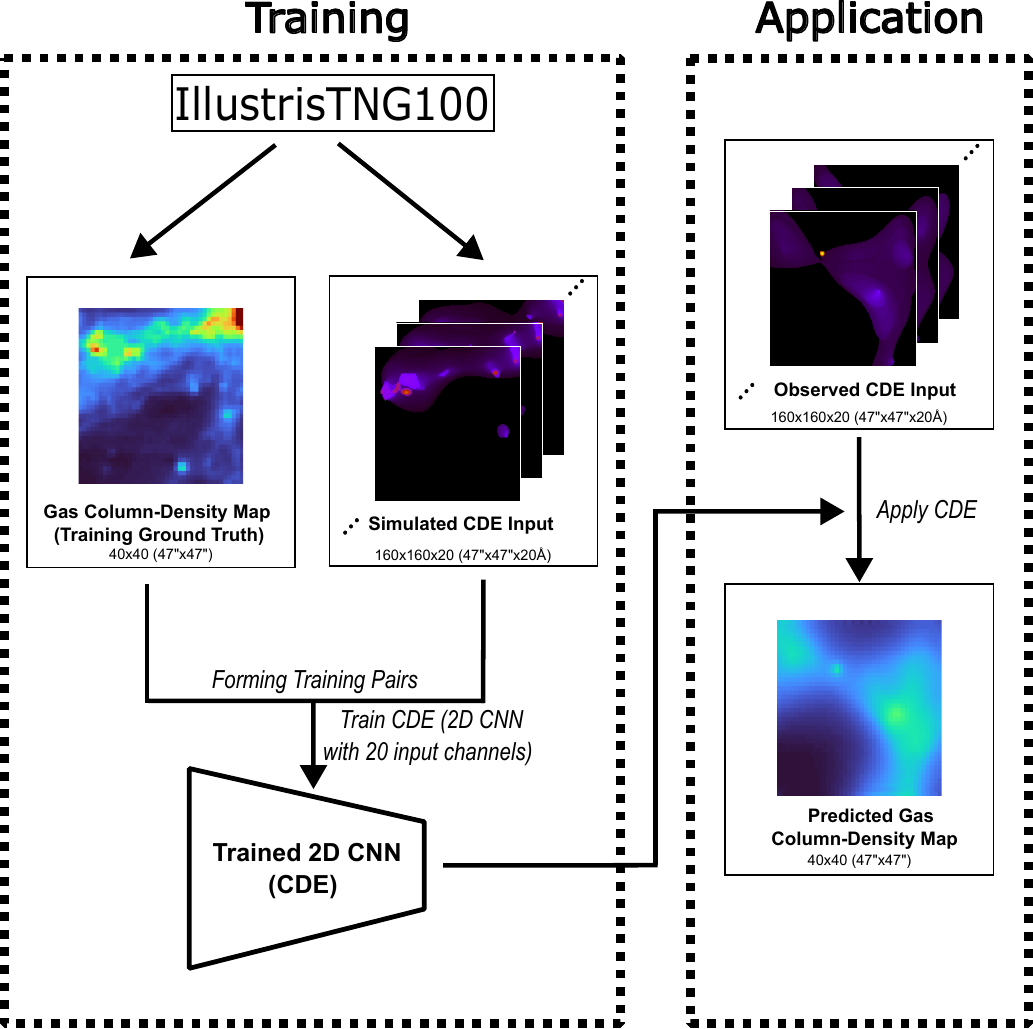}
  \caption{\textbf{Overview of the CDE framework.} Gas-cell and star-particle snapshot data from IllustrisTNG are used to generate both two-dimensional column-density maps (serving as training ground truth) and simulated CDE inputs with two spatial dimensions and one spectral dimension (serving as training input). Each input--ground-truth pair forms a training example, and the collection of such examples is used to train and test the CDE. Once trained and validated, the CDE is applied to the observed CDE inputs to predict gas column-density maps. The digital dimensions and the corresponding physical scales on the sky for each subcube and map are also indicated in the figure. The example images are included only to illustrate the workflow and do not represent the exact data products used in the final analysis.}
  \label{fig:workflow}
\end{figure*}

Guided by the workflow outlined above, we now describe the CDE and its high-level design. The CDE is a CNN designed to reconstruct coarsely sampled gas column-density maps ($40 \times 40$ pixels, with $1.16^{\prime\prime} \times 1.16^{\prime\prime}$ per pixel) from finely sampled Ly$\alpha$ CDE inputs ($160 \times 160 \times 20$ voxels, with $0.29^{\prime\prime} \times 0.29^{\prime\prime}$ per spatial pixel and $1\,\text{\AA}$ per spectral channel). Here, a voxel is the three-dimensional analogue of a pixel, corresponding to one spatial pixel along the spatial axes and one spectral channel along the wavelength axis. The network is implemented in Python using the TensorFlow deep-learning framework \citep{tensorflow2015}. As illustrated in Figure~\ref{fig:workflow}, the broader CDE framework includes training the model on simulated data designed to match the properties of the real observations and applying the trained model to observed KCWI data to reconstruct column-density maps.

The input Ly$\alpha$ subcubes and the predicted column-density maps cover the same field of view (FOV) but have different spatial samplings. In studies of Ly$\alpha$ emission from the diffuse IGM, adaptive smoothing is commonly employed to reveal low-surface-brightness structures in extremely noisy data cubes \citep[e.g.,][]{2014ApJ...786..106M,2023NatAs...7.1390M}. Such processing can introduce small-scale, noise-driven fluctuations whose detailed morphology varies between noise realizations. To focus the reconstruction on larger-scale structures, we downsample the prepared gas maps by a factor of four in each spatial dimension to create the $40 \times 40$ ground-truth maps used for training, following \citet{2021ApJ...916...42W}. The factor of four is motivated by the fact that the resulting spatial sampling is $4 \times 0.291^{\prime\prime} = 1.164^{\prime\prime}$ per pixel, slightly smaller than the adopted point-spread function (PSF) full width at half maximum (FWHM) of approximately $1.4^{\prime\prime}$. This choice reduces the map dimensions while retaining a pixel scale finer than the instrumental PSF. Gas masses are summed within each $4 \times 4$ block, equivalent to averaging the column density over the same block.

To evaluate the additional information loss introduced by spatial downsampling, we use the unmasked, PSF-convolved $160 \times 160$ gas column-density maps corresponding to all test-set regions. We downsample these maps to $40 \times 40$ and bilinearly interpolate them back to $160 \times 160$ for comparison with the original PSF-convolved maps. This test measures the combined effect of downsampling and interpolation, independently of noise injection, masking, and adaptive smoothing. We characterize the differences using the root-mean-square error (RMSE), RMSE in dex, and the structural similarity index measure (SSIM; \citealt{1284395}). SSIM measures local similarity between images, with unity indicating identical local image patches. We favor SSIM over traditional metrics such as Mean Percentage Error (MPE), which are often dominated by fluctuations in the low-signal background. By contrast, SSIM is designed to assess the local patterns of pixel intensities, making it particularly well suited for capturing the morphology of cosmic web filaments.

For each test-set region, we compute a map-level RMSE in linear units from the pixel-wise differences across the full maps. Here, the gas column density $N_{\mathrm{gas}}$ is defined as the projected gas mass per unit area divided by the proton mass, expressed as an equivalent baryonic column density in $\mathrm{cm^{-2}}$. For the logarithmic metrics, we apply a numerical floor of $10^{15}\,\mathrm{cm^{-2}}$ to both maps before taking $\log_{10}$. RMSE in dex is evaluated over pixels where the reference map has $N_{\mathrm{gas}} > 2 \times 10^{20}\,\mathrm{cm^{-2}}$. We calculate the full 2D SSIM map from the logarithmic column-density maps using a $7 \times 7$ pixel window and take its median over pixels selected by the same $2 \times 10^{20}\,\mathrm{cm^{-2}}$ reference-map threshold. We refer to this map-level statistic as \textbf{image SSIM} throughout the paper. We report the median and interquartile range (IQR) of each metric across the test set.

The resulting distributions yield a median image SSIM of $0.972^{+0.006}_{-0.007}$, a median RMSE of $6.3^{+2.8}_{-2.2} \times 10^{19}\,\mathrm{cm^{-2}}$, and a median RMSE in dex of $0.035^{+0.006}_{-0.005}$. These results indicate that fourfold downsampling preserves the broad filamentary morphology with high fidelity after PSF convolution, introducing little additional degradation of the large-scale structures of interest. Compact peaks and finer spatial details remain affected by the resolution reduction. Given that we do not aim to recover structure on scales below the instrumental PSF, these results support the adopted downsampling factor.

As discussed in Section~\ref{sec:intro}, we first train the CDE to predict the gas column density. This is because it is directly related to Ly$\alpha$ emission and effectively traces large-scale structure. Furthermore, when expressed as an equivalent line-of-sight distance assuming Hubble flow, the spectral resolution is much coarser than the transverse spatial resolution, making reconstruction of the full three-dimensional gas distribution highly degenerate. We therefore focus on inferring projected gas column-density maps from Ly$\alpha$ emission, a less degenerate inverse problem. Although the following discussion focuses on gas column density, the same network architecture and training strategy can be readily extended to other physical quantities. As an example, we will later demonstrate the application of the CDE framework to predicting dark-matter column-density maps.

While three-dimensional CNNs are commonly used for cube-based data, they are computationally expensive because convolution is performed over an additional dimension. Instead of inputting a complete $160 \times 160 \times 20$ subcube, we treat the data as 20 separate $160 \times 160$ wavelength slices. These are provided in 20 channels to a two-dimensional CNN.  To assess the computational trade-off, we performed controlled tests in which the 2D and 3D networks employed nearly identical architectures. These networks were trained using the same data and optimization settings and differed only in whether the convolutional layers operated in two or three dimensions. In these tests, the 2D approach reduced training time by approximately an order of magnitude while yielding comparable reconstruction performance. This parity is likely driven by the extreme resolution anisotropy of our IFU data: for Ly$\alpha$ at $z \approx 2.3$, the equivalent line-of-sight scale inferred from the spectral line-spread function (LSF), assuming Hubble flow, is roughly two orders of magnitude larger than the transverse physical scale corresponding to the spatial PSF. We therefore adopt the 2D CNN for this work.

During training and prediction, the input Ly$\alpha$ subcubes and the ground-truth column-density maps are both log-transformed to accommodate the high dynamic range of IGM emission and gas distribution. The Ly$\alpha$ subcubes, initially in units of $\mathrm{erg~s^{-1}~cm^{-2}~arcsec^{-2}}/\text{\AA}$, are first converted to $\mathrm{LU}/\,\text{\AA}$ and then to a logarithmic scale of $\log_{10}(x/(\mathrm{LU}/\text{\AA}) + 1)$. Here, LU denotes a line unit, defined as $1~\mathrm{LU} = 1~\mathrm{photon~s^{-1}~cm^{-2}~sr^{-1}}$. We adopt the fixed conversion $1~\mathrm{LU} \simeq 1.22 \times 10^{-22}~\mathrm{erg~s^{-1}~cm^{-2}~arcsec^{-2}}$. Similarly, the ground-truth maps are expressed as gas mass per pixel, $M$, in grams and transformed using $\log_{10}(M/(10^{40}\,\mathrm{g}) + 1)$.

This transformation is physically motivated by the expected power-law relationship between Ly$\alpha$ surface brightness and gas column density. The normalization factor of $10^{40}\,\text{g}$ is chosen to ensure that the log-transformed ground-truth values remain within an optimal range for the neural network optimization process. Furthermore, the addition of a constant unity term provides numerical stability in regions of low signal or zero flux. This acts as a floor that prevents the logarithmic function from diverging. For the final analysis and comparison with the literature, the predicted outputs are back-transformed and converted into the standard units of $\text{cm}^{-2}$.

\subsection{CNN Architecture}

\begin{figure*}[ht]
    \centering
    \includegraphics[width=0.6\linewidth]{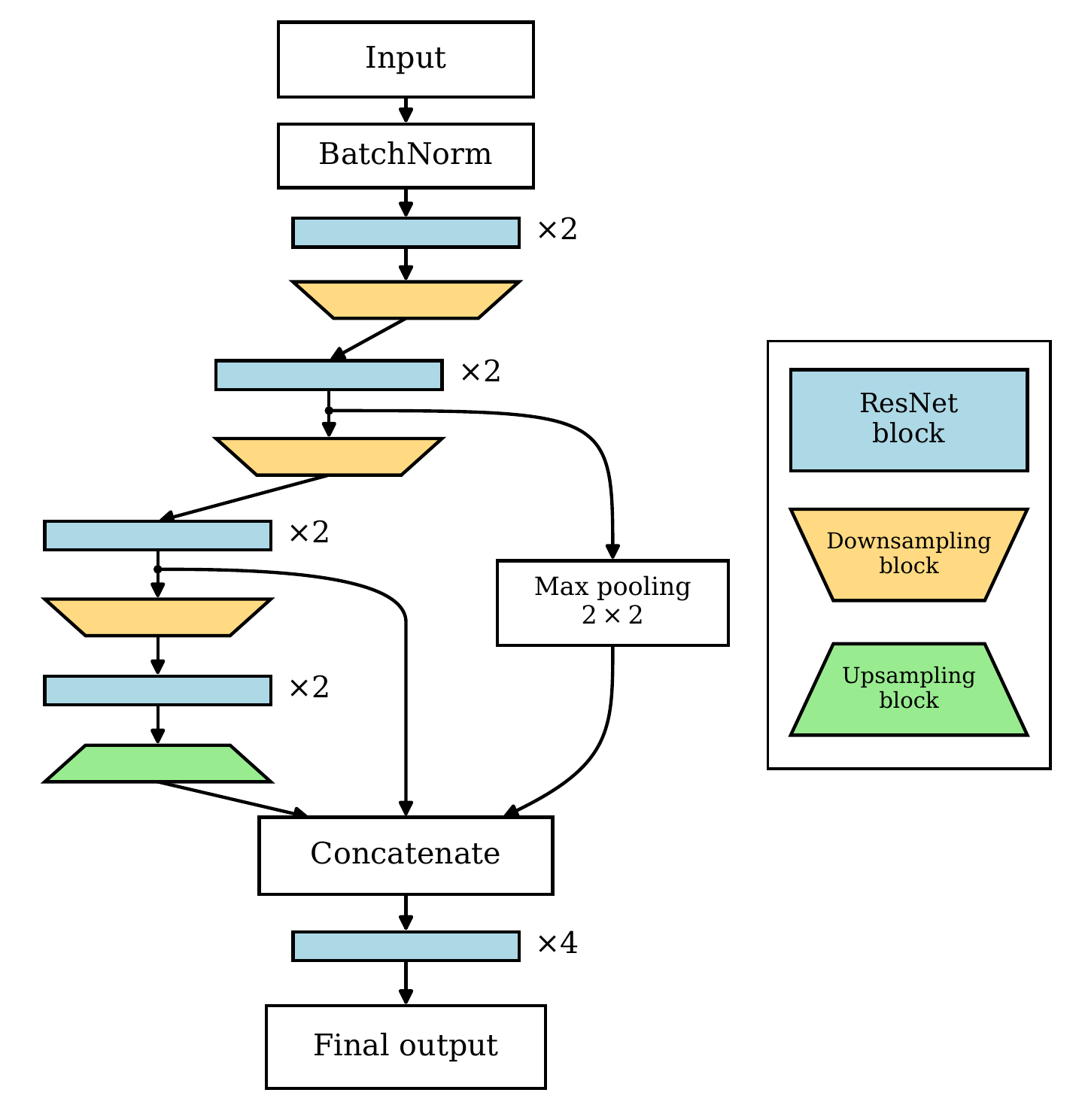}
    \caption{\textbf{CDE architecture.} The model uses a shallow U-Net configuration combining residual blocks with downsampling and upsampling pathways. Features from two earlier stages are concatenated with the upsampled features on a $40 \times 40$ pixel grid. The rightmost $80 \times 80$ skip branch is reduced to $40 \times 40$ using $2 \times 2$ max pooling before concatenation. Numbers (e.g., $\times$2) denote the count of consecutive blocks in each section.}
    \label{fig:cnn_architecture}
\end{figure*}

The general architecture of the CDE is inspired by HINet \citep{2021ApJ...916...42W} and is illustrated in Fig.~\ref{fig:cnn_architecture}. While Fig.~\ref{fig:workflow} shows the high-level schematic of how the CDE is trained and applied to the observed data, Fig.~\ref{fig:cnn_architecture} details the internal neural network design. Our design incorporates ResNet-based residual blocks \citep{2016arXiv160305027H} and downsampling blocks with average-pooling shortcuts. As shown in the schematic, the input is first processed through a batch-normalization layer before entering a series of residual blocks. We then include a shallow U-Net component to enhance the model's ability to reconstruct filamentary structures \citep{2015arXiv150504597R}.

Following the approach of \citet{2021ApJ...916...42W}, we adopt a shallow U-Net design with the aim of balancing model expressiveness and robustness. To retain spatial information from earlier stages, we introduce skip connections that concatenate earlier-layer features with the upsampled features. These connections help preserve spatial details during reconstruction. A more detailed description of the network layers is provided in Appendix~\ref{app:cde_details}. The effectiveness of this design is assessed through the noiseless baseline and subsequent applications to simulated and observed data.

\subsection{Noiseless Baseline}

We first validate the network architecture by training a variant of the CDE, hereafter referred to as CDE-noiseless. This version utilizes the same architecture and loss function as the main CDE but is trained and applied to noiseless Ly$\alpha$ emission subcubes. In this idealized case, all input subcubes and ground-truth column-density maps are unmasked to avoid introducing artificial rectangular features in the absence of adaptive smoothing. However, the PSF/LSF convolution and empirical resonant-scattering correction are applied in the same manner as in the noisy case to maintain physical consistency.

\begin{figure*}[ht]
    \centering
    \includegraphics[width=0.7\linewidth]{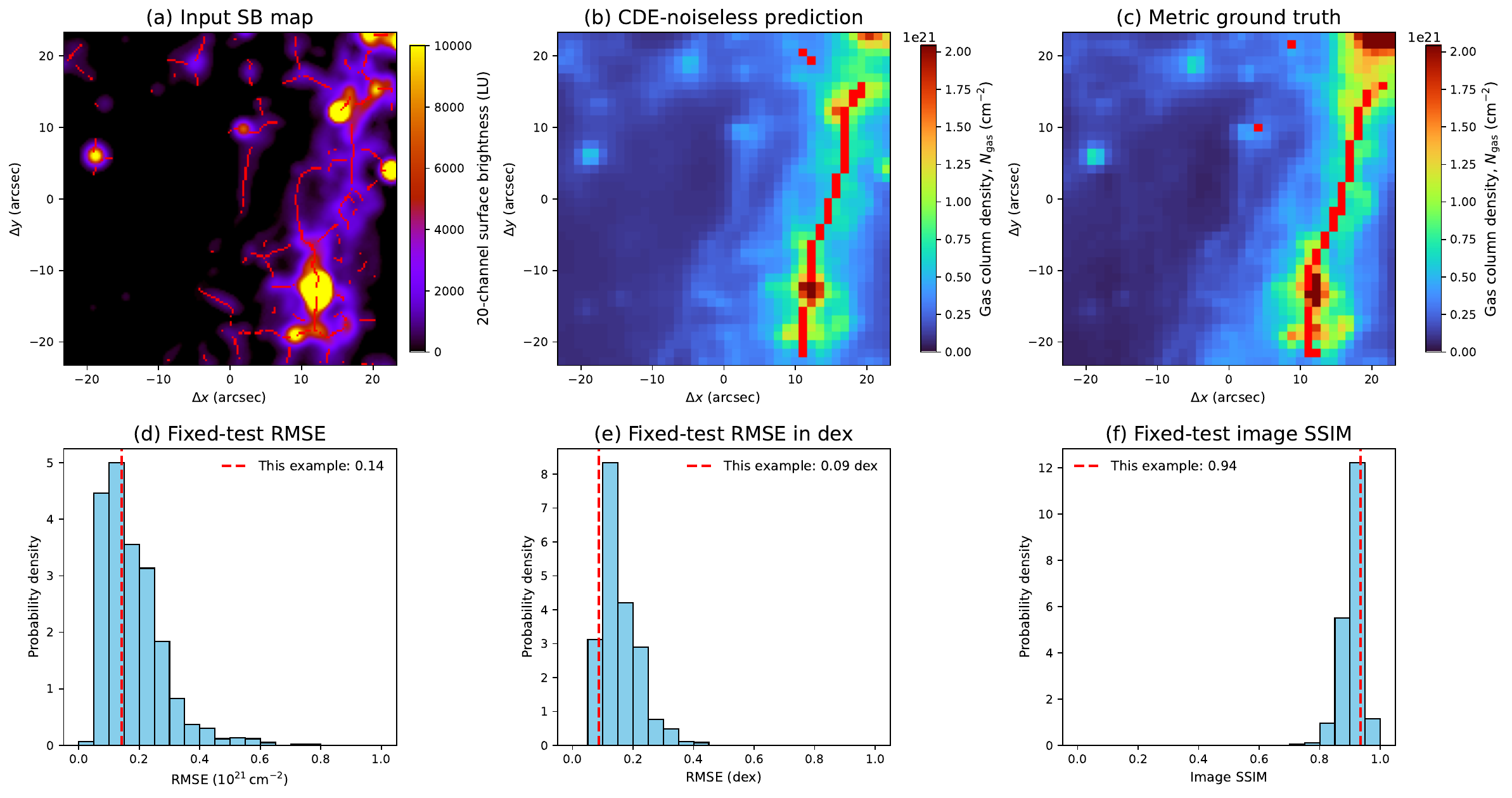}
    \caption{\textbf{CDE-noiseless.} Representative test-set example and performance statistics across all 1200 noiseless test-set regions. \textbf{Top row:} \textbf{(a)} Ly$\alpha$ surface-brightness map integrated over the 20 spectral channels of the noiseless CDE input; \textbf{(b)} CDE-noiseless-predicted gas column-density map; and \textbf{(c)} unmasked ground-truth gas column-density map. Red filament spines are obtained using SCMS. \textbf{Bottom row:} \textbf{(d)} linear RMSE; \textbf{(e)} RMSE in dex; and \textbf{(f)} image SSIM. In all three histograms, the vertical dashed line marks the value for the illustrated example. \textbf{CDE-noiseless reconstructs filamentary morphology with high fidelity under the adopted simulation and radiative-transfer assumptions.}}
    \label{fig:cde_noiseless_example}
\end{figure*}

As shown in the top row of Figure~\ref{fig:cde_noiseless_example}, the network successfully recovers fine-scale structural features in the absence of observational noise, although some underestimation remains in extremely dense regions, particularly within the nodes of the cosmic web. These remaining discrepancies may reflect degeneracies in the emission--column-density relation within these nodes or limitations of the trained CDE.

To further assess the recovery of filamentary structures, we follow the approach of \citet{2023NatAs...7.1390M} and apply the subspace-constrained mean shift (SCMS) algorithm \citep{2015MNRAS.454.1140C} to the Ly$\alpha$ surface-brightness (SB) maps integrated over the retained $20\,\text{\AA}$ interval and to both the predicted and ground-truth column-density maps. The resulting filament spines are shown in red in the top row of Figure~\ref{fig:cde_noiseless_example}. The SCMS algorithm iteratively shifts points toward ridges in a smoothed density field without prescribing the filament shape. The agreement between the spines extracted from the predicted and ground-truth maps illustrates the recovery of filamentary morphology by CDE-noiseless.

Performance statistics are summarized in panels (d)--(f) of Figure~\ref{fig:cde_noiseless_example}, which display the distributions of linear RMSE, RMSE in dex, and image SSIM over the 1200 test-set regions. We compute these metrics by comparing each predicted column-density map with its corresponding unmasked $40 \times 40$ ground-truth map, following the metric definitions introduced above. We report a median image SSIM of $0.913^{+0.017}_{-0.021}$, a median RMSE of $1.55^{+0.75}_{-0.50} \times 10^{20}\,\mathrm{cm^{-2}}$, and a median RMSE in dex of $0.138^{+0.051}_{-0.030}$. These results demonstrate that filamentary morphology can be reconstructed with high fidelity in the absence of observational noise under the adopted simulation and radiative-transfer assumptions.

These results establish a baseline for reconstruction in the idealized noiseless case under the adopted assumptions. Having established this baseline, we can now turn to more realistic observational conditions, train the model on simulated data that incorporate these effects, and finally apply the trained network to observed data.

\section{Data Preparation and Training}

\subsection{Data Preparation Overview}

To train the CDE, we generate a large set of simulated CDE inputs that resemble the observed data. It is difficult to determine a priori how many training examples are required, so we assess how performance varies with training-set size within the tested setup.

\begin{figure*}[p]
  \centering
  \includegraphics[width=0.8\linewidth]{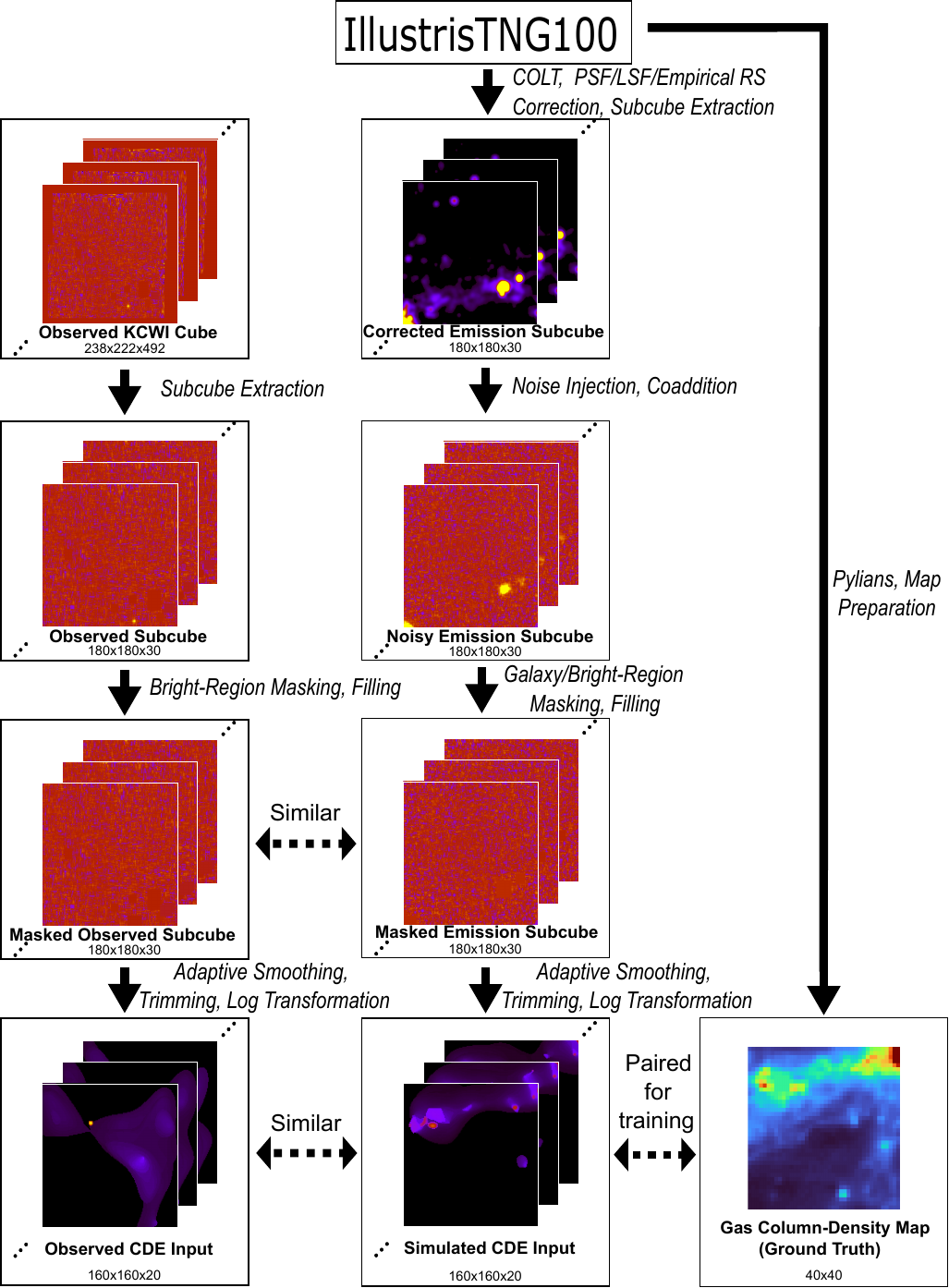}
  \caption{\textbf{Flowchart for data generation.} The left column illustrates the reduction of observed KCWI data into CDE inputs, the middle column shows the generation of simulated CDE inputs, and the right column depicts the generation of column-density maps used as ground truth. The array dimensions of the data at each step are also indicated. The example images are included only to illustrate the workflow and do not represent the exact data products used in the final analysis.}
  \label{fig:flowchart}
\end{figure*}

Figure~\ref{fig:flowchart} illustrates the overall data-preparation workflow, including observed data reduction (left), simulated data generation (middle), and ground-truth column-density map construction (right). We provide a high-level overview of the data-preparation process here, with additional details presented in Appendix~\ref{app:data_generation} and the computational cost summarized in
Appendix~\ref{app:computational_cost}.

For the observed data, we utilize observations of two protoclusters: CC2.2 (Field A, $z \sim 2.2$; \citealt{2020ApJ...892....8D}) and Hyperion (Field B, $z \sim 2.45$; \citealt{2018A&A...619A..49C}), both of which were targeted using KCWI on Keck II \citep{2018ApJ...864...93M}. We begin with rectified KCWI data cubes processed from the raw 2D detector images obtained during the observations using the KCWI data reduction pipeline (DRP; \citealt{2023ascl.soft01019N}). The final observed cubes result from coadding multiple exposures taken with various spatial offsets and position angles (PAs). We also mask galaxies identified in the zCOSMOS-deep survey \citep{2007ApJS..172...70L} to minimize foreground contamination. These cubes achieve a nominal sampling of $0.291'' \times 0.291'' \times 1\,\text{\AA}$, a full FOV of $\sim 60'' \times 60''$, and a wavelength range of 3838--4329~\AA. Because the coaddition process combines exposures taken at different PAs, significant spatial correlations are introduced between neighboring pixels, which we account for in our modeling. Nod-and-shuffle (NAS) mode was used during observation to improve sky subtraction, reducing the residual background level to a surface brightness of approximately $3\times10^{-21}\,\mathrm{erg\,cm^{-2}\,s^{-1}\,\text{\AA}^{-1}\,arcsec^{-2}}$.

From these data cubes, we extract $30\,\text{\AA}$-wide spectral subcubes centered on the emission regions examined by \citet{2023NatAs...7.1390M}. We also crop out empty padding and edge regions, reducing the spatial dimensions from $238 \times 222$ to $180 \times 180$ pixels. The resulting $180 \times 180 \times 30$ observed subcubes correspond to a FOV of $\sim 53'' \times 53''$. This reduction from the full $60'' \times 60''$ FOV is necessary because edge regions often have lower exposure times and consequently lower S/N than central regions and are therefore excluded during subcube extraction.

These subcubes are subsequently processed with per-subcube bright-source masking and then adaptively smoothed following the methodology of \citet{2023NatAs...7.1390M}. To reduce smoothing-kernel boundary effects, we then trim the spatial dimensions to $160 \times 160$ and discard the first and last five spectral channels. The resulting $160 \times 160 \times 20$ subcubes serve as the observed CDE inputs. The retained $20\,\text{\AA}$ interval corresponds to a velocity extent of approximately $1,500\,\mathrm{km\,s^{-1}}$ at $z=2.3$. This interval exceeds the typical kinematic broadening of IGM filaments ($\lesssim 500\,\mathrm{km\,s^{-1}}$), providing sufficient coverage of the emission-line profile even after accounting for an LSF with a $\sim5\,\text{\AA}$ FWHM. In total, we obtain ten CDE inputs containing significant emission, with five drawn from Field A and five from Field B. We also select one subcube without detected emission from each field as a control.

For simulated data generation, we begin with TNG100 simulation data from the IllustrisTNG family \citep{2018MNRAS.473.4077P,2018MNRAS.475..624N,2018MNRAS.475..676S,2018MNRAS.477.1206N,2018MNRAS.480.5113M,2019ComAC...6....2N} and extract gas and stellar snapshot data centered on the densest environments. Using the Cosmic Ly$\alpha$ Transfer code \citep[\colt;][]{2015MNRAS.449.4336S}, we convert these data into noiseless Ly$\alpha$ emission cubes. For each of the 25 pointing centers, we use five rotations, including the unrotated case, and generate emission cubes along the six Cartesian viewing directions ($\pm x$, $\pm y$, and $\pm z$), yielding $25 \times 5 \times 6 = 750$ full-field emission cubes. To enable large-volume data generation, most cubes are generated in intrinsic-emission mode, without resonant scattering (RS), while 30 matched intrinsic/full-RS cube pairs are used to derive an empirical voxel-level surface-brightness correction. After convolution with the KCWI-matched PSF and LSF, this correction is applied to the intrinsic cubes, yielding 750 corrected noiseless emission cubes. We then extract 6000 subcubes from these full-field cubes, each with a FOV of $\sim53^{\prime\prime}\times53^{\prime\prime}$, matching that of the observed subcubes.

For each selected subcube, we simulate observational sampling at the KCWI spatial sampling and add Poisson and read noise. A simulated coaddition procedure is then applied to reproduce the pixel-to-pixel correlations present in the observed KCWI data, yielding 6000 noisy emission subcubes. We then apply processing steps similar to those used for the observed data, including galaxy/bright-source masking, adaptive smoothing, and spatial and spectral trimming, followed by random rotation and spatial flipping as a data-augmentation strategy. This produces simulated CDE inputs with the same dimensions and broadly matched processing as the observed CDE inputs. The resulting simulated CDE inputs have dimensions of \(160\times160\times20\). Further details and differences between the two processing procedures are described in Appendix~\ref{app:data_generation}.

Ground-truth generation is illustrated on the right-hand side of the flowchart. We utilize \texttt{Pylians} \citep{Pylians} to convert the gas-cell and dark-matter particle data into two-dimensional column-density maps using the cloud-in-cell (CIC) mass-assignment algorithm. These maps serve as ground-truth targets, with each map paired with its corresponding simulated CDE input for model training and evaluation.

\subsection{Training Strategy (Bagging)}

To stabilize predictions and improve robustness, we employ bootstrap aggregating (bagging; \citealt{Breiman:1996zz}). Bagging reduces variance by training an ensemble of models on different bootstrap-resampled versions of the training set, such that each base model is exposed to a different distribution of examples. The final prediction is obtained by taking the median over the ensemble, which effectively suppresses model-specific noise and mitigates overfitting. Although the ensemble could, in principle, be trained in parallel to reduce wall-clock time (the total elapsed real-world time), we train the models sequentially due to limited computational resources. For the training of each individual base model, we use the loss function described in the following subsection and apply an early-stopping criterion to prevent overfitting.

We first define a held-out test set by randomly selecting three opposite-view pairs of full-field emission cubes per pointing center. Because cubes viewed from opposite Cartesian directions can contain very similar structures, both cubes in each pair are assigned to the same partition. To balance the sampling across viewing orientations, we select one pair along each Cartesian axis, with the three pairs drawn from three different rotations. Including all eight subcubes from each selected full-field cube gives $25 \times 6 \times 8 = 1200$ test inputs from the full set of 6000 simulated CDE inputs. This test set is excluded entirely from training and validation and is used for performance evaluation. However, the test, training, and validation sets still share the same underlying pointing centers. The remaining 4800 simulated CDE inputs constitute the pool for training and validation.

For each base model in the bagging ensemble, we randomly select two opposite-view pairs per pointing center for validation, taking one pair from each of the two rotations not used for test-set selection and requiring the pairs to lie along different Cartesian axes. These four full-field cubes per pointing center provide 800 validation inputs. We then sample the remaining 4000 subcubes \emph{with replacement} separately within each pointing center, producing 4000 training entries. For each subsequent base model, we draw a new validation selection and bootstrap sample while keeping the test set fixed.

Once all 10 base models are trained, we aggregate their outputs by taking the pixel-wise median across the ensemble in the logarithmic output space to form the final prediction. We adopt the pixel-wise median as the aggregation statistic because it is particularly robust to outliers among individual models. The resulting ensemble prediction is then evaluated on the held-out test set.

\subsection{Loss Function}

To optimize the trade-off between recovering physical scales and reconstructing filamentary structures, we utilize a composite loss function defined as:$$L = \alpha \mathcal{L}_{\mathrm{MAE}} + (1 - \alpha) (1 - \mathrm{SSIM_{unmasked}}) \, .$$ Both loss components are evaluated on the logarithmically transformed maps.

\begin{figure}[htbp]
    \centering
    \includegraphics[width=\columnwidth]{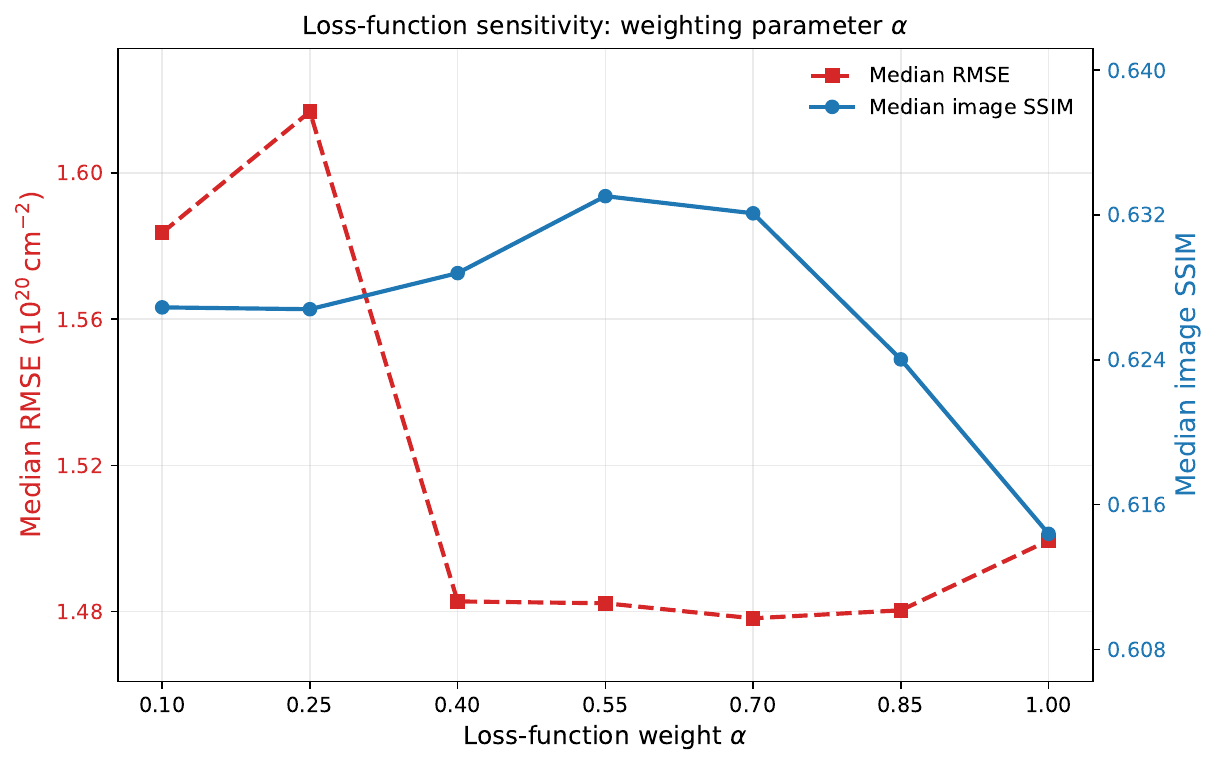}
    \caption{\textbf{Prediction error as a function of the loss function weight $\alpha$.} For each $\alpha$, we evaluate the pixel-wise median predictions of three base models on the same 1200 test inputs. The solid line represents median values for image SSIM (right axis) and the dashed line shows median RMSE (left axis), with the latter expressed in units of $10^{20}$~cm$^{-2}$. \textbf{Among the tested values, $\alpha = 0.55$ combines the highest median image SSIM with a nearly minimal median RMSE and is therefore adopted as the default.}}
    \label{fig:alpha_check}
\end{figure}

Here, $\mathrm{SSIM_{unmasked}}$ represents SSIM averaged over local image windows and then across the batch, without the additional density selection used for evaluation. We utilize this unmasked version of SSIM in the loss function to ensure the network receives continuous gradient feedback across the entire FOV. While the evaluation image SSIM is summarized only over high-density regions, the training loss incorporates low-density regions to prevent background artifacts and maintain the integrity of the local window statistics essential for SSIM calculations. We employ Mean Absolute Error (MAE) rather than RMSE for the pixel-wise component, as the combination of $\mathcal{L}_1$ loss and SSIM is widely preferred in image reconstruction tasks to prevent the blurring of fine structural details (e.g., \citealt{2015arXiv151108861Z}). This composite approach penalizes the model for both absolute pixel-value deviations and morphological mismatches. The sensitivity of model performance to the $\alpha$ parameter is shown in Fig.~\ref{fig:alpha_check}. To highlight the primary trends, we plot only the median values for RMSE and image SSIM, omitting the IQR for visual clarity. For each tested value of $\alpha$, we train three base models and combine their predictions using the pixel-wise median. We then evaluate the resulting ensemble predictions on the same fixed 1200 test inputs using the metrics described in Section 2.1. Corresponding ensemble members use the same training and validation selections across different values of $\alpha$.

Among the tested values, $\alpha = 0.55$ yields the highest median image SSIM, while $\alpha = 0.7$ yields the lowest median RMSE, although the median RMSE varies little over $\alpha = 0.55$--$0.85$. We adopt $\alpha = 0.55$ because it combines the highest median image SSIM with a nearly minimal median RMSE. Unless otherwise specified, all CDE models in this paper are trained using this value.

\subsection{Training Set Size Test}

We evaluate the sensitivity of model performance to training-set size using two subsampling procedures. For each combination of subsampling procedure and training-set size, we train a ten-model ensemble with $\alpha = 0.55$ and evaluate its pixel-wise median prediction on the same fixed set of 1200 test inputs. The standard CDE uses 4000 bootstrap training entries and 800 validation inputs per base model, drawn from the remaining 4800 simulated CDE inputs available for training and validation.

In the first ``crop thinning'' procedure, we retain all 24 remaining full-field cubes per pointing center but select only 4, 2, or 1 of the 8 subcubes from each cube, compared with all 8 in the standard configuration. We use a fixed random ordering of the 8 subcubes so that the smaller subsets are contained within the larger ones. For each base model, the same two opposite-view pairs (4 full-field cubes) per pointing center are assigned to validation across all sizes, while the number of retained subcubes per cube varies with training-set size. Following the bootstrap procedure described above, these configurations provide 2000, 1000, and 500 bootstrap training entries per model, with 400, 200, and 100 validation inputs, respectively.

In the second ``group thinning'' procedure, we retain all 8 subcubes from each selected full-field cube but reduce the number of selected full-field cubes per pointing center. For the configuration with 2000 bootstrap training entries, we retain 5 opposite-view pairs (10 full-field cubes) per pointing center for training, with 1 pair per rotation, and 1 opposite-view pair (2 full-field cubes) for validation. For the configuration with 1000 bootstrap training entries, we retain 2 complete opposite-view pairs plus 1 direction from a third pair selected from the preceding training selection, giving 5 full-field cubes per pointing center. For validation, we retain 1 direction from the previously selected validation pair. The unused opposite-view cubes are excluded, so no opposite-view pair crosses partitions. The smaller selections remain subsets of the larger ones. Both procedures share the standard configuration with 4000 bootstrap training entries and 800 validation inputs per base model.

\begin{figure}
    \centering
    \includegraphics[width=\columnwidth]{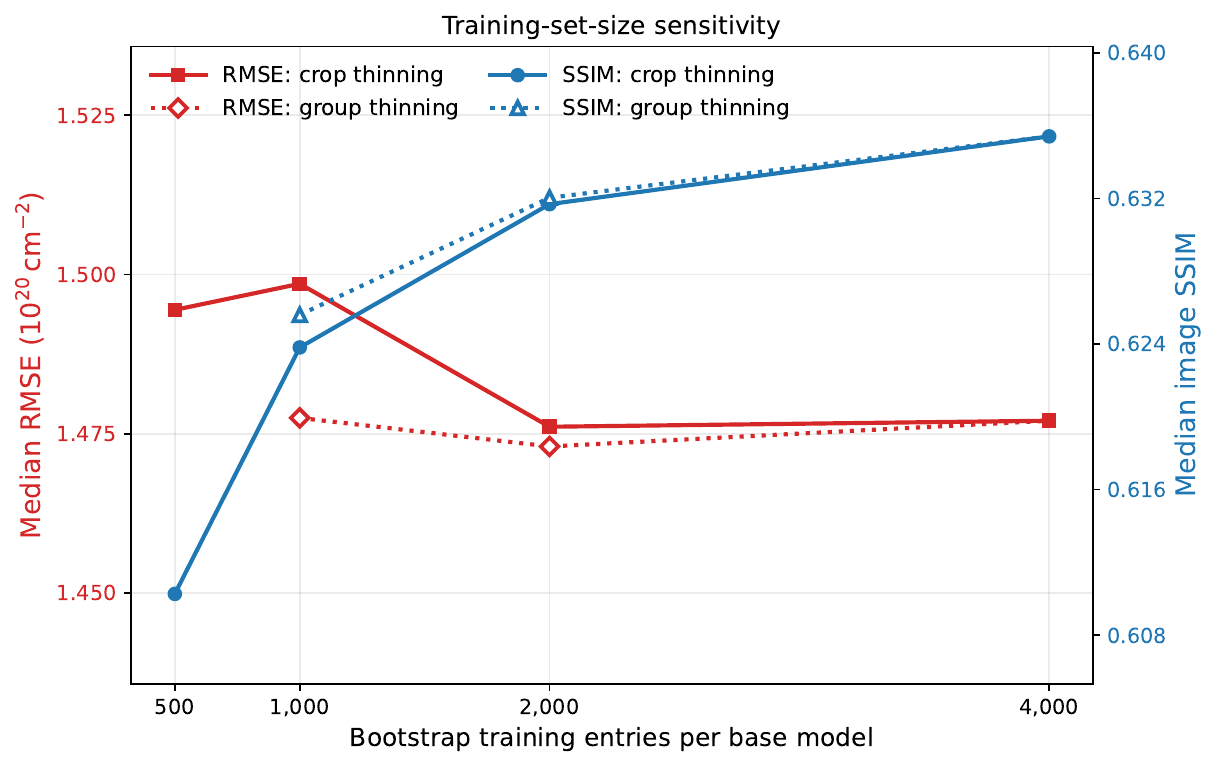}
    \caption{\textbf{Prediction error as a function of training-set size.} For each configuration, we evaluate the pixel-wise median predictions of ten base models on the same 1200 test inputs. The curves show median image SSIM (right axis) and median RMSE (left axis), with the latter expressed in units of $10^{20}$~cm$^{-2}$. Solid and dotted lines represent crop thinning and group thinning, respectively. Training-set size counts bootstrap training entries per base model. \textbf{The two procedures yield similar performance at matched training-set sizes, with only modest gains in image SSIM between 2000 and 4000 bootstrap training entries.}}
    \label{fig:training_size_check}
\end{figure}

The median values for image SSIM and RMSE are shown in Fig.~\ref{fig:training_size_check}. Image SSIM improves monotonically with increasing training-set size, while RMSE varies only slightly. The two subsampling procedures yield similar performance at matched training-set sizes. Given the modest gains in image SSIM between 2000 and 4000 bootstrap training entries and the computational cost of further doubling the training-set size, we adopt the standard configuration of 4000 bootstrap training entries per base model as a practical choice for this proof-of-concept study.

\begin{figure*}[!t]
    \centering
    \includegraphics[width=0.7\linewidth]{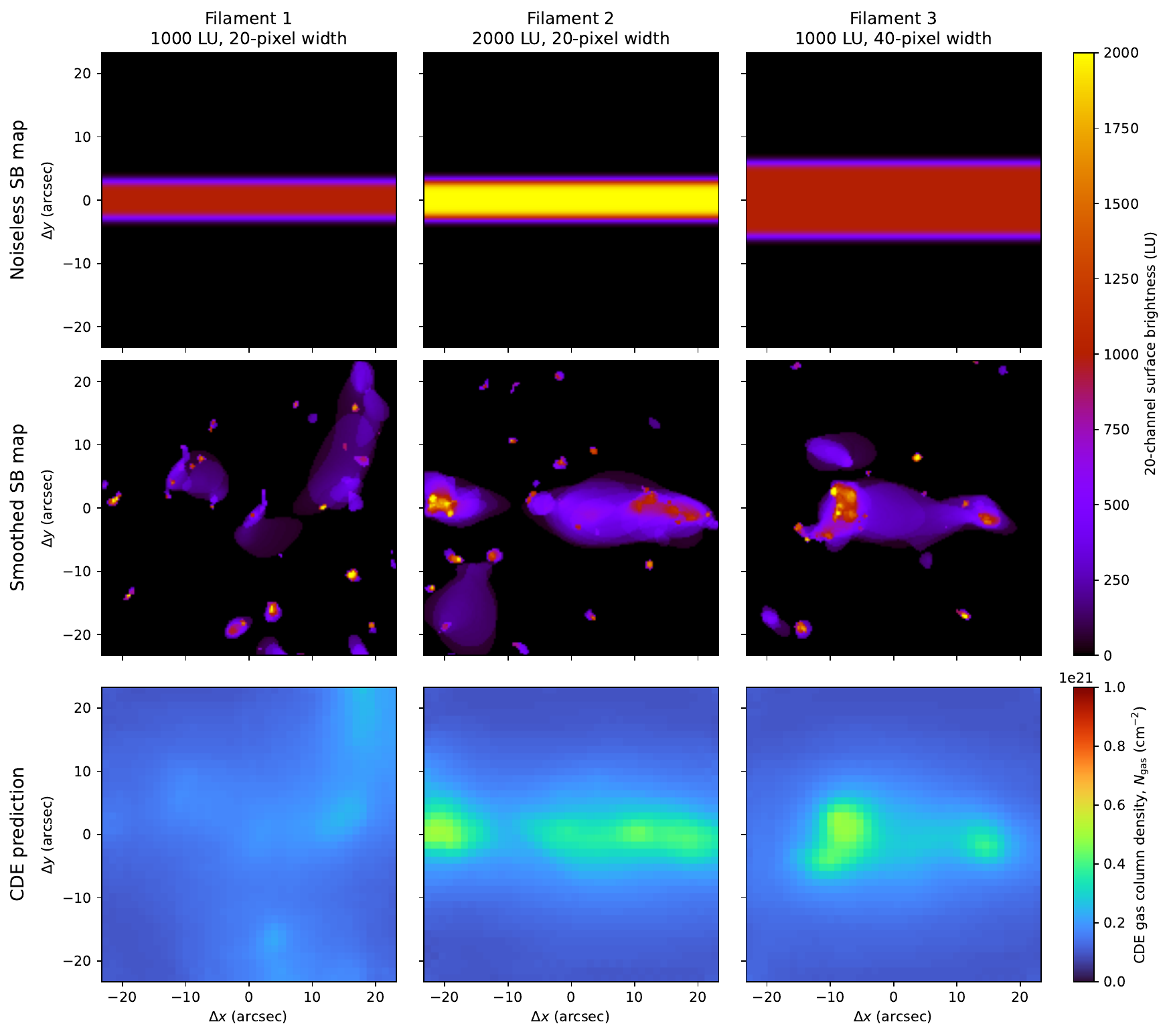}
    \caption{\textbf{Simple Structure Test: }tests of adaptive smoothing and the CDE on toy filament subcubes. Rows show (top to bottom): the noiseless surface-brightness map after PSF and LSF convolution, integrated over the central 20 spectral channels; the noise-injected, adaptively smoothed surface-brightness map integrated over the same channels; and the CDE-predicted gas column-density map. Columns represent three toy cases: \textbf{Filament 1}, a narrow filament ($1000~\mathrm{LU}$ brightness, $20~\mathrm{pixel}$ width); \textbf{Filament 2}, the same structure with doubled surface brightness; and \textbf{Filament 3}, the same structure with doubled width. The adaptively smoothed map for Filament 1 appears noisy and distorted, while the maps for Filament 2 and Filament 3 show clearer but still patchy and distorted filamentary structures. Correspondingly, the CDE reconstructions only yield clearly visible filamentary structures for Filaments 2 and 3, with some distortions introduced during the adaptive smoothing process. \textbf{Broader or brighter input filaments produce clearer filamentary structures after adaptive smoothing and in the CDE predictions.} }
    \label{fig:simple_structure}
\end{figure*}

\section{Applications I: Simulated Tests}

\subsection{Simple Structure Tests}

To assess the behavior of the complete CDE framework, we first evaluate adaptive smoothing and the CDE on simple input subcubes containing toy filamentary structures. These tests provide a baseline for the model's sensitivity to surface brightness and spatial scale.

As shown in Fig.~\ref{fig:simple_structure}, Filament 1, a narrow toy filament with a uniform line-integrated surface brightness of $\sim$1000~LU across a width of 20~pixels and a Gaussian spectral profile with intrinsic FWHM $7\,\text{\AA}$, is barely visible after PSF/LSF convolution, noise injection, and adaptive smoothing. Doubling the surface brightness (Filament 2) or filament width (Filament 3) results in a much cleaner and more continuous filamentary structure after adaptive smoothing, although the recovered structures in both Filaments 2 and 3 remain patchy.

We then apply the CDE to these adaptively smoothed subcubes. As expected, the CDE predicts clear but distorted filamentary structures for Filament 2 and Filament 3, while the reconstructed structure in Filament 1 remains weak and does not clearly trace the input filament. These examples illustrate the sensitivity of both adaptive smoothing and the CDE to filament surface brightness and width, with broader or brighter input filaments producing clearer filamentary structures at both stages.

\subsection{Application to Held-out Test Set}

We first provide a general assessment of the CDE performance on the held-out test set, with representative examples shown in Appendix~\ref{app:cnn_examples}. To ensure clarity regarding our terminology, we note that for the masked column-density and surface-brightness maps, ``masked'' refers to the masking of bright compact sources followed by local-median filling. In the definition of the image SSIM, ``masked'' means that the ground-truth column-density threshold selects which local SSIM values enter the per-map median.

To further evaluate the reliability of the CDE predictions on noisy simulated inputs, we illustrate a representative reconstruction of a filamentary structure in the top row of Fig.~\ref{fig:cde_noisy_example}. This example corresponds to the same test-set example shown in Fig.~\ref{fig:cde_noiseless_example}, enabling a direct comparison between noiseless and noisy inputs. Although the prediction derived from noisy, masked inputs is coarser than the noiseless case, it successfully reproduces filamentary features that align well with the ground truth, as demonstrated by the red filament spines calculated with SCMS. We observe an underestimation of dense nodes similar to that seen in the noiseless case.

We further quantify the performance using RMSE, RMSE in dex, and image SSIM, computed by comparing the predicted and ground-truth column-density maps across the noisy, masked test set (Fig.~\ref{fig:cde_noisy_example}(d)--(f)). We report a median RMSE of $1.48^{+0.42}_{-0.27} \times 10^{20}$~cm$^{-2}$, a median RMSE in dex of $0.206^{+0.055}_{-0.038}$, and a median image SSIM of $0.635^{+0.072}_{-0.088}$. Notably, the linear RMSE is marginally lower than in the noiseless case, potentially due to the masking and local-median filling which could suppress high-density peaks. While the image SSIM is lower than in the noiseless case, broad filamentary structures remain clearly visible in the example prediction. These results demonstrate the recovery of broad filamentary structures in noisy simulated data under the adopted assumptions.

\subsection{Comparison to Empirical Scaling Model}

We compare predictions from the CDE with those from a simple empirical scaling model based on pixel-wise power-law fitting. In this empirical approach, we first retain 8-connected emission regions with surface brightness $\geq 100$~LU containing at least 50 spatial pixels, and average the surface-brightness maps over $4 \times 4$ blocks. For each of the 1200 test maps, we fit a linear relation between the retained surface brightness and the corresponding gas column density in logarithmic space. We then average the fitted parameters across the maps to obtain a single relation, which is applied independently to each pixel with retained emission.

\onecolumngrid
\begin{figure}[!b]
    \setlength{\hsize}{\textwidth}
    \setlength{\linewidth}{\textwidth}
    \centering
    \includegraphics[width=0.7\linewidth]{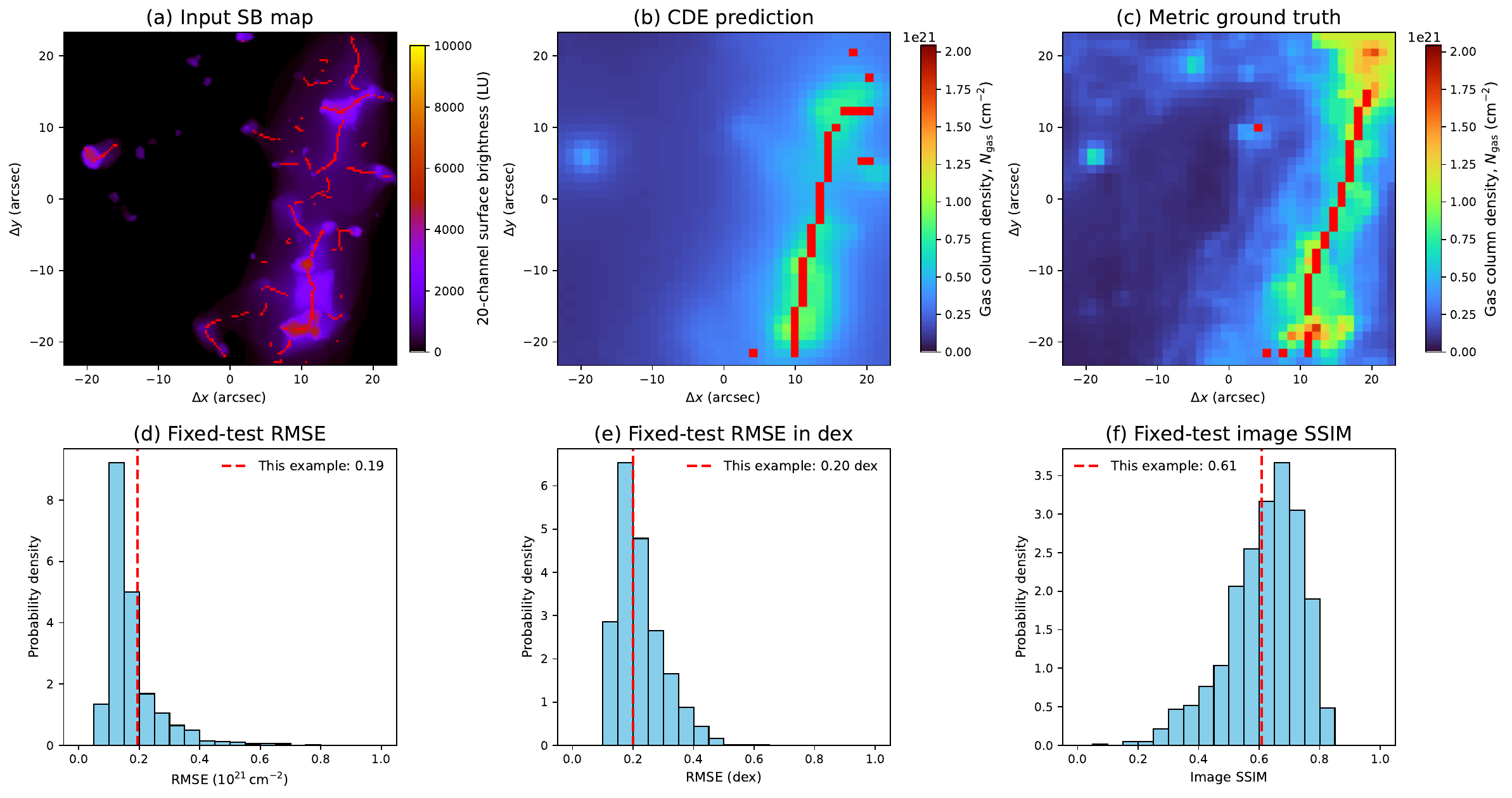}
    \caption{\textbf{CDE on Test Set:} representative test-set example for the CDE with noisy, masked inputs and error statistics over the full test set. 
    \textbf{Top row}: 
    \textbf{(a)} the input Ly$\alpha$ surface-brightness map; 
    \textbf{(b)} the CDE-predicted gas column-density map; 
    \textbf{(c)} the masked ground-truth column-density map. Red filament spines are calculated using the SCMS algorithm. 
    \textbf{Bottom row}:
    \textbf{(d)} the RMSE histogram for the full test set;
    \textbf{(e)} the RMSE-in-dex histogram for the full test set;
    \textbf{(f)} the image SSIM histogram for the full test set. In all three histograms, the vertical red dashed line marks the value for the example shown above. For the column-density and surface-brightness maps, ``masked'' refers to source masking followed by local-median filling; for image SSIM, the ground-truth column-density threshold selects which local SSIM values enter the per-map median. \textbf{The CDE reconstructs broad filamentary structures from noisy simulated inputs under the adopted assumptions.}}
    \label{fig:cde_noisy_example}
\end{figure}

\begin{figure*}[!t]
    \setlength{\hsize}{\textwidth}
    \setlength{\linewidth}{\textwidth}
    \centering
    \includegraphics[width=0.7\linewidth]{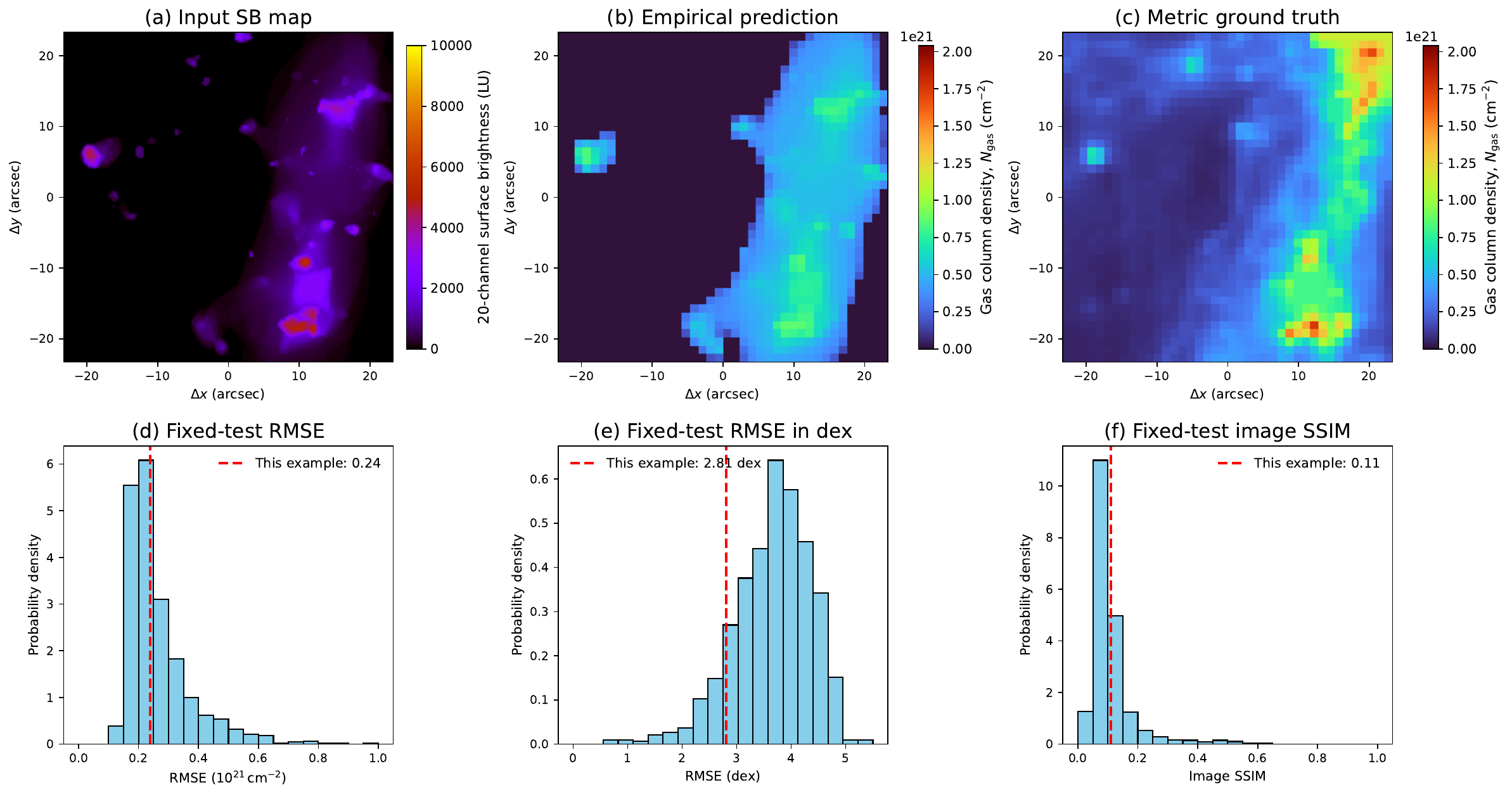}
    \caption{\textbf{Empirical Scaling Model on Test Set:} representative test-set example for the empirical scaling model with noisy, masked inputs and error statistics over the full test set. \textbf{Top row}:
    \textbf{(a)} the input Ly$\alpha$ surface-brightness map; 
    \textbf{(b)} the gas column-density map predicted by the empirical model; 
    \textbf{(c)} the masked ground-truth column-density map. 
    \textbf{Bottom row}:
    \textbf{(d)} the RMSE histogram for the full test set;
    \textbf{(e)} the RMSE-in-dex histogram for the full test set;
    \textbf{(f)} the image SSIM histogram for the full test set. In all three histograms, the vertical red dashed line marks the value for the example shown above. For the column-density and surface-brightness maps, ``masked'' refers to source masking followed by local-median filling; for image SSIM, the ground-truth column-density threshold selects which local SSIM values enter the per-map median. \textbf{The CDE achieves lower RMSE and RMSE in dex, and higher image SSIM, than this pixel-wise empirical baseline on the simulated test set.}}
    \label{fig:simple_example}
\end{figure*}
\newpage
\twocolumngrid

The top row of Fig.~\ref{fig:simple_example} shows a test-set example, chosen to match the examples in Figs.~\ref{fig:cde_noiseless_example} and \ref{fig:cde_noisy_example}. Pixels without retained emission are assigned a column-density floor of $10^{15}\,\mathrm{cm^{-2}}$ regardless of the surrounding structure, producing sharp, unphysical transitions at the boundaries of the retained emission regions. In contrast, the CDE avoids this issue by leveraging spatial context and produces coherent filamentary predictions that align more closely with the ground truth.

The corresponding error statistics are shown in Fig.~\ref{fig:simple_example}(d)--(f). The empirical scaling model yields a median RMSE of $2.29^{+0.68}_{-0.34} \times 10^{20}$~cm$^{-2}$, a median RMSE in dex of $3.71^{+0.44}_{-0.49}$, and a median image SSIM of $0.090^{+0.027}_{-0.020}$. These results stand in sharp contrast to the CDE performance, which yields significantly lower RMSE and higher image SSIM. The RMSE in dex is also much larger for the empirical model, as it is dominated by pixels assigned the column-density floor despite having appreciable ground-truth column density. These results demonstrate improved predictive performance for the CDE relative to this pixel-wise empirical baseline on the simulated test set.

\begin{figure*}[!t]
    \centering
    \includegraphics[width=0.75\linewidth]{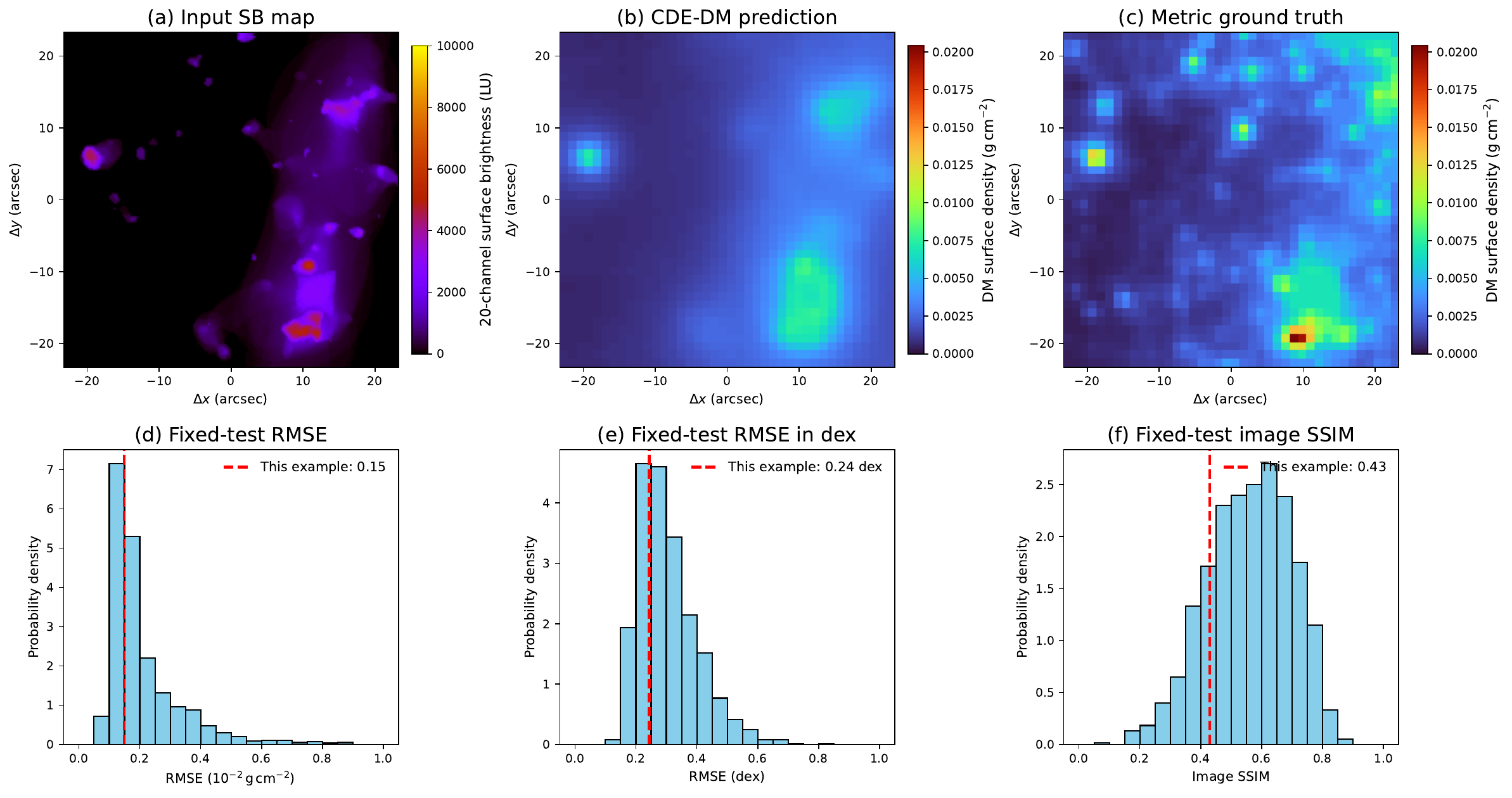}
    \caption{\textbf{CDE-DM on Test Set:} representative test-set example for the CDE trained to predict dark matter with noisy, masked inputs and error statistics over the full test set.
    \textbf{Top row}:
    \textbf{(a)} the input Ly$\alpha$ surface-brightness map;
    \textbf{(b)} the CDE-predicted DM column-density map; 
    \textbf{(c)} the masked ground-truth DM column-density map. 
    The simulated CDE input shown here is identical to that used for the gas reconstruction in Fig.~\ref{fig:cde_noisy_example}.
    \textbf{Bottom row}:
    \textbf{(d)} the RMSE histogram for the full test set;
    \textbf{(e)} the RMSE-in-dex histogram for the full test set;
    \textbf{(f)} the image SSIM histogram for the full test set. In all three histograms, the vertical red dashed line marks the value for the example shown above. For the column-density and surface-brightness maps, ``masked'' refers to source masking followed by local-median filling; for image SSIM, the ground-truth column-density threshold selects which local SSIM values enter the per-map median. \textbf{These results illustrate the flexibility of the CDE framework to predict DM column-density maps within the adopted simulation.}}
    \label{fig:cde_noisy_example_DM}
\end{figure*}

\subsection{Extension to Dark Matter}

We evaluate the versatility of the CDE framework by training an additional ten-model ensemble to predict dark matter (DM) column-density maps from Ly$\alpha$ emission subcubes. For this application, we utilize the same simulated CDE inputs but generate training labels from the TNG100 dark matter snapshot data. The architecture and loss function remain identical to the gas-prediction model. This demonstration illustrates the flexibility of the framework rather than providing a direct constraint on the dark matter distribution from Ly$\alpha$ observations.

It should be noted that the connection between dark matter and Ly$\alpha$ luminosity is significantly more uncertain and model-dependent than that between gas and Ly$\alpha$ luminosity. On large (cMpc) scales, gas overdensity is known to trace dark-matter overdensity closely, a relationship consistently recovered across modern cosmological simulations including IllustrisTNG, Simba, and COLIBRE \citep{2018MNRAS.475..676S,2019MNRAS.486.2827D,2025arXiv250821126S}. However, at smaller scales within halos, these simulations also show that the gas-dark-matter connection is loosened by baryonic processes, such as cooling, feedback, and shock-heating. These processes could substantially alter the gas distribution relative to the dark matter and can also modify the dark-matter distribution through gravitational effects. Consequently, even in the filamentary IGM regime, baryonic physics can make dark matter prediction from gas Ly$\alpha$ emission model-dependent.

The top row of Fig.~\ref{fig:cde_noisy_example_DM} shows a representative DM column-density reconstruction for the same test-set example analyzed in Figs.~\ref{fig:cde_noiseless_example} and \ref{fig:cde_noisy_example}. Morphologically, the CDE-predicted DM map is broadly similar to the gas prediction, exhibiting large-scale filamentary structures aligned with the Ly$\alpha$ emission. However, the ground-truth maps differ in morphology: in this example, the gas map shows a more continuous filament, while the DM map contains compact clumps distributed across the field.

We quantify the performance of the dark matter reconstructions using RMSE, RMSE in dex, and image SSIM, as presented in Fig.~\ref{fig:cde_noisy_example_DM}(d)--(f). For this DM test set, we find a median RMSE of $0.166^{+0.075}_{-0.035} \times 10^{-2}\,\mathrm{g\,cm^{-2}}$, a median RMSE in dex of $0.285^{+0.073}_{-0.050}$, and a median image SSIM of $0.569^{+0.094}_{-0.103}$. Both RMSE in dex and image SSIM use a ground-truth selection threshold of $2.5 \times 10^{-3}\,\mathrm{g\,cm^{-2}}$ and a logarithmic floor of $10^{-6}\,\mathrm{g\,cm^{-2}}$. Looking at the example prediction, it captures the broad distribution of overdense DM regions, but the ground-truth DM map contains compact clumps that are smoothed or underestimated by the CDE.

\subsection{Cross-simulation Stress Test with FABLE}
\label{sec:fable_test}

To further assess the transferability of the CDE, we conduct a cross-simulation stress test using the FABLE simulation suite \citep{2018MNRAS.479.5385H}. FABLE and IllustrisTNG both use the moving-mesh code AREPO but adopt different subgrid prescriptions. The FABLE realization used here has approximately ten times coarser mass resolution than TNG100. We use one zoom simulation cutout of a protocluster at $z \simeq 2.75$ \citep[the standard FABLE resolution simulation described in][]{2026OJAp....965188B}. The zoom targets one of the most massive protoclusters in the Millennium-XXL parent simulation \citep{2012MNRAS.426.2046A}, whose much larger cosmological volume than TNG100 allows more extreme overdense environments to be sampled. The cutout itself is also larger in comoving extent than the volumes extracted around the TNG pointing centers. We rescale the gas and stellar catalogs to an effective redshift of $z = 2.3$ and use \colt{} to generate intrinsic Ly$\alpha$ emission cubes, retaining the supplied FABLE gas ionization states without performing an additional ionization calculation. We then apply the same PSF/LSF convolution and empirical resonant-scattering correction as for the TNG dataset. Under the adopted radiative-transfer configuration, the total intrinsic Ly$\alpha$ luminosity of the FABLE cutout exceeds that of every TNG region used to construct the training dataset.

\begin{figure*}[!t]
    \centering
    \includegraphics[width=0.75\linewidth]{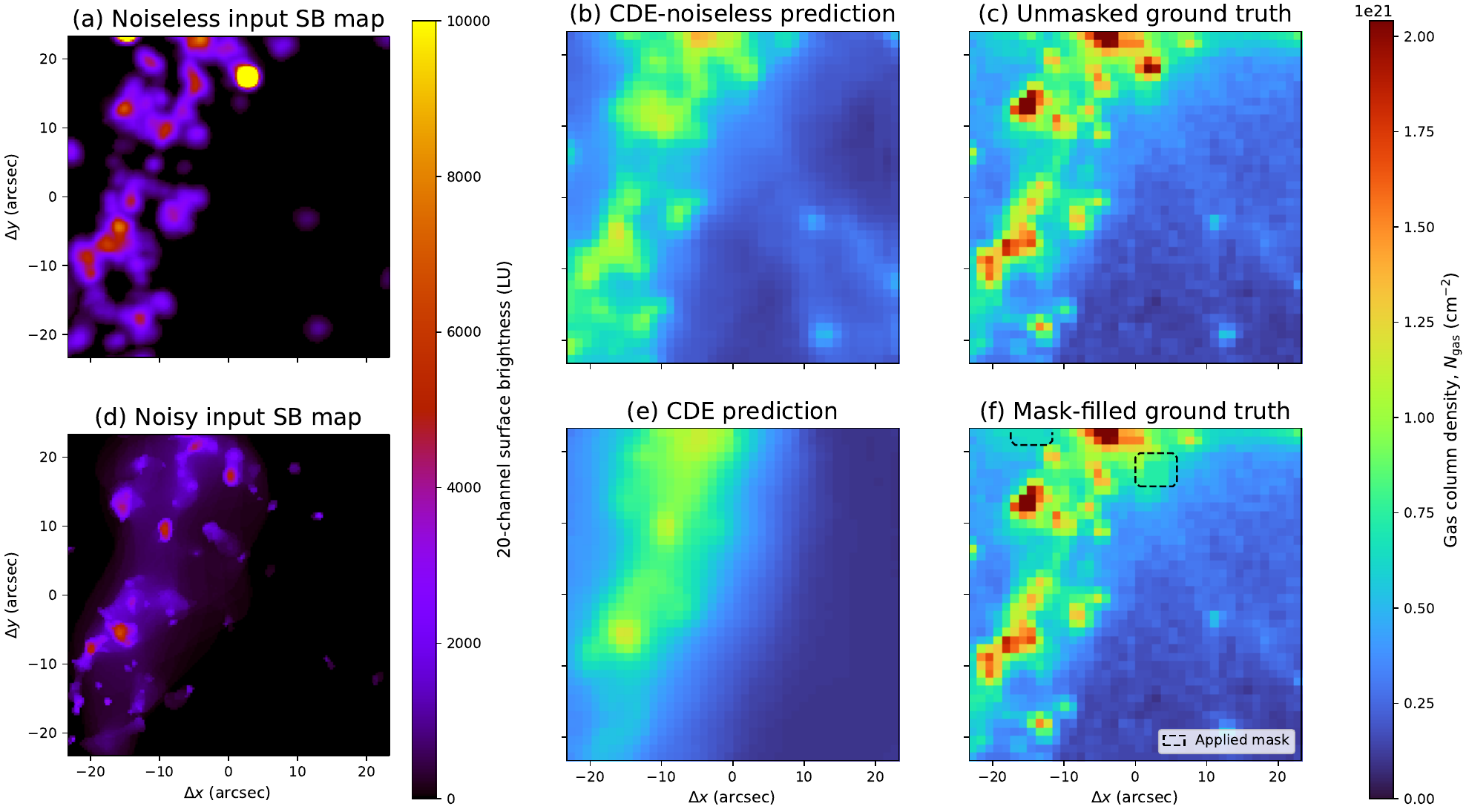}
    \caption{\textbf{Cross-simulation stress test with FABLE.}
    A representative example from the 30 matched FABLE subcubes.
    \textbf{Top row:}
    \textbf{(a)} noiseless surface-brightness map integrated over the central 20 spectral channels;
    \textbf{(b)} prediction from the CDE-noiseless trained on noiseless TNG inputs;
    \textbf{(c)} unmasked ground-truth gas column-density map.
    \textbf{Bottom row:}
    \textbf{(d)} corresponding surface-brightness map after noise injection, masking, and adaptive smoothing;
    \textbf{(e)} prediction from the CDE trained on noisy TNG inputs;
    \textbf{(f)} ground-truth map after source masking and local-median filling, with black dashed outlines marking the affected regions. The noiseless and noisy predictions are evaluated against the corresponding ground-truth maps shown in panels~(c) and~(f), respectively. \textbf{The noiseless prediction captures more morphological detail, while the noisy prediction retains the main extended structure.}}
    \label{fig:fable_test}
\end{figure*}

We extract five emission subcubes from each of the six full-field cubes corresponding to the Cartesian viewing directions, yielding 30 subcubes. Selection follows the same surface-brightness and masking criteria as the TNG dataset. Given the substantially higher intrinsic Ly$\alpha$ luminosity of the FABLE cutout, we additionally restrict the comparison to regions with mean gas column densities comparable to those represented during training. Specifically, we require the mean gas column density of each FABLE ground-truth map after source masking and local-median filling to lie between the 10th and 90th percentiles of the corresponding map means in a fixed reference sample of 4800 TNG input--ground-truth pairs. We then construct matched noisy and noiseless CDE inputs at the same subcube locations using their respective processing procedures, and apply the corresponding ten-model ensembles trained on TNG without retraining.

We evaluate the noisy predictions against the source-masked and locally filled FABLE gas column-density maps, and the noiseless predictions against the corresponding unmasked maps, using the gas-reconstruction metrics defined above. For the noiseless case, the median RMSE is $2.21^{+0.64}_{-0.66} \times 10^{20}$~cm$^{-2}$, the median RMSE in dex is $0.140^{+0.038}_{-0.009}$, and the median image SSIM is $0.840^{+0.029}_{-0.066}$. The corresponding noisy values are $2.38^{+0.27}_{-0.30} \times 10^{20}$~cm$^{-2}$, $0.290^{+0.031}_{-0.027}$, and $0.477^{+0.058}_{-0.090}$, respectively. The quoted offsets indicate the interquartile ranges around the medians.

Figure~\ref{fig:fable_test} shows a representative paired example. The noiseless prediction captures much of the broad gas morphology, while the noisy prediction retains the main extended structure but smooths over smaller features and underestimates the prominent compact peaks. These results suggest that the mapping learned from TNG retains useful predictive information for this selected FABLE sample. Differences in subgrid physics, mass resolution, and the sampled environment may contribute to the reconstruction discrepancies, although this test does not isolate their individual effects. Because the subcubes come from one volume and are selected using ground-truth gas properties, we interpret these results as a restricted cross-simulation stress test rather than a representative assessment of performance across FABLE.

\section{Applications II: Real KCWI Data}

\label{sec:results}

\subsection{Filament Detection}

\begin{figure*}[!t]
    \centering
    \includegraphics[width=0.49\linewidth]{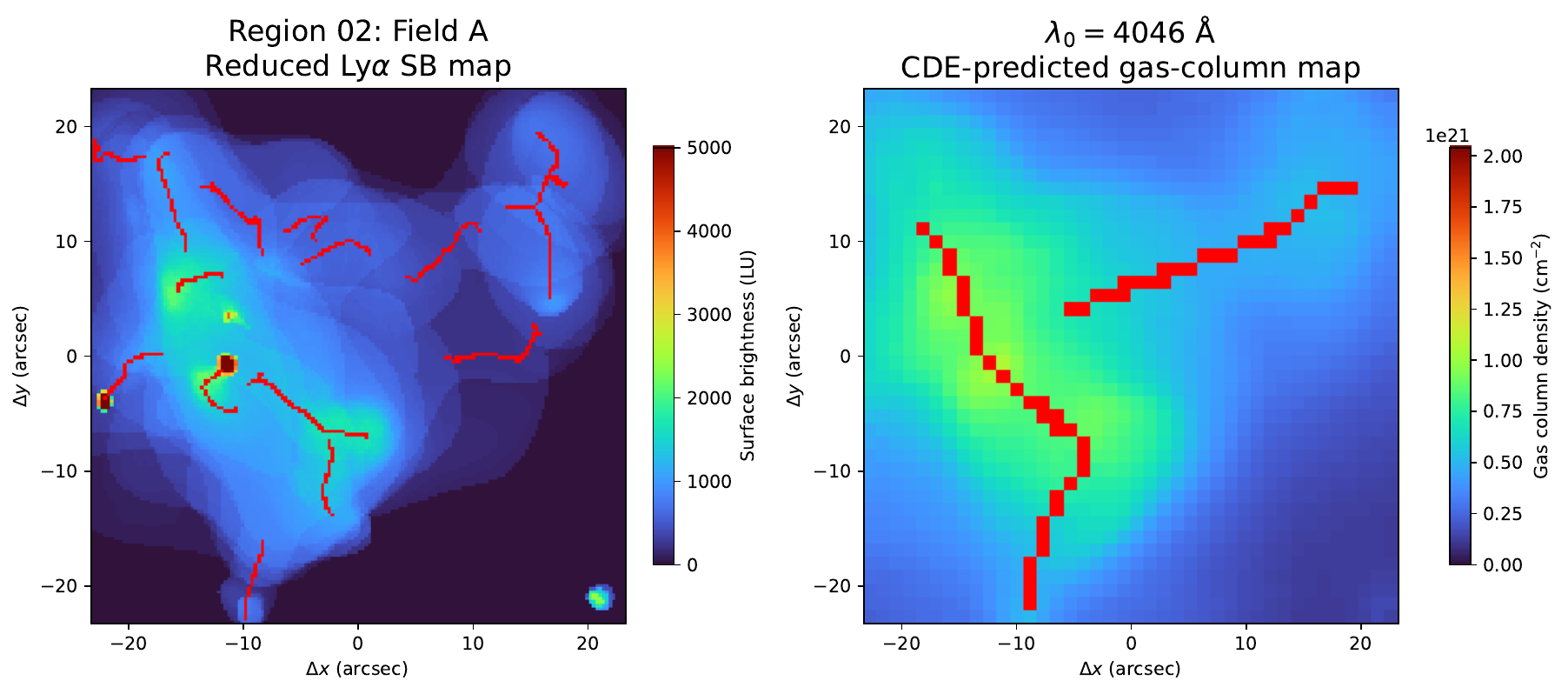}
    \includegraphics[width=0.49\linewidth]{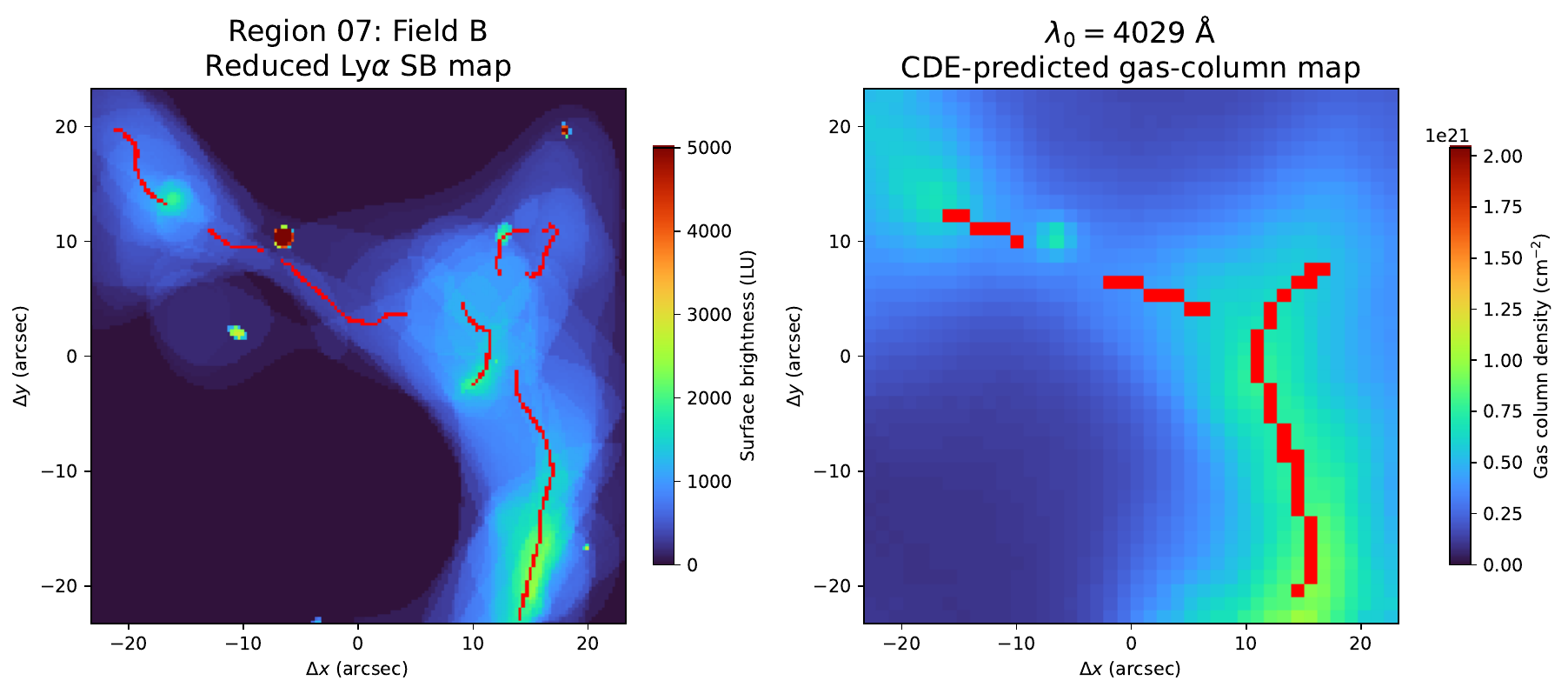}\\[1ex]
    \includegraphics[width=0.49\linewidth]{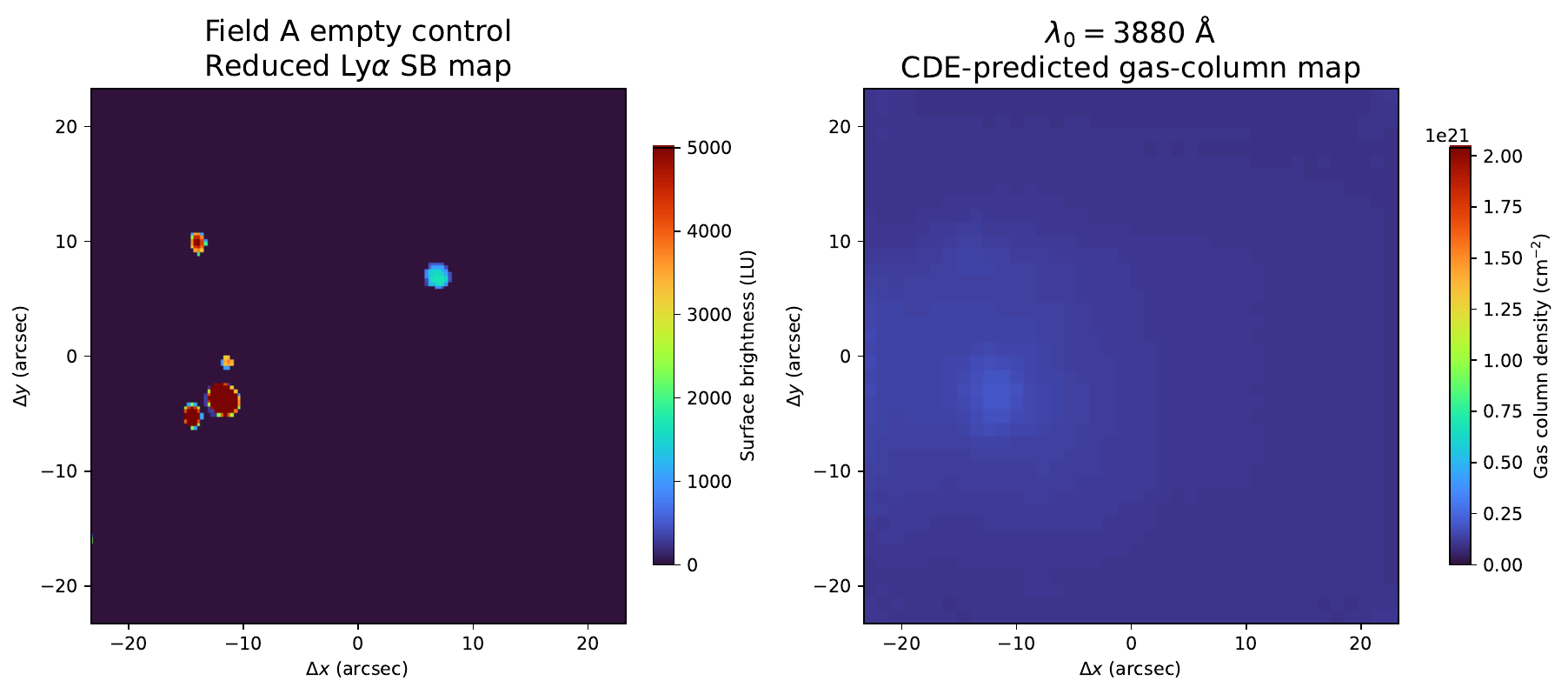}
    \includegraphics[width=0.49\linewidth]{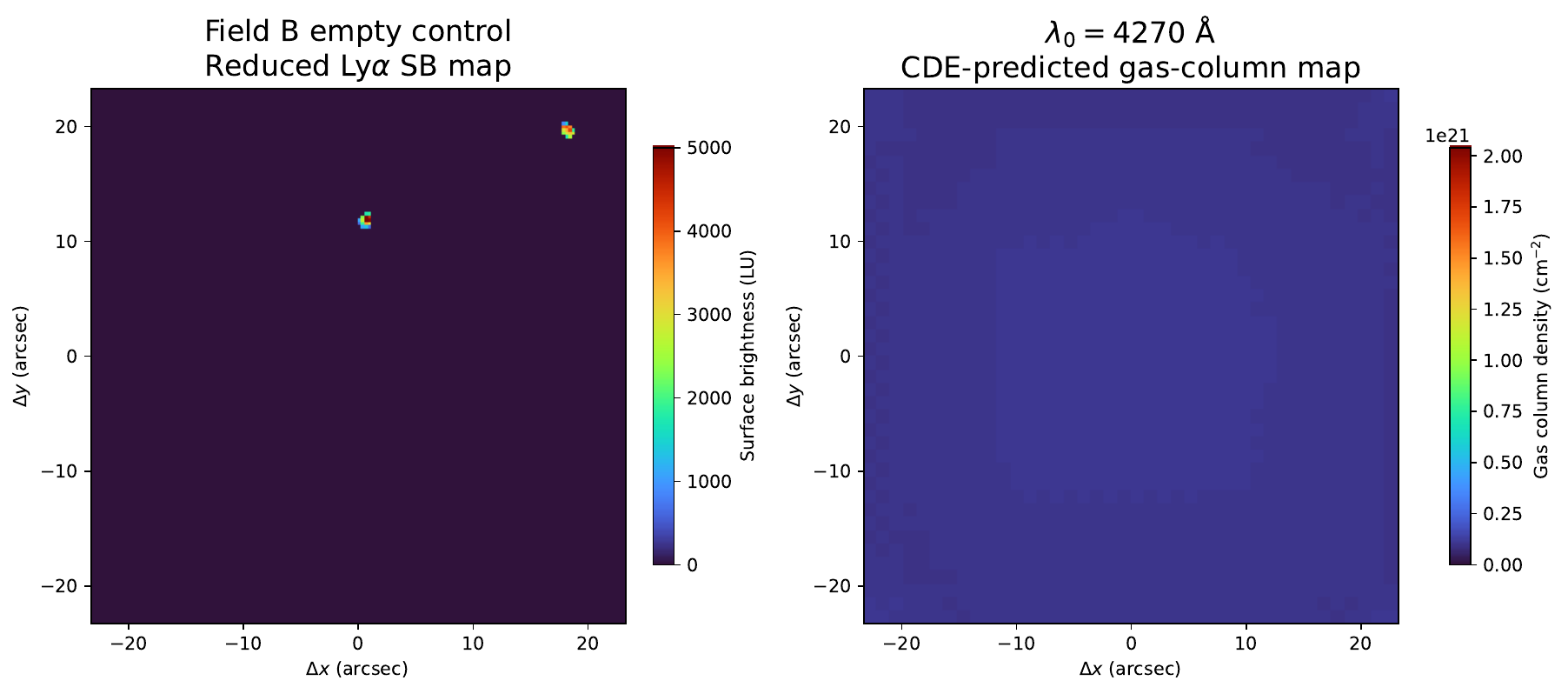}
    \caption{\textbf{CDE predictions and filament identification in observed KCWI data.}
    \textbf{Top row:} Region 02 from Field A (left pair) and region 07 from Field B (right pair).
    \textbf{Bottom row:} Observed empty controls covering $3870$--$3889\,\text{\AA}$ in Field A (left pair) and $4260$--$4279\,\text{\AA}$ in Field B (right pair).
    In each pair, the left panel shows the adaptively smoothed Ly$\alpha$ surface-brightness map integrated over the retained 20 spectral channels, and the right panel shows the corresponding CDE-predicted gas column-density map. Filamentary structures identified using SCMS are overlaid in red. The same surface-brightness and column-density color scales are used for all regions and controls.
    \textbf{The emission regions show extended filamentary structures in the predicted maps, while neither empty control yields an SCMS filament under the adopted settings.}}
    \label{fig:real_example}
\end{figure*}

We apply the trained CDE to observed CDE inputs of the protocluster CC2.2 (Field A) at $z \sim 2.2$ and the protocluster Hyperion (Field B) at $z \sim 2.45$. Representative examples of column-density reconstruction and filament detection are presented in Fig.~\ref{fig:real_example}. The top row shows region 02 from Field A (left pair) and region 07 from Field B (right pair), while the bottom row shows the two observed empty controls. In each pair, the left panel displays the input Ly$\alpha$ surface-brightness map integrated over all 20 spectral channels of the observed CDE inputs, and the right panel shows the corresponding CDE-predicted column-density map, with SCMS-identified filamentary structures overlaid as red curves.

Filamentary features are clearly visible in both the SB and predicted column-density maps in the top row. However, those detected in the SB maps tend to be shorter and more curved, which may reflect the limitations of adaptive smoothing and wavelength summation that can fragment continuous emission into irregular, isolated blobs. The CDE predictions show smoother and more extended filamentary structures, whose morphology remains conditional on the simulation-based mapping learned during training.

Similar filamentary structures are identified in the CDE-predicted maps for the other observed regions (see Appendix~\ref{app:more_filament_detection}). We emphasize, however, that the CDE and SCMS primarily highlight the dense filamentary backbone. The diffuse outskirts of the IGM, which fall below the effective sensitivity threshold, remain largely undetected and poorly constrained.

We additionally apply the same CDE and SCMS procedures to two observed empty-control subcubes, covering $3870$--$3889\,\text{\AA}$ in Field A and $4260$--$4279\,\text{\AA}$ in Field B. These controls are processed using the same masking, adaptive-smoothing, and trimming procedures as the emission regions. Although some isolated features remain in their surface-brightness maps, the CDE predictions show very low, nearly uniform column densities, and SCMS identifies no filamentary structures under the adopted settings.

\subsection{Uncertainty Estimation (Monte Carlo Realizations)}

To assess the robustness of the CDE predictions against observational noise, we generated multiple Monte Carlo (MC) realizations for each observed KCWI subcube. As described earlier, we have 10 regions in total: five (regions 00--04) from Field A (CC2.2) and five (regions 05--09) from Field B (Hyperion).

We first draw Gaussian noise perturbation fields based on the raw variance subcubes. The corrected variance subcubes used for adaptive smoothing are eight times the raw variance to account for voxel correlations and remain unchanged during these perturbations. Each perturbation is both added to and subtracted from the corresponding masked unsmoothed observed subcube, producing a pair of realizations with symmetric Gaussian noise perturbations. This paired construction ensures that the perturbations average to zero at each input voxel across all realizations. All perturbed subcubes are then processed using the standard adaptive-smoothing algorithm, and the trained CDE is applied to reconstruct column-density maps. For each region, we generate 10 such perturbation pairs, yielding a total of 20 realizations per region.

For each pixel in a given region, we compute the median and standard deviation $\sigma$ of the predicted column density across the 20 realizations and define the fractional uncertainty as $\sigma/\mathrm{median}$. This procedure yields a single fractional-uncertainty map for each region. We further characterize the overall uncertainty by examining the distribution of fractional-uncertainty values across all pixels in each map.

\begin{figure}
    \centering
    \includegraphics[width=0.9\columnwidth]{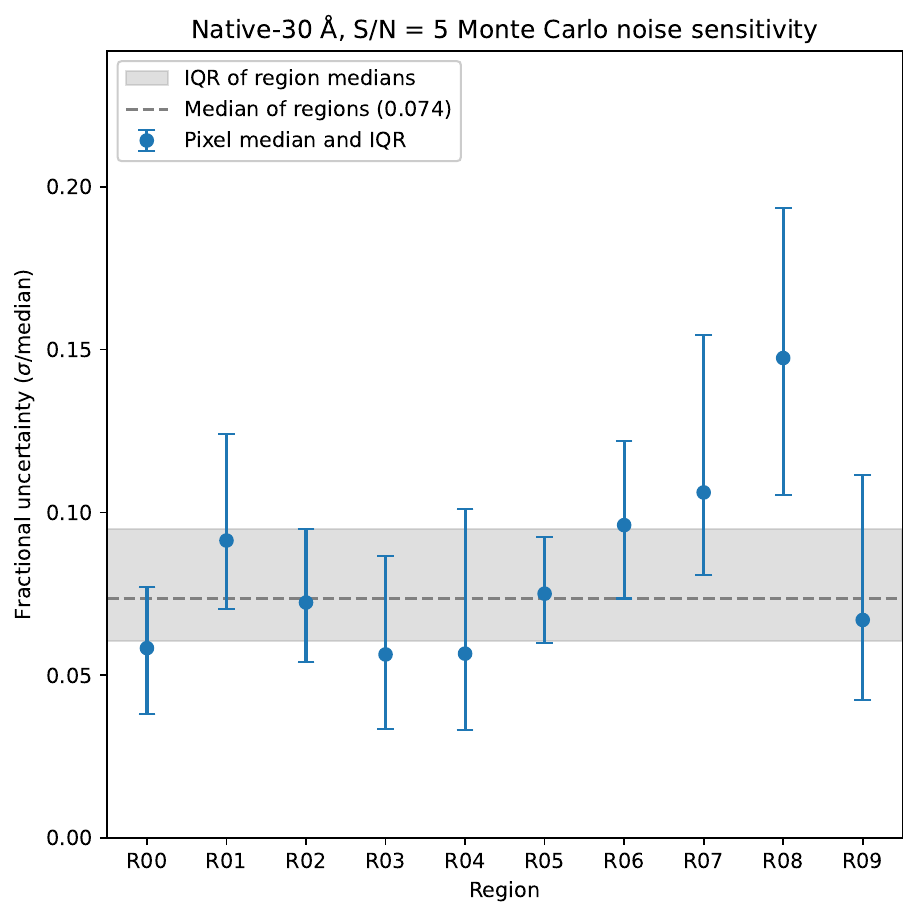}
    \caption{\textbf{MC Uncertainty:} fractional uncertainty of CDE predictions estimated from paired Gaussian noise perturbations of observed KCWI subcubes. Points show the median pixel-wise fractional uncertainty within each region, and error bars show the corresponding interquartile range. The horizontal dashed line marks the median of the ten region medians ($\sim7.4\%$), and the shaded band indicates their interquartile range. \textbf{The predicted column densities show only modest sensitivity to the applied Gaussian noise perturbations.}}
    \label{fig:MC_realization}
\end{figure}

The resulting statistics are illustrated in Fig.~\ref{fig:MC_realization}. The median fractional uncertainties within individual regions range from $5.6\%$ to $14.7\%$, with a median across the ten regions of $(7.4^{+2.1}_{-1.3})\%$, where the offsets indicate the interquartile range of the region medians. These results indicate only modest sensitivity of the predicted column densities to the applied Gaussian noise perturbations, although this procedure does not capture the full uncertainty associated with correlated observational residuals or the adopted physical models.

\subsection{Overdensity Normalization}

When applying the CDE to observed KCWI data, we perform a normalization check to assess whether the predicted gas overdensities $\delta_{\rm gas}$ are consistent with the simulation-based SB--overdensity relation. To carry out this check, we use the relation between the mean Ly$\alpha$ surface brightness and the corresponding CDE-predicted mean gas overdensity in the simulated test set as a calibration curve, and assess whether the SB-normalized overdensity defined below is consistent with zero. This assesses consistency between the normalization implied by the observed predictions and the adopted simulations.

In the simulated test set, each subcube is characterized by both a directly computed average Ly$\alpha$ surface brightness $\langle{\rm SB}\rangle$ and a CDE-predicted gas overdensity $\delta_{\rm gas}$, which is defined as the mean gas column density within the subcube divided by the cosmic mean column density, minus one. These two subcube-integrated quantities exhibit a clear monotonic correlation spanning more than one order of magnitude in surface brightness. We model this relation using a quadratic fit of the form

\begin{equation}
\log_{10}\langle {\rm SB} \rangle
= a\,\left[\log_{10}\!\left(1+\delta_{\rm gas}\right)\right]^2
+ b\,\log_{10}\!\left(1+\delta_{\rm gas}\right)
+ c \, .
\end{equation}

This provides an improved description of the simulated data compared to a purely linear relation. We adopt this calibrated relation to infer an expected gas overdensity, $\delta_{\rm gas}^{\rm exp}$, for each observed CDE input based solely on its measured $\langle{\rm SB}\rangle$.

To compare the CDE prediction with this simulation-based expectation, we define a normalized overdensity:
\begin{equation}
\delta_{\rm gas}^{\rm norm}=\frac{1+\delta_{\rm gas}^{\rm pred}}{1+\delta_{\rm gas}^{\rm exp}}- 1 \, .
\end{equation}
where $\delta_{\rm gas}^{\rm pred}$ is the gas overdensity inferred from the CDE-predicted gas column-density map. Values of $\delta_{\rm gas}^{\rm norm}$ near zero indicate agreement with the simulation-based normalization at the measured surface brightness. This procedure checks the normalization without imposing any correction on the predicted maps: it simply checks whether the CDE outputs are consistent with the SB--$\delta$ relation that holds in the simulation. We use $(1+\delta)$ rather than $\delta$ because it is guaranteed to be positive and corresponds directly to the gas density inside each subcube.

\begin{figure}
    \centering
    \includegraphics[width=1\columnwidth]{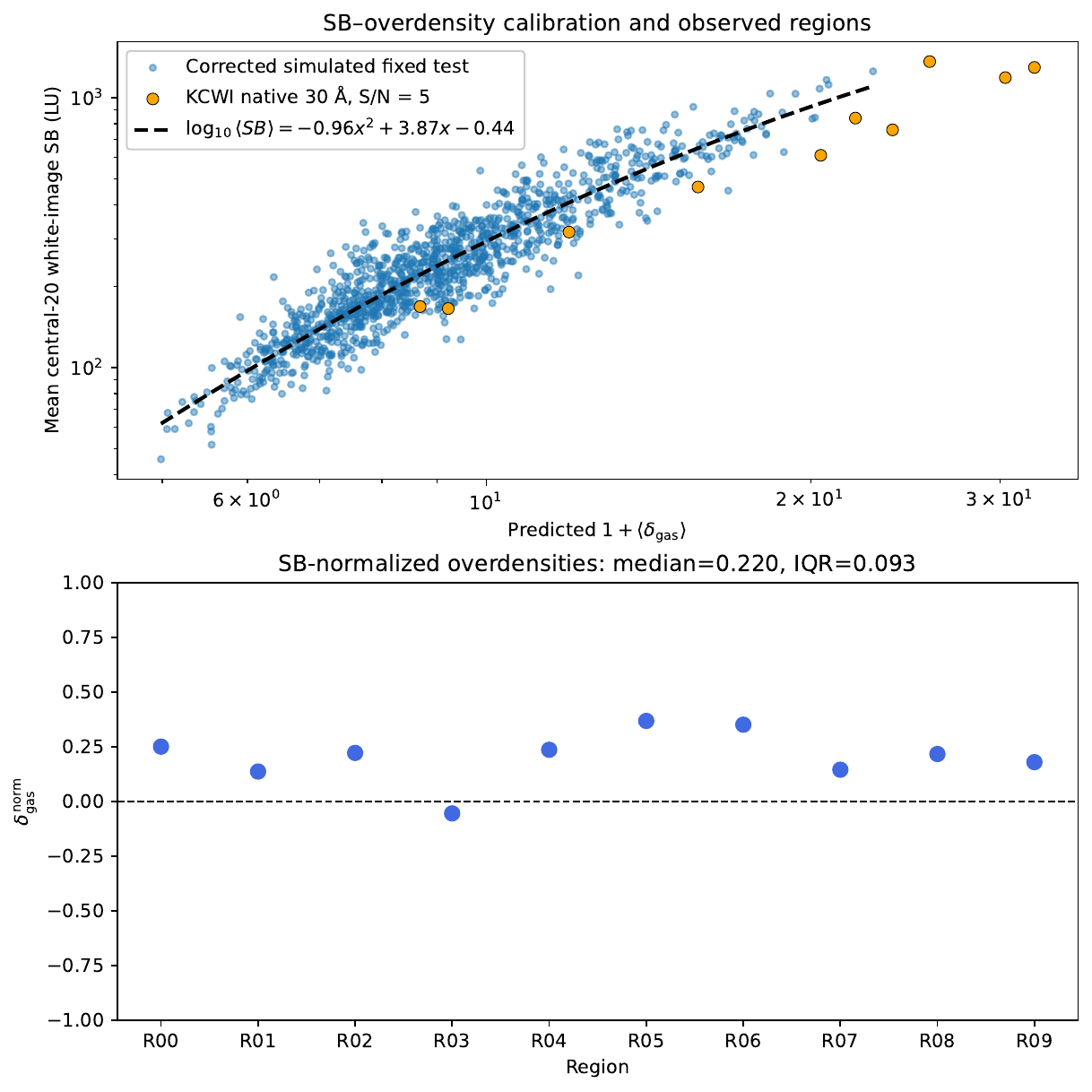}
    \caption{\textbf{Gas overdensity normalization.}
    \textbf{Top:} Spatially averaged Ly$\alpha$ surface brightness integrated over the retained 20 spectral channels versus CDE-predicted $1+\delta_{\rm gas}$ for the simulated test set (blue) and observed regions (orange). The black dashed curve shows the quadratic relation fitted in logarithmic coordinates exclusively to the simulated data.
    \textbf{Bottom:} SB-normalized overdensities $\delta_{\rm gas}^{\rm norm}$ for the ten KCWI regions (R00--R09). The horizontal dashed line marks $\delta_{\rm gas}^{\rm norm}=0$, corresponding to agreement with the simulation-based normalization at the measured surface brightness.
    \textbf{The observed predictions show a modest positive offset from the simulation-based SB expectation.}}
    \label{fig:overdensity}
\end{figure}

Figure~\ref{fig:overdensity} shows the fitted SB--$\delta$ relation (top panel) and the resulting normalized overdensities, $\delta_{\rm gas}^{\rm norm}$, for the ten KCWI regions (bottom panel). The observed regions broadly follow the simulated trend, although the predicted overdensities of regions 00, 03, 04, and 05 slightly exceed the range represented in the simulated test set. The SB-based estimates for regions 00, 03, and 04 additionally require slight extrapolation of the fitted relation. The best-fitting parameters are indicated in the figure. The normalized overdensities show a modest positive offset, with a median $\delta_{\rm gas}^{\rm norm}=0.22^{+0.03}_{-0.07}$, where the offsets indicate the interquartile range. This normalization check reveals a modest positive offset relative to the simulation-based expectation, and no additional rescaling is applied to the final CDE predictions.

We also considered using galaxy overdensity instead of surface brightness for normalization. However, the galaxy catalog used in \citet{2023NatAs...7.1390M} is not sufficiently complete on the small spatial scales relevant for our sliced KCWI subcubes. In that work, galaxy overdensities were computed within spheres of radius $8\,h^{-1}\,\mathrm{cMpc}$, where catalog incompleteness has minimal impact. In contrast, the emitting regions in our subcubes span only $\sim 0.9 \times 0.9 \times 3.3 \, h^{-3}\,\mathrm{cMpc}^3$, requiring much smaller apertures for a meaningful $\delta_{\rm gal}$ estimate. At these scales, incompleteness becomes severe and leads to unrealistically inflated galaxy overdensities, sometimes exceeding several hundred. For this reason, we rely on surface-brightness-based normalization for the present analysis and defer galaxy-overdensity-based normalization to future work once more complete catalogs become available.

\section{Discussion}

Our results show that the CDE framework recovers gas filaments in simulated test data and predicts spatially coherent filamentary structures from observed KCWI data. Compared with our empirical scaling model, the CDE yields smoother and more spatially continuous reconstructions, avoiding the abrupt cutoffs produced by naive pixel-wise column-density estimates. The stress test on the selected FABLE sample shows that the TNG-trained CDE retains useful predictive information without retraining, suggesting some transferability to this different simulation. Neither of the two observed empty controls yields SCMS-identified filamentary structures under the adopted settings, providing a check against false-positive filament detections under the tested conditions. Demonstrating these capabilities with the CDE is an important first step toward robust reconstruction of IGM gas column-density maps from diffuse Ly$\alpha$ observations.

As illustrated in Fig.~\ref{fig:real_example} and Appendix~\ref{app:more_filament_detection}, the inferred gas filaments in the observed KCWI data extend over $\gtrsim1$~cMpc in projection, beyond typical circumgalactic scales. Such a significant spatial extent suggests an association with the IGM rather than individual host galaxies, although this interpretation remains strongly dependent on simulation-based priors. The potential to infer physical IGM information from KCWI observations, even under current noise and sensitivity limitations, highlights Ly$\alpha$ emission as a promising probe complementary to Ly$\alpha$ absorption tomography. Taken together, our findings support the CDE as an extensible framework that could be adapted to other IFU instruments and emission lines.

Several caveats must be considered. The most direct technical limitation is the systematic underprediction of compact, dense sources, as CDE tends to favor smoother reconstructions. This underestimation may reflect the difficulty in capturing complex subgrid physics within high-density regions. Additionally, the low-S/N nature of the data prevents the CDE from capturing diffuse outskirts and the finest structural details. Nonetheless, these shortcomings primarily affect small-scale or low-surface-brightness features, while our validation tests indicate that the large-scale filamentary backbone is generally better preserved.

We emphasize that this framework was developed specifically for massive protocluster fields characterized by unusually bright Ly$\alpha$ emission. In lower-density environments, such as typical galaxy groups or cosmic voids, the emission is expected to be substantially fainter and may remain undetectable without extremely long exposure times. While the current application to dense protocluster fields provides a necessary first step, extending the framework to more typical IGM environments will require both advancements in machine-learning methods and improvements in telescope sensitivity.

A more subtle but fundamental caveat is the framework's dependence on the underlying simulations. Most emission cubes are generated in intrinsic-emission mode with a voxel-level empirical correction calibrated against a limited set of full resonant-scattering calculations, while dust attenuation is disabled. Paired tests show that this correction brings the voxel surface-brightness distributions of the simulated CDE inputs closer to their full-RS counterparts for the selected surface-brightness range. However, such a local correction cannot reproduce the nonlocal spatial and spectral redistribution of photons and has not been independently validated using pointing centers beyond the five used for its calibration. Although dust is expected to have limited direct influence on emission produced in the nearly dust-free IGM, Ly$\alpha$ photons originating in the interstellar medium of galaxies may be attenuated by galactic dust before escaping and resonantly scattering into the surrounding medium.

Furthermore, the training data are derived from a single hydrodynamical simulation (TNG100) coupled to one radiative-transfer configuration implemented with \colt. These choices encode physical assumptions and approximations that may limit the generalizability of the features learned by the CDE. Although simulations broadly agree on some properties of the large-scale IGM and its intrinsic Ly$\alpha$ emission \citep[e.g.,][]{2018MNRAS.475..676S,2019MNRAS.486.2827D,2025arXiv250821126S}, differences in hydrodynamics, feedback, and radiative-transfer prescriptions may alter the detailed relationship between Ly$\alpha$ emission and gas column density. Although we do not aim to recover spatial structure below the instrumental PSF, the numerical resolution of the simulations can still affect the predicted Ly$\alpha$ surface brightness, even after PSF convolution \citep[e.g.,][]{2026OJAp....965188B}. The observed reconstructions should therefore be interpreted as conditional on these assumptions. The cross-simulation stress test on the selected FABLE sample suggests some transferability of the CDE, although the noisy predictions have larger errors and lower image SSIM than those on the TNG test set. Because the comparison uses selected subcubes from one FABLE volume, it provides only a limited assessment of transferability. Future applications should train and evaluate the CDE across different cosmological simulations and radiative-transfer configurations, while larger observational samples may ultimately help distinguish among and calibrate their underlying physical prescriptions.

Moreover, the 6000 simulated CDE inputs are derived from volumes extracted around only 25 pointing centers in TNG100. Although each subcube samples only a small portion of the corresponding extracted volume, generating multiple subcubes from the same physical environment may introduce correlations between the training and test sets. To reduce direct overlap among the subsets, subcubes from a given full-field emission cube and its opposite-view counterpart are assigned to at most one of the training, validation, or test sets for each base model. We also apply random rotations and spatial flipping after adaptive smoothing to increase orientation diversity. However, these measures cannot completely remove correlations between subcubes derived from the same pointing center, which may lead to optimistic estimates of CDE performance. Given the scarcity and novelty of the KCWI observations and the inherent uncertainties in numerical simulations, the present study focuses on establishing feasibility at a tractable computational cost. A more comprehensive analysis should separate the training, validation, and test samples at the pointing-center level and include a larger number of physically independent environments drawn from multiple cosmological simulations, which will require substantially greater computational investment. Our results should therefore be viewed as a proof-of-concept feasibility study and a foundation for future work rather than a definitive measurement of the IGM.

Looking ahead, we aim to improve the inference of dark-matter distributions and extend the CDE to additional physical properties, such as gas kinematics, using larger and more diverse simulated training sets. We are also conducting a new multi-night KCWI observing campaign targeting high-redshift overdensities in the COSMOS field. These observations provide higher spectral resolution and broader wavelength coverage using both the blue and red KCWI channels. Preliminary analysis of the data obtained so far suggests the presence of Ly$\alpha$ filaments connecting known galaxies. Extending the data-generation procedure to higher redshifts and developing a more sophisticated observational model for the red channel, where sky emission lines and telluric absorption pose greater challenges than in the blue channel, will enable a more stringent test of the CDE framework.

\section{Conclusion}

We have developed a proof-of-concept convolutional neural-network framework to reconstruct gas column-density maps from low-S/N KCWI Ly$\alpha$ emission data. Benchmark tests show that the CDE recovers filamentary column-density structures in simulated data and produces plausible, spatially coherent filamentary structures when applied to observed KCWI data. The cross-simulation stress test on the selected FABLE sample suggests some CDE transferability, and the two observed empty controls provide a check against false-positive filament detections under the adopted settings. The observed reconstructions are consistent with the intergalactic filaments previously reported by \citet{2023NatAs...7.1390M}, suggesting that Ly$\alpha$ observations near the current KCWI sensitivity limit retain useful information about the large-scale gas distribution.

To the best of our knowledge, this is the first application of a deep neural network to realistic IFU emission data for high-redshift IGM reconstruction. Mapping structure through Ly$\alpha$ emission provides a valuable complement to Ly$\alpha$ absorption tomography, which becomes increasingly challenging at high redshift. Nevertheless, our results should be viewed as a proof-of-concept feasibility study rather than a definitive measurement. Although our work suggests that useful structural information can be inferred from current KCWI observations, the reconstructions remain conditional on the TNG100 training data, simplified radiative-transfer calculations, and approximate observational modeling. As larger and more diverse simulation suites and higher-quality observations become available, the CDE can be retrained, evaluated, and refined into a more thoroughly tested tool for quantitative IGM studies. More broadly, this work illustrates the potential of simulation-trained inverse modeling for other low-S/N, multidimensional astronomical datasets.

\section*{Acknowledgments}

This work is based on observations obtained with the KCWI on the Keck II telescope. The authors wish to recognize and acknowledge the very significant cultural role and reverence that the summit of Mauna Kea has always had within the Hawaiian community. We are most fortunate to have the opportunity to conduct observations from this mountain.

This work was supported in part by the Caltech Carver Mead New Adventures Fund Seed Grant 25550038 and NASA APRA Grant 80NSSC22K1649.

This research made use of the publicly available IllustrisTNG simulations. Radiative-transfer calculations were performed using the \colt{} code, and the machine-learning models and analysis procedures were implemented in Python using TensorFlow, NumPy \citep{harris2020array}, SciPy \citep{2020SciPy-NMeth}, and Matplotlib \citep{Hunter:2007}. We also made use of NASA's Astrophysics Data System (ADS).

Generative AI tools, including ChatGPT and Codex (OpenAI) and Gemini (Google), were used for language editing, manuscript refinement, and programming assistance. This assistance included converting Jupyter notebooks into Python scripts, debugging and checking the resulting code and numerical outputs, and refining figure formatting. The adopted physical assumptions were retained throughout these revisions. The authors reviewed and verified the AI-assisted material and take full responsibility for all scientific content and conclusions presented in this work.

\section*{Data Availability}

The raw KCWI observations analyzed in this work are available through the Keck Observatory Archive (KOA). The IllustrisTNG simulation data are publicly available at \url{https://www.tng-project.org}. The accompanying software release (version 0.8.0; \citealp{Deng2026CDE}), identified by the Zenodo DOI \dataset[10.5281/zenodo.23328516]{https://doi.org/10.5281/zenodo.23328516}, contains the CDE data-preparation, training, and analysis code, trained model weights, and processed example data. The release includes three ten-member CDE ensembles for noisy gas, noiseless gas, and noisy dark matter, simulated test subcube 1451, and the ten observed CDE inputs. Reference predictions are included for simulated test subcube 1451 and the ten observed CDE inputs. Metric tables for the 1200 simulated test subcubes are also provided. The complete simulated training dataset, FABLE data, and full observed mosaics are not included in this release. Researchers interested in additional data products or collaboration may contact the corresponding author.


\clearpage
\appendix
\section{Details of CDE Architecture}
\label{app:cde_details}

Each residual block in Figure~\ref{fig:cnn_architecture} uses a sequence of three convolutions. It begins with a $1\times1$ convolution, batch normalization, and LeakyReLU activation to mix channel-wise information, followed by a $3\times3$ spatial convolution with the same normalization and activation layers. A final $1\times1$ convolution followed by batch normalization produces processed features with the required output channel count. The shortcut of the residual block is an identity connection when the input and output channel counts match. If the channel counts differ, a $1\times1$ convolution followed by batch normalization is used to adjust the channel count. The processed features and shortcut features are added, and a final LeakyReLU activation produces the block output.

The downsampling block employs two parallel pathways whose outputs are summed prior to a LeakyReLU activation. The primary pathway, which halves the spatial dimensions and doubles the number of feature channels, contains three convolutions: $1\times1$, $3\times3$ with stride 2, and $1\times1$. Each convolution is followed by batch normalization and dropout, and the first two also have LeakyReLU activation between batch normalization and dropout. The secondary pathway applies $2\times2$ average pooling with a stride of 2, followed by dropout and a $1\times1$ convolution to match the channel count of the primary pathway. Another dropout layer follows this convolution.

The upsampling block consists of a single $2\times2$ transposed convolution with a stride of 2. This operation doubles the spatial dimensions while reducing the number of feature channels by half, effectively reconstructing the high-resolution features from the latent representation.

\section{Details of Data Preparation}
\label{app:data_generation}

We describe each step of data generation and reduction in detail in the following subsections. As noted above, the bright-source masking and adaptive-smoothing procedures are similar for both observed and simulated data. To avoid repetition, we defer a unified discussion of these shared steps until after introducing the construction of the simulated noisy emission subcubes.

\subsection{Observed KCWI Cube}

In the study by \citet{2023NatAs...7.1390M}, two massive protoclusters, CC2.2 \citep{2020ApJ...892....8D} (Field A, at $z \sim 2.2$) and Hyperion \citep{2018A&A...619A..49C} (Field B, at $z \sim 2.45$), were observed with the KCWI on Keck II \citep{2018ApJ...864...93M}. Protoclusters are ideal environments for detecting IGM Ly$\alpha$ emission because of their high gas densities and the intense ultraviolet (UV) radiation fields produced by young galaxies and active galactic nuclei (AGNs) \citep{2014Natur.506...63C,2019Sci...366...97U}.

Observations were conducted using the large slicer, blue low-dispersion (BL) grating, and NAS mode. The large slicer yields a spatial sampling of $0.291^{\prime\prime} \times 1.35^{\prime\prime}$ and an effective FOV of $\sim 20^{\prime\prime} \times 33^{\prime\prime}$ per exposure. To improve spatial resolution and ensure contiguous coverage, exposures were repeated at multiple spatial offsets and PAs of $0^\circ$ and $90^\circ$ \citep{2023NatAs...7.1390M}. Each PA--offset combination was observed for approximately 1.5 hours, resulting in a total average integration time of 3--4 hours across the combined FOV. The spectral sampling of the BL grating is 1 \AA, with a wavelength coverage of 3838--4329 \AA\ in this configuration.

The KCWI DRP \citep{2023ascl.soft01019N} was utilized to remove strong sky-lines, subtract the sky background, and transform raw two-dimensional detector images into data cubes with two spatial dimensions and one wavelength dimension. We then coadded the exposures from multiple spatial offsets and PAs to produce the KCWI observed cubes for Field A and Field B. To minimize foreground contamination, we also masked galaxies identified in the zCOSMOS-deep survey \citep{2007ApJS..172...70L}. The resulting coadded and galaxy-masked cubes achieve a nominal spatial sampling of $0.291^{\prime\prime} \times 0.291^{\prime\prime}$ and a total FOV of $\sim 60^{\prime\prime} \times 60^{\prime\prime}$. Notably, the extensive resampling required during the coaddition process introduces significant spatial-pixel correlations, which must be carefully reproduced in the simulations.

In NAS mode, the telescope alternates between CC2.2 and Hyperion every two minutes, recording each target's spectra in separate CCD regions during a 40-minute cycle. This strategy enables precise sky subtraction by differencing the alternating exposures \citep{1996AJ....112..797S}, under the assumption of minimal source overlap and a stable sky background. NAS mode can reduce low-order residuals correlated with the sky background to below $0.07\%$, representing an exceptionally clean subtraction compared to the $\sim$1--2\% residuals typical of standard observing modes in MUSE \citep{2020A&A...641A..28W}. In our KCWI observations with NAS mode, sky residuals correspond to a surface brightness of approximately $3\times10^{-21}\,\mathrm{erg\,cm^{-2}\,s^{-1}\,\text{\AA}^{-1}\,arcsec^{-2}}$. In these observations, NAS mode reduces sky residuals to a level comparable to the expected Ly$\alpha$ surface brightness of the IGM \citep{2018MNRAS.475.3854G}. While NAS mode preserves the full FOV, it reduces the usable wavelength coverage by $\sim75\%$ relative to standard mode.

\subsection{Observed Subcubes}

Because the KCWI observed cubes are largely empty, we follow \citet{2023NatAs...7.1390M} and focus on spectral intervals containing significant emission. Specifically, we extract subcubes of size $180 \times 180 \times 30$ voxels, corresponding to a FOV of $\sim 53^{\prime\prime} \times 53^{\prime\prime}$ and a wavelength interval of $30\,\text{\AA}$, centered on the bright regions shown in Fig.~1 of their paper. The additional five observed channels on each side serve as spectral padding for adaptive smoothing and are discarded afterward, leaving the central $20\,\text{\AA}$ interval in the observed CDE inputs. While their figure presents $5\,\text{\AA}$ surface-brightness slices, our broader retained $20\,\text{\AA}$ interval allows us to capture more of the total emission associated with each structure. This consistent wavelength centering also facilitates a direct comparison between our predicted gas column-density maps and the observed surface-brightness morphology in these regions.

\subsection{IllustrisTNG Data}

We base our simulated data on the IllustrisTNG project \citep{2018MNRAS.473.4077P,2018MNRAS.475..624N,2018MNRAS.475..676S,2018MNRAS.477.1206N,2018MNRAS.480.5113M,2019ComAC...6....2N}, a state-of-the-art suite of publicly available cosmological hydrodynamical simulations that self-consistently model the evolution of dark matter, gas, stars, and black holes. At the time we began constructing our training set, the IllustrisTNG suite comprised three flagship simulations: TNG50, TNG100, and TNG300, which are named according to the side length of the simulation volume in cMpc. Among these, TNG50 provides the highest spatial and mass resolution, while TNG300 offers the largest cosmological volume. To date, full Ly$\alpha$ emission cubes have primarily been generated from TNG50, most notably by \citet{2021MNRAS.506.5129B} and \citet{2023MNRAS.523.5248B}, where its superior resolution enables detailed modeling of small-scale circumgalactic and sub-galactic structures.

Our goal is to generate mock Ly$\alpha$ emission cubes suitable for comparison with KCWI observations of CC2.2 and Hyperion, located at redshifts $z \sim 2.2$ and $z \sim 2.45$, respectively. To enable a consistent comparison with the emission regions selected from the two fields and to avoid unnecessary duplication in the simulation stage, we adopt an intermediate redshift of $z = 2.3$ as the effective emission redshift for our simulated Ly$\alpha$ cubes. This choice is motivated by the slow evolution of large-scale IGM structure in comoving coordinates over this redshift interval: by $z \sim 2$, the cosmic web is already well established and subsequently evolves primarily through gravitational collapse rather than dramatic large-scale rearrangement \citep{2014MNRAS.441.2923C}. Consequently, generating separate simulated datasets at the central redshift of each selected emission region would not qualitatively alter the IGM structures relevant to this work, while significantly increasing the computational cost of data generation.

The public IllustrisTNG data release provides full simulation snapshots only at discrete redshifts, with no full snapshot available exactly at $z = 2.3$; the nearest full snapshots are at $z = 2$ and $z = 3$. We therefore adopt Snapshot 33 ($z = 2$) to define the underlying gas and galaxy distribution, using gas-cell and star-particle positions in comoving coordinates as a proxy for the large-scale IGM morphology at $z = 2.3$. Only the gas and stellar positions are rescaled to the adopted redshift. Stellar ages remain those evaluated at $z = 2$ and are not recalculated at $z = 2.3$, which preserves the correspondence between stellar ages and the surrounding gas state (star-forming or non-star-forming) in the source snapshot. Recalculating stellar ages at the earlier epoch while retaining the later gas states could assign artificially young ages to stars residing in non-star-forming gas cells, introducing an inconsistency between the stellar sources and their gas environments.

The cosmic-time difference between $z = 2.3$ and $z = 2$ is approximately 0.5 Gyr, an interval over which overdensities grow gradually without significantly modifying the large-scale structure of the IGM \citep{2014MNRAS.441.2923C}. The spatial rescaling changes the physical gas densities inferred from the retained cell masses and rescaled volumes, which can affect the predicted emissivity, particularly in high-density regions. Our analysis focuses on the diffuse IGM, and we explicitly mask galaxies to minimize their contribution. Although greater care regarding galaxy modeling should be taken in future studies, Snapshot 33 currently provides a reasonable spatial proxy for the IGM at $z = 2.3$. In subsequent radiative-transfer calculations, we evaluate Ly$\alpha$ emissivities and observational effects assuming an effective redshift of $z = 2.3$, which enables direct comparison with both CC2.2 and Hyperion.

We initially considered adopting TNG50, following earlier IllustrisTNG-based Ly$\alpha$ studies. However, the relatively small simulation volume of TNG50 yields very few halos with masses comparable to our observational fields at these redshifts. In contrast, TNG100 contains dozens of sufficiently massive halos, enabling us to construct a statistically meaningful training dataset while sampling a range of environments. Although TNG100 has lower spatial resolution than TNG50, it resolves dense, star-forming gas on scales well below the adopted KCWI PSF FWHM of $1.4^{\prime\prime}$, while its adaptive spatial resolution could be coarser in more diffuse environments. Given our focus on broad intergalactic structures at the KCWI observational resolution, we adopt TNG100 as a practical balance between resolution and volume, while recognizing the possible dependence of the predicted Ly$\alpha$ emission on numerical resolution discussed above.

In the KCWI observations, the instrument is intentionally pointed toward known massive structures within CC2.2 and Hyperion. To emulate this observational strategy in the simulations, we first identify the densest regions within the full TNG100 volume. To do this, we partition the simulation volume into selection volumes with a side length of $729\,\mathrm{ckpc}\,h^{-1}$, corresponding to $\sim 38.8^{\prime\prime}$ at $z = 2.3$, or $133.3$ pixels per side at the observed cube sampling resolution. Here, ckpc denotes comoving kiloparsecs. For each selection volume, we compute the total mass by summing the masses of all subhalos whose centers of mass fall within it. We then rank all selection volumes by their total mass and designate the most massive ones as the densest regions in the simulation. This density-evaluation scale is chosen to be modestly smaller than the FOV of a KCWI observed cube, providing a well-defined pointing center while remaining sufficiently large to avoid selecting a center dominated by a very local mass maximum.

Around each pointing center, we then define a larger parent volume that encompasses not only the central dense structure but also its surrounding environment. Because halos and their associated gas distributions typically extend well beyond the spatial extent of the extracted KCWI observed subcubes ($\sim 53^{\prime\prime}$ at $z = 2.3$, corresponding to $180$ pixels), we select all gas cells and star particles within a cubic region of side length $4374\,\mathrm{ckpc}\,h^{-1}$. This corresponds to $\sim 232.6^{\prime\prime}$ at $z = 2.3$, or $800$ pixels per side at the observed cube sampling resolution. This larger cubic region defines the parent volume from which the full mock gas and star catalogs are constructed.

From each parent volume, we extract the properties of gas cells, including positions, velocities, masses, neutral hydrogen fractions, internal energies, electron fractions, star formation rates, and metallicities, to construct a mock gas catalog. We likewise extract stellar properties, including positions, velocities, masses, formation times, initial masses, and metallicities, to generate the corresponding mock star catalog.

We initially selected the 26 pointing centers associated with the most massive selection volumes, each with a total mass of at least $1/20$ that of the most massive selection volume, and generated five pairs of mock gas and star catalogs for each: for every selected pointing center, we applied five distinct three-dimensional rotations to simulate different viewing orientations. The first rotation is the identity transformation, which preserves the native simulation coordinate system and allows for straightforward interpretation of structural and kinematic features. The remaining four rotations are randomly drawn from a uniform distribution over three-dimensional rotation space. After applying each rotation, we re-center and re-clip the region to the same fixed-size box. This procedure produces physically distinct projections, as different gas cells and star particles fall within the extraction volume, thereby introducing both projection diversity and physical variation across the rotated realizations.

However, due to an out-of-memory (OOM) error encountered during the geometry calculations for the pointing center associated with the most massive selection volume, we excluded this case from the final dataset. We therefore retain 25 pointing centers, each with five rotated realizations, including the unrotated case. This yields a total of $25 \times 5 = 125$ pairs of TNG-based mock gas and star catalogs at an effective emission redshift of $z = 2.3$ for subsequent analysis.

\subsection{Simulated Emission Cubes}

\subsubsection{Radiative Transfer with COLT}

We use \colt{} \citep{2015MNRAS.449.4336S,2019MNRAS.484...39S,2022MNRAS.517....1S,2025arXiv251013952M} to post-process the TNG100 data and generate mock surface-brightness cubes with two spatial dimensions and one wavelength dimension. For each of the 125 gas and star catalog pairs, we generate emission cubes along the six Cartesian viewing directions ($\pm x$, $\pm y$, and $\pm z$), yielding $125 \times 6 = 750$ full-field emission cubes.

\colt{} is configured as follows. For non-star-forming gas, \colt{} computes the Ly$\alpha$ emissivity from spatially resolved hydrogen recombination and collisional excitation, based on the local temperature and the number densities of electrons, neutral hydrogen, and ionized hydrogen in the TNG-based mock gas catalogs. For star-forming gas, only an effective temperature is assigned by the simulation's subgrid model. We therefore do not calculate the Ly$\alpha$ luminosity from gas-phase recombination or collisional excitation in these cells. Instead, the unresolved nebular contribution is calculated from star particles using the BPASS--CHAB population-synthesis model \citep{2018MNRAS.479...75S,2003PASP..115..763C}, based on their ages, initial masses, and metallicities. This treatment motivates retaining consistent stellar ages and gas-cell states from the same source snapshot. The initial line profiles of these nebular photons are assigned a thermal width corresponding to a temperature of $10^4$~K and are centered on the bulk velocities of the associated star-forming gas cells. Because our analysis focuses on diffuse gas, we assume a dust-free environment during the radiative-transfer calculations and subsequently mask galaxies and other bright regions in the emission cubes.

This treatment is a deliberately simplified radiative-transfer configuration that neglects dust attenuation, stellar continuum emission, and continuum-photon pumping by the UV background. The impact of these processes on the IGM Ly$\alpha$ surface brightness remains uncertain, and theoretical predictions can vary by up to an order of magnitude depending on the adopted assumptions. We therefore use this configuration as a controlled starting point for the present analysis, while noting that the CDE can be retrained using data generated with updated radiative-transfer models in the future.

We disable resonant scattering (RS) for most \colt{} calculations to reduce the computational cost. During RS, Ly$\alpha$ photons emitted in optically thick regions may undergo many absorption and re-emission events before escaping, redistributing their propagation directions and frequencies. Full MCRT calculations model this diffusion in both frequency and physical space \citep[see][]{2014PASA...31...40D}, but are computationally expensive. We therefore perform full-RS calculations only for the unrotated case for five pointing centers, with indices 3, 8, 13, 18, and 23. These pointing centers sample a range of selection-volume masses because the indices follow the selection-volume mass ranking. With six Cartesian viewing directions per pointing center, this produces 30 matched intrinsic/full-RS cube pairs, which we use to calibrate and assess the empirical correction described below.

For the production dataset, we instead use the intrinsic-emission ``Monte Carlo line profile'' (MCLP) mode. In this mode, photon packets are recorded without subsequent resonant-scattering propagation, allowing \colt{} to calculate the intrinsic Ly$\alpha$ emissivity and source line profiles substantially faster than in the full-RS mode. Under the adopted configurations, MCLP uses five times as many photon packets as the full-RS calculations but remains several hundred times faster in wall-clock time. The larger number of MCLP photon packets improves the Monte Carlo sampling of low-surface-brightness regions, which can additionally receive photons redistributed by resonant scattering in the full-RS calculations. Instrumental effects are then applied by convolving the cubes with a Gaussian PSF with FWHM $\sim1.4^{\prime\prime}$ \citep{2023NatAs...7.1390M} and an LSF modeled as the convolution of a 5.105~\AA\ top-hat with a Gaussian of $\sigma=0.627$~\AA\ \citep{2023ApJ...945L..35L}.

\subsubsection{Empirical Resonant-Scattering Correction}

For the intrinsic-emission cubes, we apply an empirical resonant-scattering correction to approximate the impact of resonant scattering on the emergent Ly$\alpha$ emission. We apply the correction over the surface-brightness range $10^{-21}$--$10^{-17}\,\mathrm{erg\,s^{-1}\,cm^{-2}\,arcsec^{-2}}\,\text{\AA}^{-1}$, which encompasses the regime most relevant to our observations. Fainter voxels lie well below the nominal sensitivity, while brighter regions are typically masked. Voxels outside this interval are left unchanged.

We compare 30 matched intrinsic/full-RS cube pairs after applying the same PSF and LSF. In the middle spectral channel, we select matched voxels for which both the intrinsic and full-RS surface brightnesses lie within the stated interval. We calculate their full-RS/intrinsic surface-brightness ratios and group them into 20 logarithmic bins according to intrinsic surface brightness. We then construct a smooth correction curve by applying a shape-preserving piecewise cubic Hermite interpolation (PCHIP) to the median ratios as a function of $\log_{10}$ intrinsic surface brightness. This curve is applied voxel by voxel to the PSF/LSF-convolved intrinsic-emission cubes across all 30 spectral channels.

Within the correction interval, the correction factor exceeds unity at low surface brightness and falls below unity at high surface brightness. This behavior is consistent with resonant scattering redistributing Ly$\alpha$ photons from intrinsically bright regions into fainter surrounding regions.

\begin{figure}[htbp]
  \centering
  \includegraphics[width=0.9\linewidth]{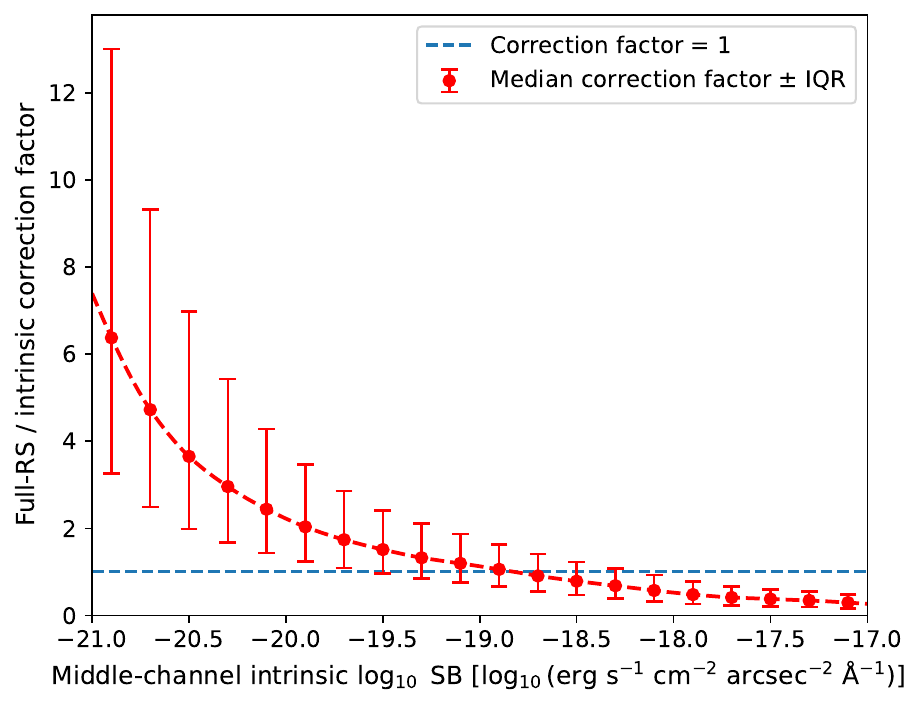}
  \caption{\textbf{Empirical resonant-scattering correction.}
  Full-RS/intrinsic correction factor as a function of intrinsic voxel surface brightness. Points show the median correction factors measured from 30 matched intrinsic/full-RS cube pairs, and the error bars show their interquartile ranges. The red dashed line shows the PCHIP interpolation, while the blue horizontal dashed line denotes a correction factor of unity. \textbf{The correction boosts fainter voxels and suppresses brighter voxels within the adopted correction interval.}}
  \label{fig:boosting_curve}
\end{figure}

\begin{figure*}[ht]
    \centering
    \includegraphics[width=0.8\linewidth,trim=0 24bp 0 0,clip]{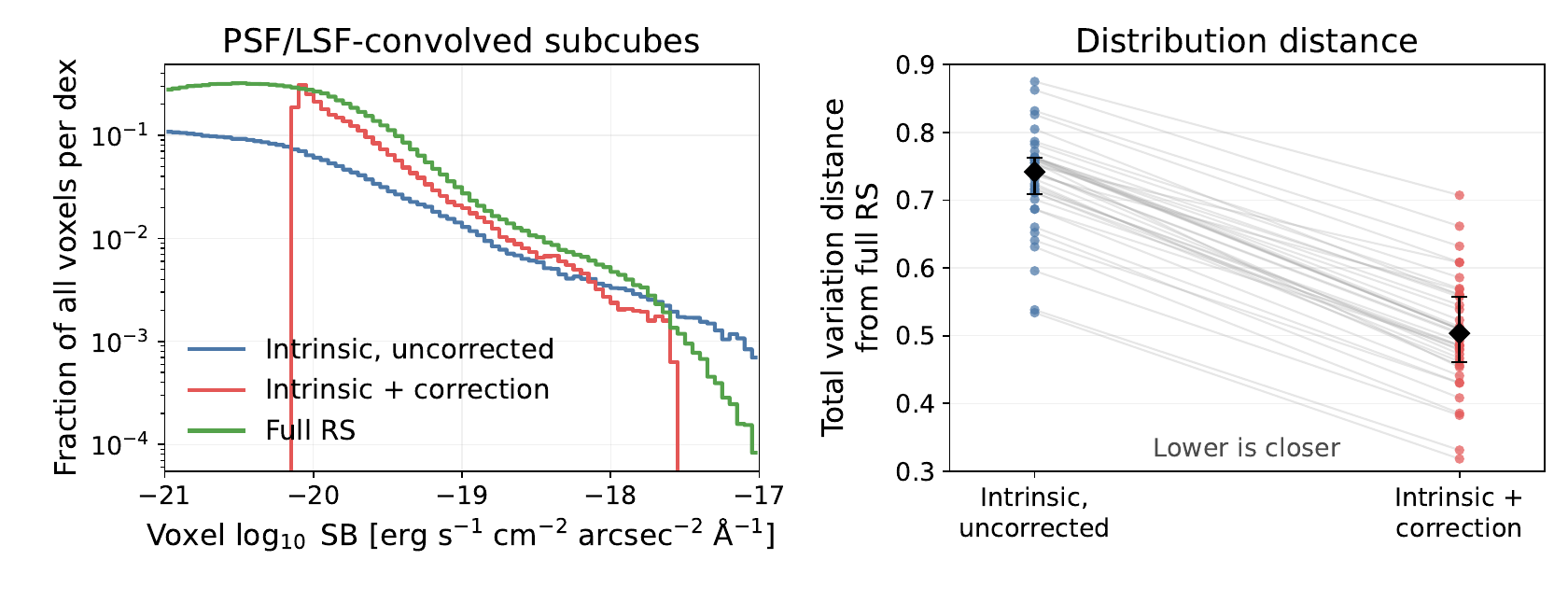}
    \includegraphics[width=0.8\linewidth]{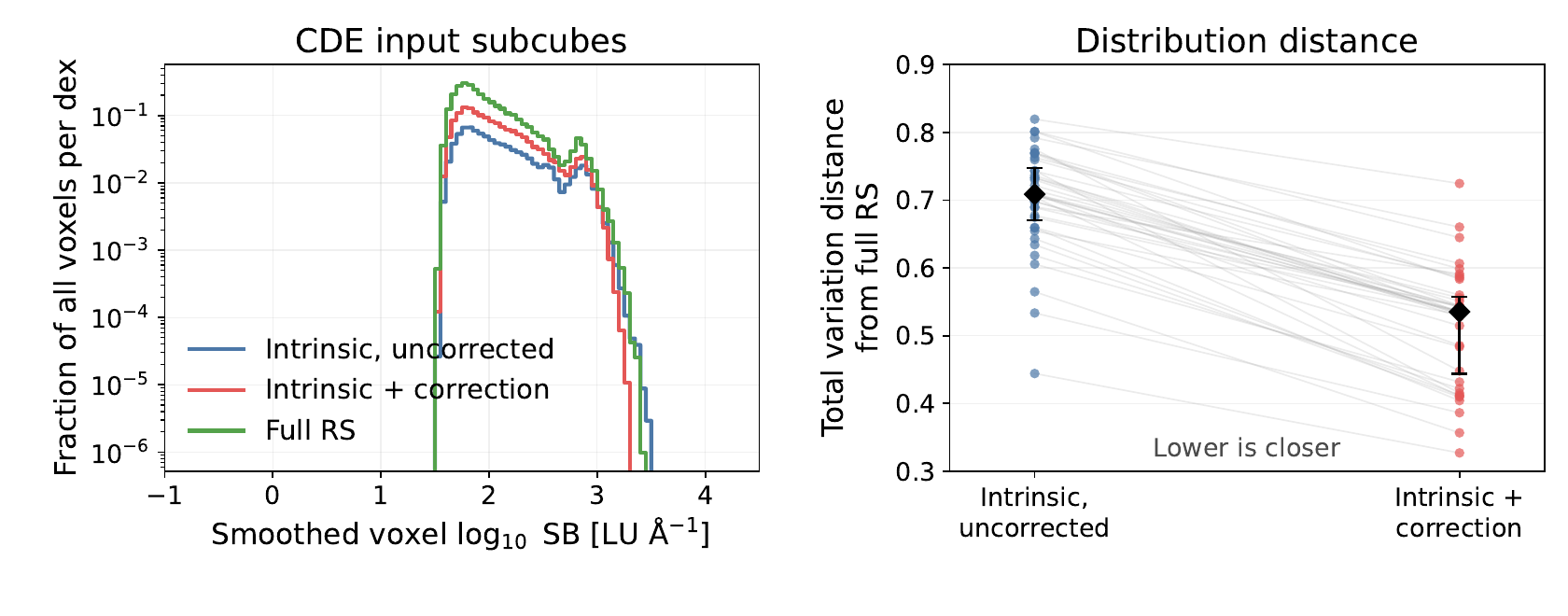}
    \caption{\textbf{Empirical resonant-scattering correction comparison.} Results for 40 matched subcubes extracted from five intrinsic/full-RS cube pairs, using the central 20 spectral channels. \textbf{Top row:} PSF/LSF-convolved subcubes before noise injection, masking, and adaptive smoothing. \textbf{Bottom row:} corresponding final CDE inputs. \textbf{Left column:} voxel surface-brightness distributions. \textbf{Right column:} total variation distance (TVD) between the normalized voxel surface-brightness distributions of each intrinsic-emission treatment and the corresponding full-RS subcube. TVD is half the summed absolute difference between corresponding bins of the normalized histograms, including categories for values below and above the adopted histogram range, with zero indicating identical distributions. For both right-hand panels, voxel positions are selected where the full-RS surface brightness lies between $100$ and $10^4$\,LU\,\AA$^{-1}$, evaluated after PSF/LSF convolution for the upper-right panel and after noise injection, masking, and adaptive smoothing for the lower-right panel. This selection is not applied to the left-hand distributions. Gray lines connect the same subcube across conditions, while black diamonds and error bars show the median and interquartile range. \textbf{The correction reduces the distribution discrepancy from full RS for the selected voxels both before and after noise injection, masking, and adaptive smoothing.}}
    \label{fig:rs_no_rs}
\end{figure*}

We assess the correction using 40 matched subcubes extracted from five intrinsic/full-RS cube pairs. Figure~\ref{fig:rs_no_rs} compares the uncorrected intrinsic, corrected intrinsic, and full-RS subcubes before noise injection, masking, and adaptive smoothing, together with their corresponding final CDE inputs. The correction reduces the discrepancy from the full-RS voxel surface-brightness distribution at both stages for voxels in the surface-brightness range of interest. However, the local voxel-level correction does not explicitly model the spatial and spectral redistribution of photons by resonant scattering. Noise injection, masking, and adaptive smoothing could further modify the final input distributions. Because these 40 matched subcubes are extracted from five cube pairs also used to derive the correction, this comparison assesses the correction's behavior within the calibration sample rather than its generalization to independent pointing centers.

\subsubsection{Creating Masks}

In \citet{2023NatAs...7.1390M}, galaxies were removed via PSF subtraction after the KCWI DRP was applied, and it was noted that subtracting the PSF or masking the affected pixels yields nearly identical results. We therefore adopt a simpler approach by defining rectangular masks around galaxies and bright regions using projected stellar-mass maps and corrected emission cubes, respectively. These masks are constructed over the full spatial field of each $600 \times 600 \times 30$ cube and used to determine whether candidate $180 \times 180 \times 30$ subcubes retain sufficient unmasked area for analysis. The corresponding cropped masks are stored with the subcubes and applied after noise injection and coadding. Galaxies are masked to maintain consistency with the observed KCWI cubes, while additional bright regions are excluded because similarly bright features are not seen in the observed cubes used in this study.

We define two types of masks: (i) bright regions and (ii) regions of high projected stellar density. At our current noise level, sources with $SB > 2 \times 10^{4}$~LU in maps integrated over the central $20\,\text{\AA}$ interval are readily visible after noise injection and coadding. We therefore define bright-region masks using the corresponding integrated surface-brightness maps constructed from the corrected emission cubes before noise injection. We identify contiguous pixels above this threshold and enclose each connected region in a padded rectangular mask.

To identify regions of high projected stellar density, we construct projected stellar-mass maps from the extracted star particles using \texttt{Pylians} \citep{Pylians}, at the same spatial sampling as the simulated surface-brightness cubes. To approximate the removal of galaxies identified in zCOSMOS-deep \citep{2007ApJS..172...70L}, we adopt a lower stellar-mass threshold of $10^{9}\,M_\odot$ for galaxy masking, using the GOODS NICMOS stellar-mass completeness estimates of \citet{2011MNRAS.413.2845M} as an approximate guide. Extrapolating the late-type galaxy size--mass relation of \citet{2014ApJ...788...28V} to this mass gives an effective radius of approximately $1.5$~kpc at $z \sim 2.3$. This is comparable to our spatial sampling of $0.291^{\prime\prime} \simeq 2.4$~kpc per pixel, suggesting that the central stellar distribution is sampled by only a few pixels. We first flag pixels containing at least $10^{8}\,M_\odot$ of stellar mass. We then identify 8-connected components of flagged pixels and sum the stellar masses within each component. Components with summed stellar masses of at least $10^{9}\,M_\odot$ are masked using squares centered on their stellar-mass-weighted positions. The masks have sizes of $11 \times 11$ pixels for component masses of $10^{9} \leq M_\star/M_\odot < 10^{10}$ and $13 \times 13$ pixels for component masses of $M_\star \geq 10^{10}\,M_\odot$.

\subsubsection{Selecting Bright Subcubes}

After defining the masks, we randomly extract subcubes of size $180 \times 180 \times 30$ voxels from each $600 \times 600 \times 30$ corrected full-field cube. From each full-field cube, we retain eight such subcubes, yielding a total of $750 \times 8 = 6000$ emission subcubes.

To ensure that the selected regions are sufficiently bright and are not affected by large masked patches, we require each candidate subcube to satisfy three criteria, evaluated within its central $160 \times 160$ spatial region corresponding to the spatial extent of the final CDE input. First, the mean surface brightness, integrated over the central $20\,\text{\AA}$ interval and averaged over spatial pixels outside the union of the brightness and stellar masks, must be at least $200$~LU. Second, no single 4-connected component of this combined mask within the central region may cover more than $5\%$ of that region's area. Third, at least $85\%$ of the spatial pixels in the central region must remain unmasked. Random extraction is repeated until eight subcubes satisfying all three criteria are obtained for each full-field cube. At this stage, the masks are used only for subcube selection and are not yet applied to the data.

\subsection{Noisy Emission Subcubes}

\subsubsection{Adding Poisson Noise and Coadding}

In the relevant wavelength range, we adopt a representative sky level of $\sim 4 \times 10^{-18}\,\mathrm{erg\,s^{-1}\,cm^{-2}\,arcsec^{-2}}\,\text{\AA}^{-1}$, broadly consistent with the dark-sky measurements of \citet{2003A&A...400.1183P}. Although the sky background in KCWI observations varies with time and wavelength, we approximate its continuum level as constant across each $30\,\text{\AA}$ window. In the sky-noise-dominated regime, the S/N scales approximately as the inverse square root of the sky level, so moderate fractional inaccuracies in the adopted background produce smaller fractional changes in the simulated S/N. We therefore add a uniform sky contribution to each simulated exposure.

To convert surface brightness into expected electron counts, we use the KCWI inverse-sensitivity function (\texttt{invense}), together with the detector-pixel area and exposure time. Based on the ensemble of calibration curves recorded during our observing run, we adopt a representative value of $0.3 \times 10^{-16}\,\mathrm{erg\,cm^{-2}}\,\text{\AA}^{-1}\,\mathrm{e}^{-1}$ near $4000\,\text{\AA}$ for the BL and large-slicer configuration. We treat this value as constant over each $30\,\text{\AA}$ window. In the sky-noise-dominated regime, the S/N also scales approximately as the inverse square root of the adopted \texttt{invense}, reducing its sensitivity to moderate inaccuracies in this approximation.

To mimic exposures at PA=$0^\circ$ and $90^\circ$, the KCWI NAS mode, and the subsequent coadding of multiple exposures, we sample the corrected full-field cube around each selected subcube position onto $66 \times 24 \times 30$ exposure grids, with the spatial axes exchanged for PA=$90^\circ$. These grids have a spatial sampling of $0.291^{\prime\prime} \times 1.35^{\prime\prime}$. We simulate six nominal spatial offsets at each PA, with six $1200$~s exposures per offset, including random pointing offsets. Each exposure grid is converted into expected electron counts using the adopted sky level, \texttt{invense}, detector-pixel area, and exposure time. For every source-plus-sky exposure, we also construct a corresponding sky-only exposure with the same sky level and calibration parameters. We draw independent Poisson realizations for the two exposures and add independent Gaussian read noise with a standard deviation of $2\,\mathrm{e}^{-}$ to each. Subtracting the sky-only exposure from the source-plus-sky exposure approximates the differencing step in NAS mode.

Finally, we convert the sky-subtracted, noise-injected exposures back to surface-brightness units and coadd them onto a $180 \times 180 \times 30$ grid with $0.291^{\prime\prime} \times 0.291^{\prime\prime}$ spatial pixels. Coaddition includes random registration errors and weights contributions by pixel overlap and exposure time, introducing spatial-pixel correlations. We compare the resulting noise properties with those of the observed KCWI subcubes below.

\subsubsection{Noise Model Assessment}

\begin{figure*}
    \centering
    \includegraphics[width=0.9\linewidth]{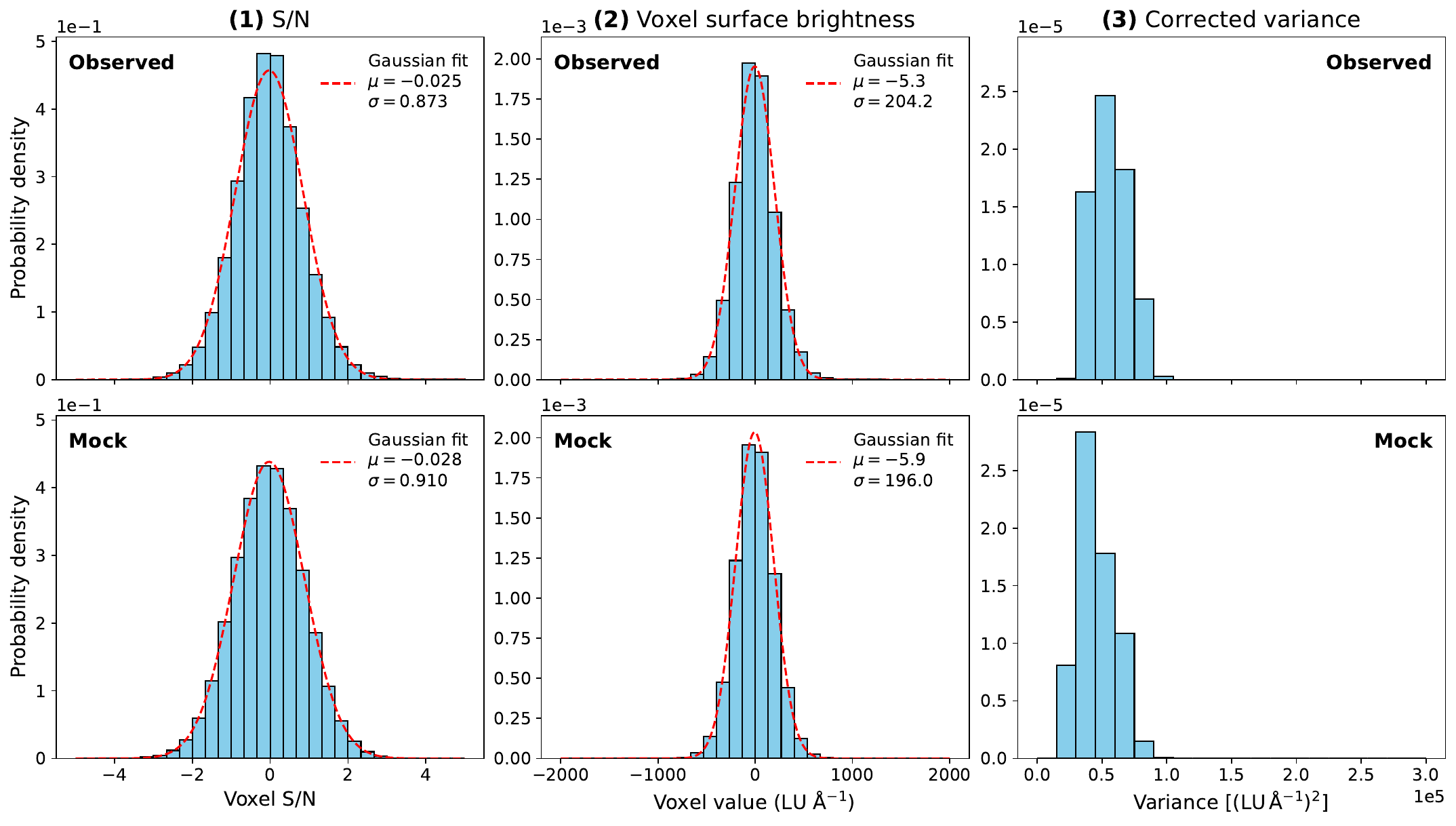}
    \caption{
    \textbf{Noise-statistics comparison.} Comparison of noise statistics between the low-surface-brightness observed subcube from Field A and the noise-injected simulated empty subcube after variance correction, the first smoothing round, and boundary trimming. Columns show, from left to right, the voxel-value distributions of \textbf{(1)} S/N, \textbf{(2)} surface brightness, and \textbf{(3)} corrected variance. The smoothing uses a spatial Gaussian kernel with FWHM $=5$ pixels followed by a spectral boxcar kernel with a width of four channels. \textbf{The simulated and observed distributions show broadly similar noise properties.}}
    \label{fig:noise_distribution}
\end{figure*}

We assess the noise model by comparing a noise-injected simulated empty subcube with an observed low-surface-brightness subcube from Field A, using 20 spectral channels centered at wavelengths from $3858$ to $3877\,\text{\AA}$. Figure~\ref{fig:noise_distribution} shows the voxel-value distributions of S/N, surface brightness, and variance for the observed and simulated subcubes after variance correction, the first smoothing round, and boundary trimming. To account for spatial-pixel correlations introduced during coadding that are not captured by the propagated variance, \citet{2023NatAs...7.1390M} multiplied the variance by a factor of 8. The variance subcubes shown here include this correction. For this comparison, we apply a spatial Gaussian kernel with FWHM $=5$ pixels followed by a spectral boxcar kernel with a width of four channels.

The simulated and observed distributions show broadly similar statistical properties in this comparison. The same sky level and instrumental-noise prescription are used throughout the simulated dataset.

\begin{figure*}
    \centering
    \includegraphics[width=0.9\linewidth]{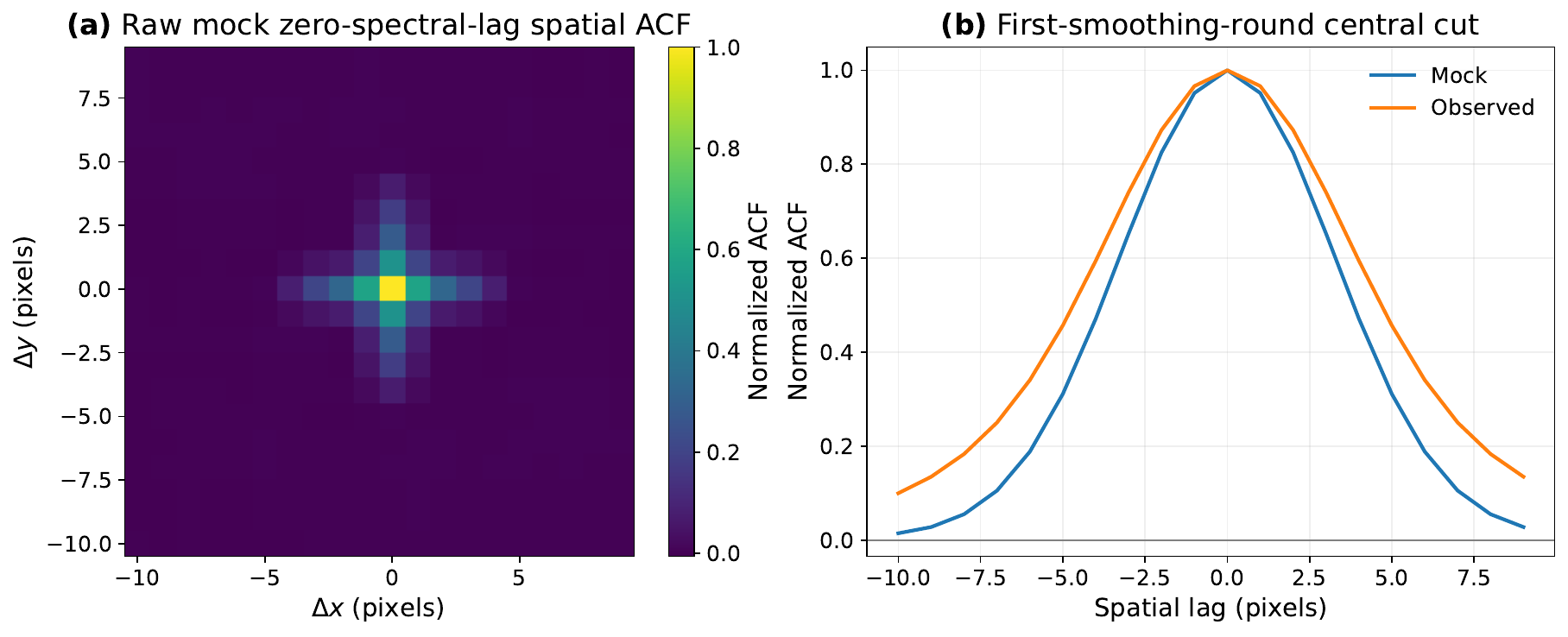}
    \caption{
    \textbf{Spatial ACF comparison.} Both panels use spatial ACFs evaluated at zero spectral lag by combining the within-channel spatial correlations across all 20 retained channels and normalizing to unity at zero spatial lag. \textbf{(a)} 2D spatial ACF of the noise-injected simulated empty subcube before smoothing, showing the characteristic plus-sign pattern induced by coadding exposures at PA$=0^\circ$ and PA$=90^\circ$. \textbf{(b)} Central 1D cuts through the spatial ACFs after the first smoothing round and boundary trimming, comparing observed and simulated subcubes. Each cut holds the lag in the orthogonal spatial direction at zero. \textbf{The central profiles are similar, while the observed ACF has broader wings.}}
    \label{fig:ACF}
\end{figure*}

In Figure~\ref{fig:ACF}, we additionally compare the spatial autocorrelation function (ACF), evaluated at zero spectral lag by combining the within-channel spatial correlations across all 20 channels. The left panel shows the 2D spatial ACF of the simulated empty subcube before smoothing, which exhibits a characteristic plus-sign pattern arising from coadding exposures at PA$=0^\circ$ and PA$=90^\circ$. The right panel compares central 1D cuts through the spatial ACFs of the observed and simulated subcubes after the first smoothing round and boundary trimming. Each ACF is normalized to unity at zero spatial lag. The central profiles are similar, while the observed ACF has broader wings.

These two diagnostics indicate broad agreement between the simulated and observed noise properties in the comparison presented here. Differences remain, and the current noise model omits effects such as flat-field residuals, calibration errors, detector nonlinearities, and structured sky-subtraction residuals.

\subsection{Masking and Adaptive Smoothing}

At this stage, we have both observed subcubes and noisy simulated emission subcubes. While galaxies are already masked in the observed subcubes, the masks for the simulated subcubes are defined but not yet applied. Both types of subcube are dominated by noise because KCWI observations are conducted near the instrumental limit, and the simulated subcubes are designed to reproduce this noise-dominated regime. Such strong noise makes it difficult to visually identify meaningful structures. Adaptive smoothing is therefore required to recover large-scale features from the noisy background.

We apply very similar adaptive-smoothing procedures to the observed and simulated subcubes, with different S/N thresholds as described below. Bright sources are masked before smoothing to prevent their light from spreading into surrounding regions. For the simulated subcubes, we apply the stellar masks and bright-region masks defined earlier. Galaxies are already masked in the observed subcubes, and we additionally identify regions with $|SB| > 2 \times 10^4$~LU in the map integrated over the central $20\,\text{\AA}$ interval and flag 8-connected structures containing at least 25 spatial pixels. These regions occupy only a small fraction of the observed subcube area, and we mask them for consistency with the simulated bright-region masking. We apply rectangular masks to these regions to prevent strong positive or negative features from spreading during smoothing.

When the masks are applied, masked pixels are replaced with the local-background level, estimated separately in each spectral channel from the median of surrounding unmasked pixels, to avoid introducing artificial voids or unphysical discontinuities in the data.

As summarized in \citet{2016ApJ...824L...5M}, \citet{2019NatAs...3..822M}, and \citet{2023NatAs...7.1390M}, adaptive smoothing proceeds iteratively: the surface-brightness subcube is first smoothed spatially, then along the wavelength axis, followed by another round of spatial smoothing, and so on.

We adopt the same variance correction described above and otherwise broadly follow the adaptive-smoothing algorithm in \citet{2023NatAs...7.1390M}. The variance correction is applied to the simulated subcubes and is already included in the supplied observed variance. We adopt detection thresholds of S/N $\geq 3$ for the simulated subcubes and S/N $\geq 5$ for the observed subcubes. The higher observed threshold provides a more conservative criterion in the presence of residual instrumental and sky-subtraction effects that are not fully captured by the simulated noise model. Both datasets use 30 spectral channels during smoothing, with five channels on each side serving as spectral padding. For the observed subcubes, the padding channels are actual neighboring KCWI data and variance channels.

Because edge voxels are affected differently from central voxels during the smoothing process, we perform spatial trimming to exclude all voxels within 10 pixels of the subcube boundaries in both spatial directions, as well as the five additional spectral channels on each side. Specifically, we retain the subregion \texttt{[10:170, 10:170, 5:25]} from each simulated and observed smoothed subcube. The resulting data products have dimensions $160 \times 160 \times 20$ and, after the logarithmic transformation described above, are referred to as the ``simulated CDE inputs'' and ``observed CDE inputs.'' Random spatial rotations by multiples of $90^\circ$ and spatial flipping are applied consistently to the simulated subcubes and their corresponding column-density maps for data augmentation. Surface brightness is expressed in $\mathrm{LU}\,\text{\AA}^{-1}$ before logarithmic transformation. In total, we generate 6000 simulated CDE inputs for training, validation, and testing, and 10 observed CDE inputs for predicting column density in the observed emission regions, together with two observed CDE inputs serving as empty controls.

\subsection{Column Density Maps}

The column density maps serve as the ground truth for supervised training and are calculated using \texttt{Pylians} \citep{Pylians} from the corresponding TNG100 gas-cell catalog. The final maps have dimensions $40 \times 40$ and share the same FOV as the simulated CDE inputs. To construct them, we first generate $600 \times 600$ full-field maps and convolve them with the same spatial PSF as the emission cubes and then extract the corresponding $180 \times 180$ maps. For the noisy simulated dataset, each map is masked and filled using the same masks and local-median-filling procedure as the corresponding emission subcube, ensuring consistency between input and ground truth. The maps are then trimmed to $160 \times 160$ and downsampled to $40 \times 40$. This masking step also helps the CDE focus on extended large-scale structures rather than isolated dense pixels. Furthermore, the same random spatial rotation and flipping used for data augmentation are applied to the ground-truth maps to maintain spatial alignment.

\section{Computational Cost}
\label{app:computational_cost}

Table~\ref{tab:dataset_summary} summarizes the number of data products created at each stage of the preparation procedure and the corresponding computational cost. Based on the measured runtimes and estimates for individual stages, the main data-preparation steps require approximately 60~hours of wall-clock time on an AMD Ryzen~9~7900 (12 cores) and an NVIDIA RTX~4070~Ti GPU. This approximate total excludes the initial construction of the TNG100 catalogs.

For clarity, CPU time denotes the cumulative time spent by all processor cores, while costs for GPU-based stages are reported as wall-clock time using one GPU. The intrinsic-emission runs with \colt{}  produce 750 cubes and require approximately 12~CPU~hours, corresponding to 2~hours of wall-clock time using 12 cores. The cumulative CPU time is lower than the wall-clock time multiplied by the core count because input reading and other partially parallel steps do not fully utilize all 12 cores. The full-RS calculations produce 30 cubes and require approximately 600~CPU~hours, corresponding to 50~hours of wall-clock time, making them the most computationally expensive step. Applying the empirical correction and masks adds negligible computational cost. Noise injection and coaddition use 10 CPU workers and require approximately 1.3~hours of wall-clock time, and we estimate the corresponding cumulative CPU cost to be $\sim$10~hours. Adaptive smoothing requires approximately 5~hours of wall-clock time on one GPU. Overall, the computational cost of the CDE framework remains manageable on a single high-end desktop workstation.

During the bagging training phase, we use a batch size of 5. With early stopping applied (patience = 25), training each base model requires approximately 10--20 minutes, resulting in a total training time of approximately 2.5~hours for the 10-model ensemble. This training cost is modest compared to the approximately 60~hours required for the main data-preparation stages.

\begin{table*}[htbp]
\centering
\caption{Summary of dataset sizes and approximate computational costs. Costs are rounded from recorded runtimes or estimated from execution and file timestamps. The noise-injection CPU cost is estimated from the wall-clock runtime and worker count.}
\label{tab:dataset_summary}
\setlength{\tabcolsep}{3pt} 
\begin{tabular}{@{}lccc@{}}
\hline
Stage & \# Data Products & Approximate Total Cost & Notes \\
\hline
Mock gas/star catalog pairs & 125 & -- & Starting point \\
Noiseless intrinsic-emission cubes & 750 & $\sim$12 CPU hr ($\sim$2 wall hr) & \colt{} MCLP mode \\
Noiseless full-RS cubes & 30 & $\sim$600 CPU hr ($\sim$50 wall hr) & \colt{} with full RS \\
Corrected noiseless emission cubes & 750 & negligible & Empirical correction \\
Noisy emission subcubes & 6000 & $\sim$10 CPU hr ($\sim$1.3 wall hr) & Noise injection and coaddition \\
Masked emission subcubes & 6000 & negligible & Stellar and bright-region masks \\
Adaptively smoothed emission subcubes & 6000 & $\sim$5 wall hr & Adaptive smoothing on one GPU \\
Column-density maps & 6000 & $\sim$1 wall hr & Pylians and map preparation \\
Final input--ground-truth pairs & 6000 & -- & Simulated CDE input + target map \\
\hline
\end{tabular}
\end{table*}

\section{Illustrative CNN Examples}
\label{app:cnn_examples}

\begin{figure*}[htbp]
    \centering
    \includegraphics[width=0.75\linewidth]{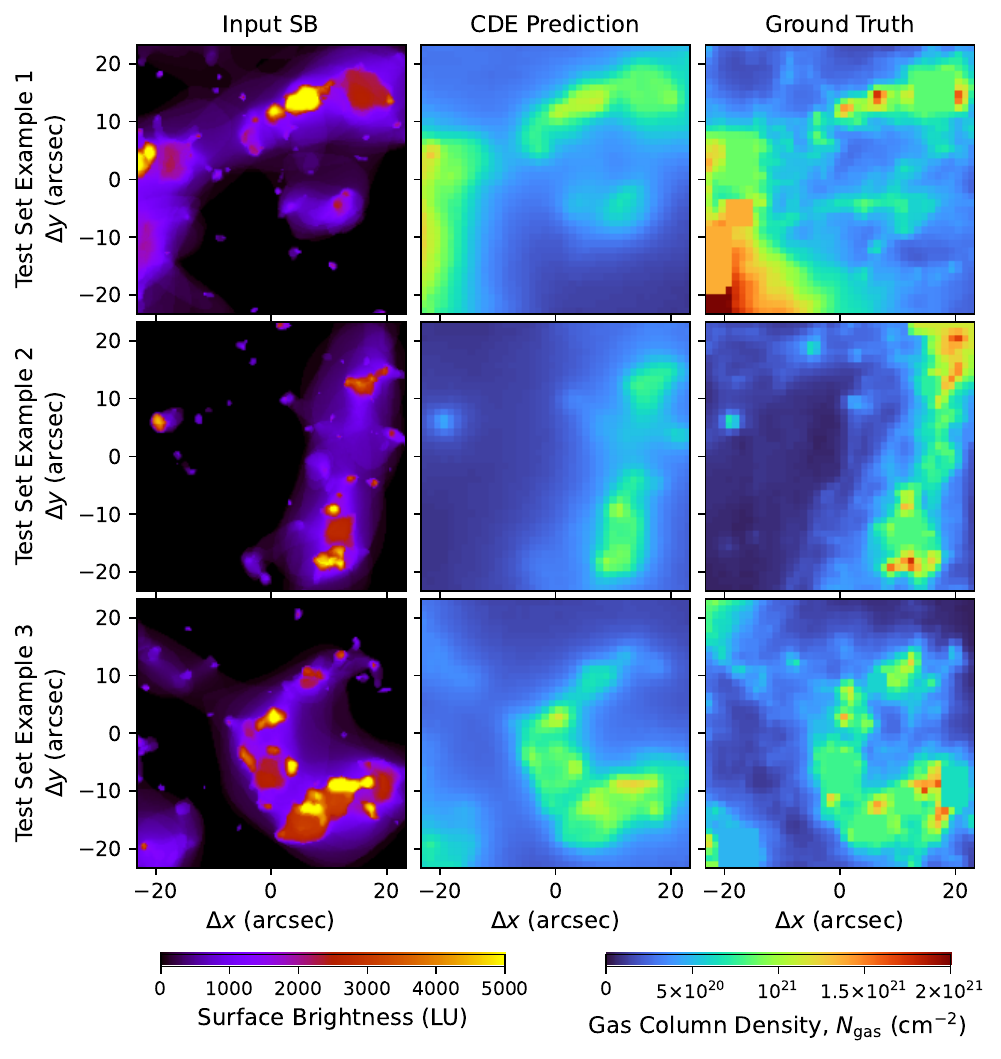}
    \caption{\textbf{More representative test-set examples for the CDE}. 
    Each row corresponds to a different system: 
    \textbf{left}, the input Ly$\alpha$ surface-brightness map integrated over the retained 20 spectral channels; 
    \textbf{middle}, the CDE-predicted gas column density map; 
    and \textbf{right}, the ground-truth masked column density from the simulation. 
    \textbf{The CDE consistently recovers the broad filamentary structures, while compact, high-density regions remain under-predicted.}}
    \label{fig:cnn_examples}
\end{figure*}

Figure~\ref{fig:cnn_examples} shows representative test-set examples, including the input Ly$\alpha$ surface brightness maps, the column density maps predicted by the CDE, and the corresponding ground-truth (masked) column density distributions. In these test-set examples, the CDE successfully reconstructs the broad filamentary structures but systematically underestimates compact, high-density regions. Given that our primary interest lies in the large-scale filamentary structures, underestimation of small, dense spots is acceptable at this stage.

\section{More Filament Detections}

\label{app:more_filament_detection}

\begin{figure*}[htbp]
  \centering
  \begin{minipage}[t]{0.45\textwidth}
    \centering
    \includegraphics[width=\linewidth]{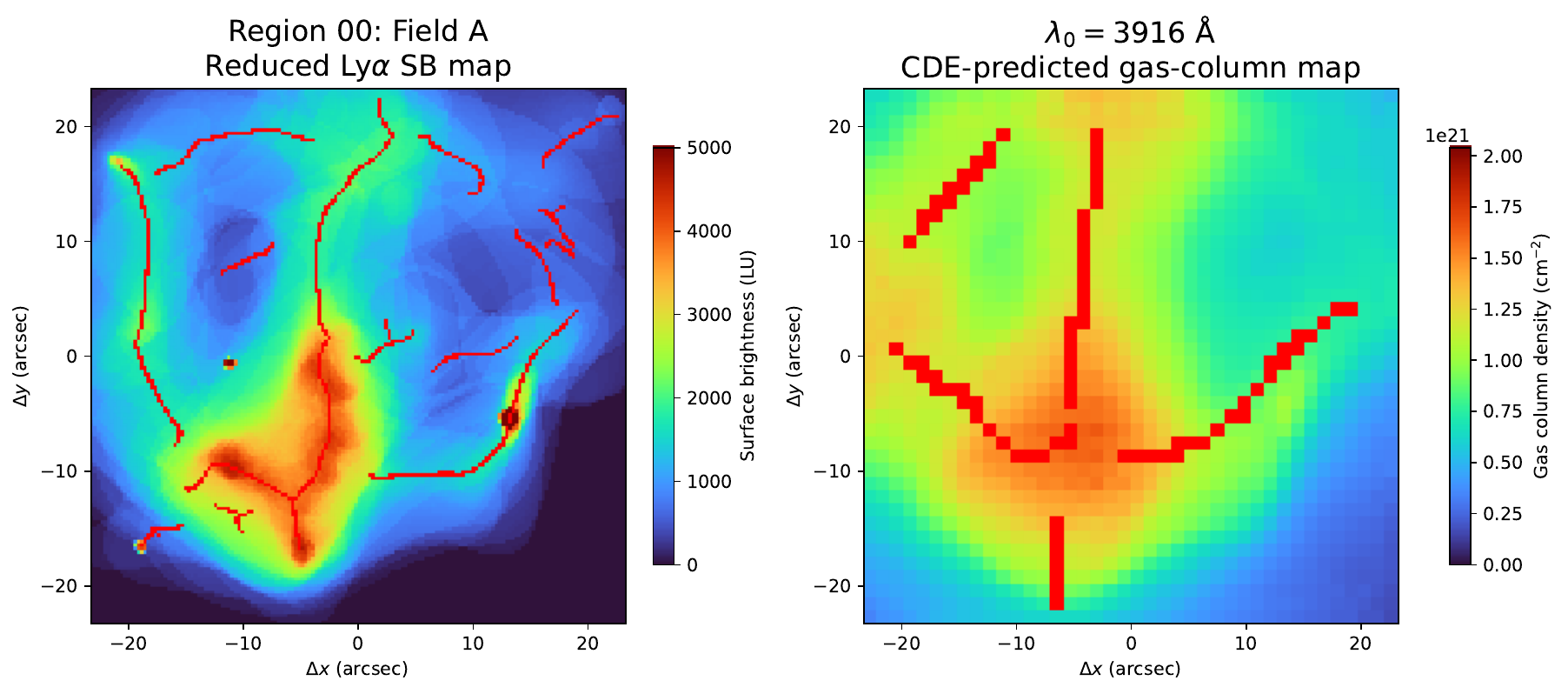}\vspace{0.6ex}
    \includegraphics[width=\linewidth]{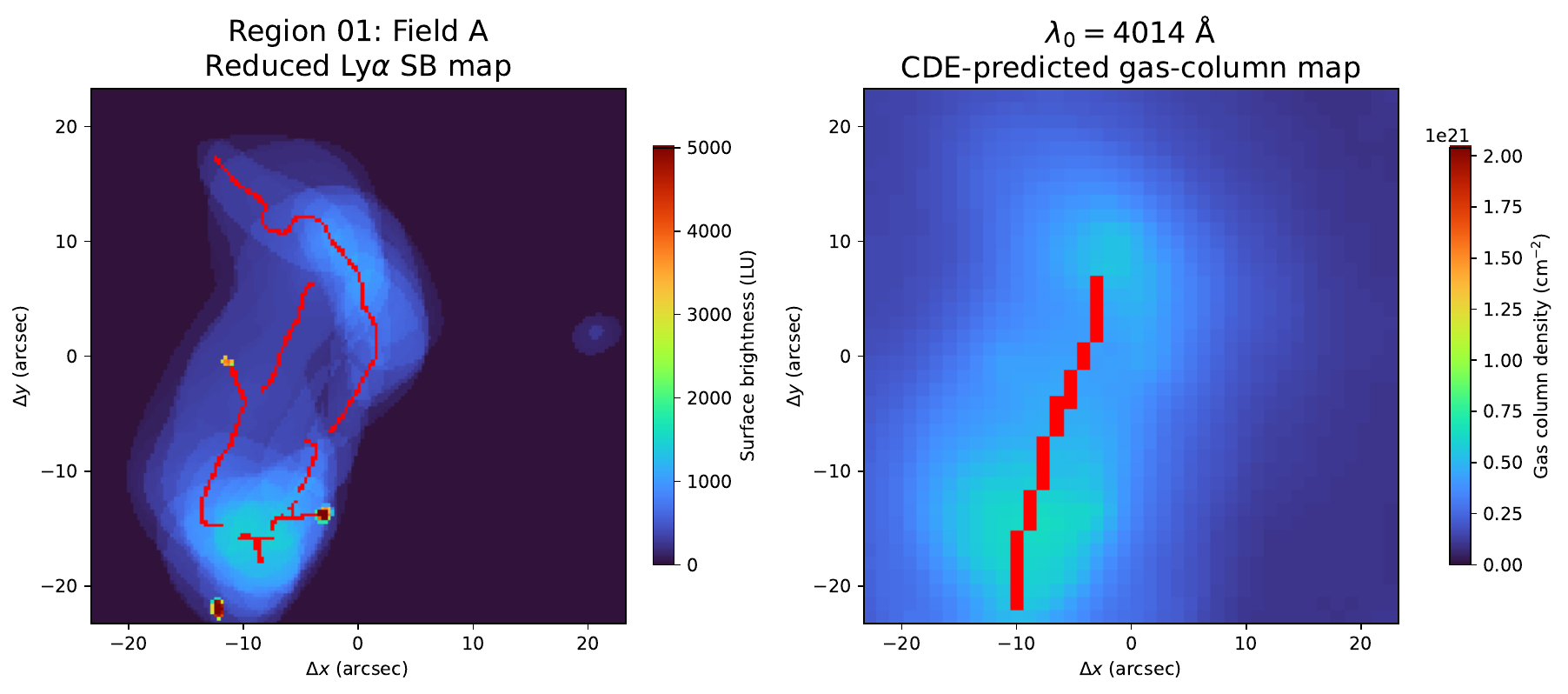}\vspace{0.6ex}
    \includegraphics[width=\linewidth]{target_2_gas.pdf}\vspace{0.6ex}
    \includegraphics[width=\linewidth]{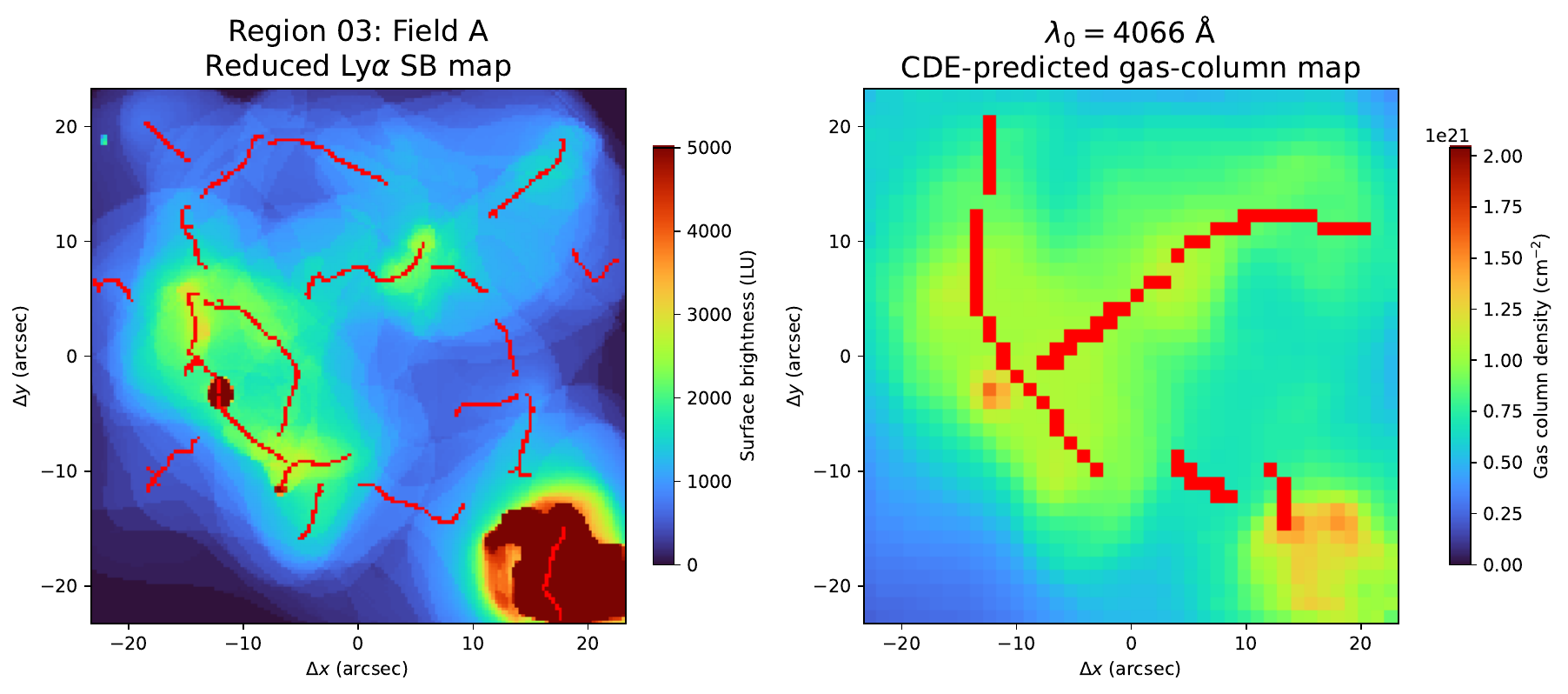}\vspace{0.6ex}
    \includegraphics[width=\linewidth]{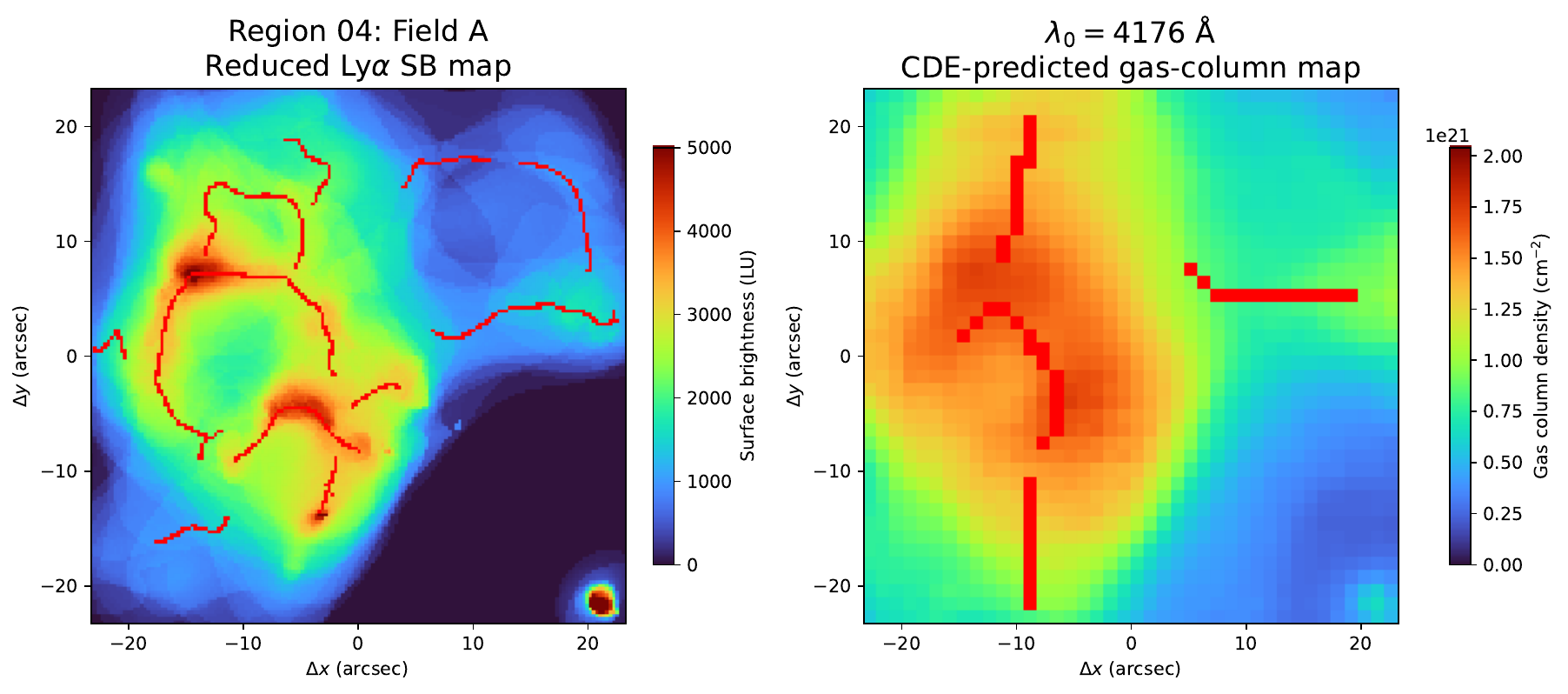}
  \end{minipage}
  \hfill
  \begin{minipage}[t]{0.45\textwidth}
    \centering
    \includegraphics[width=\linewidth]{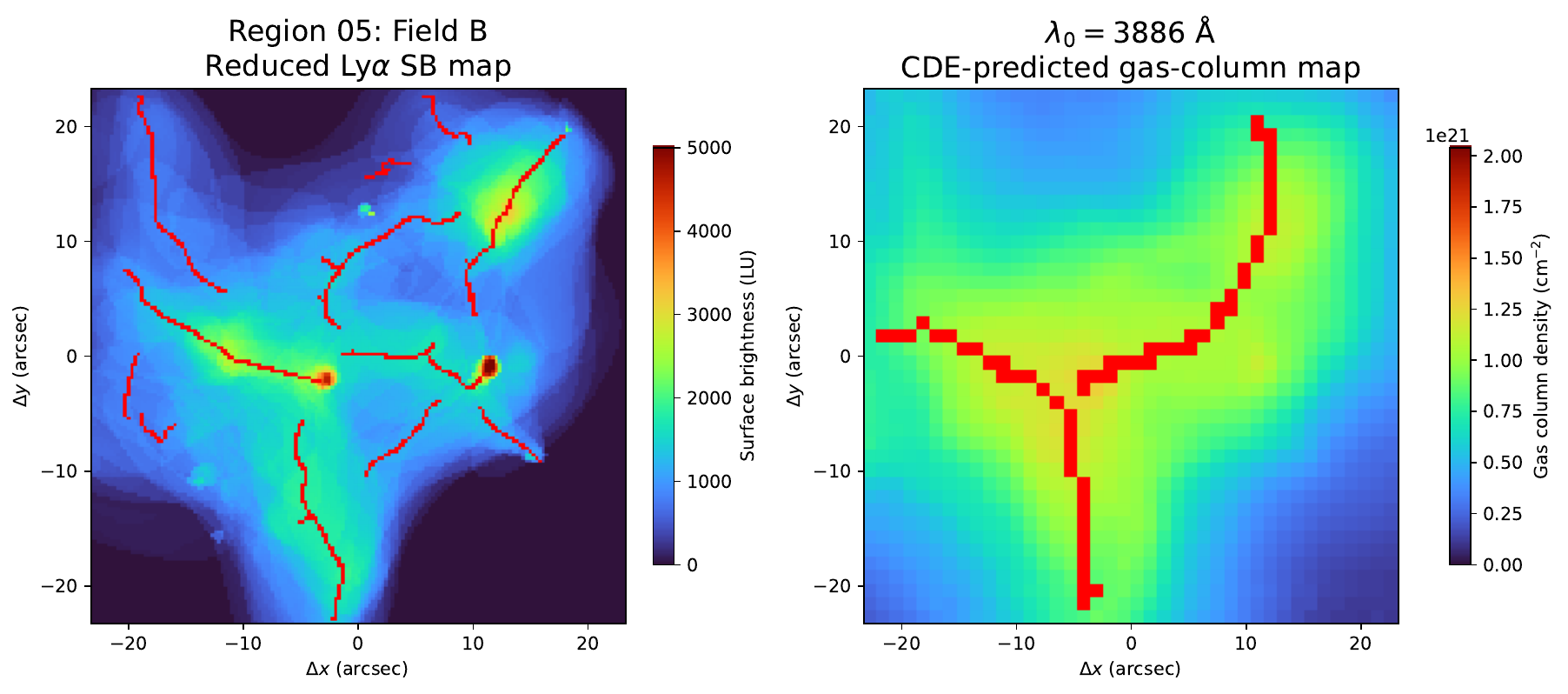}\vspace{0.6ex}
    \includegraphics[width=\linewidth]{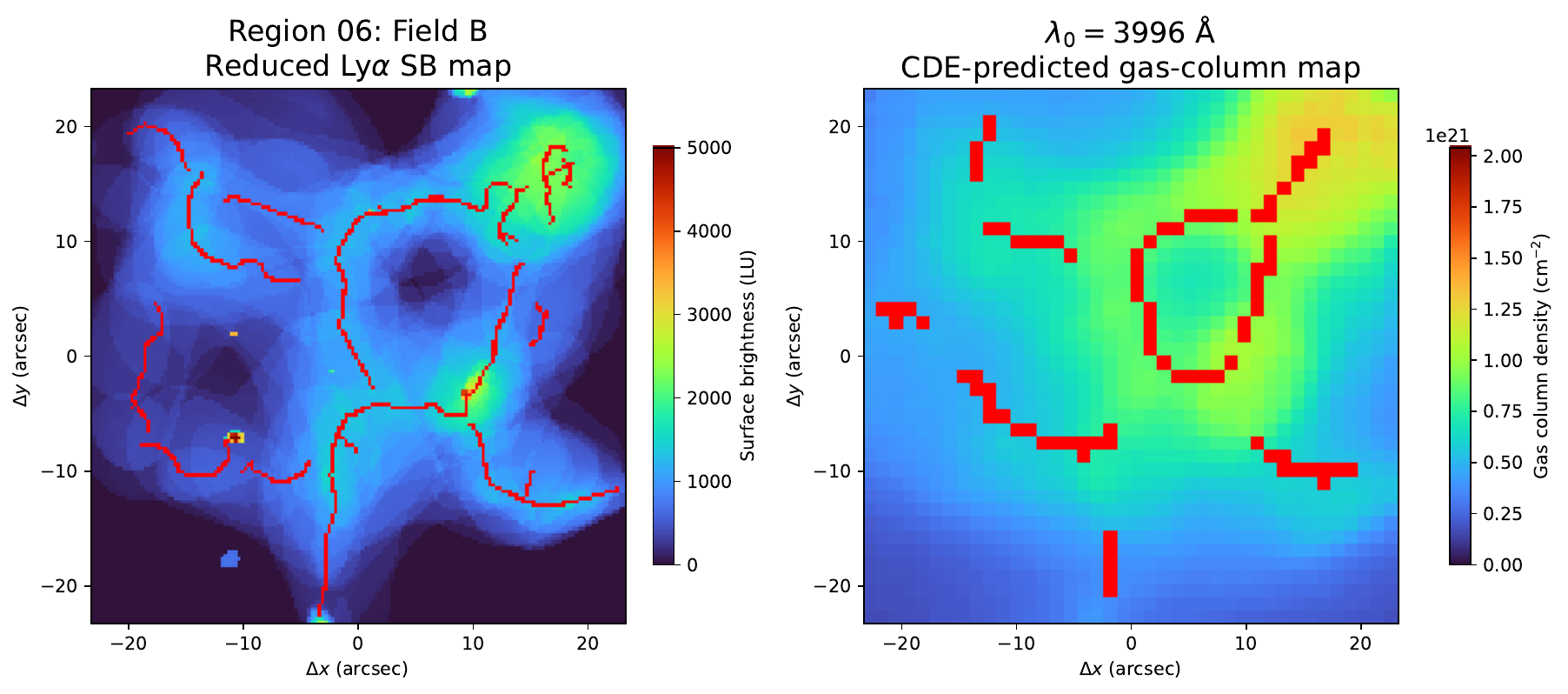}\vspace{0.6ex}
    \includegraphics[width=\linewidth]{target_7_gas.pdf}\vspace{0.6ex}
    \includegraphics[width=\linewidth]{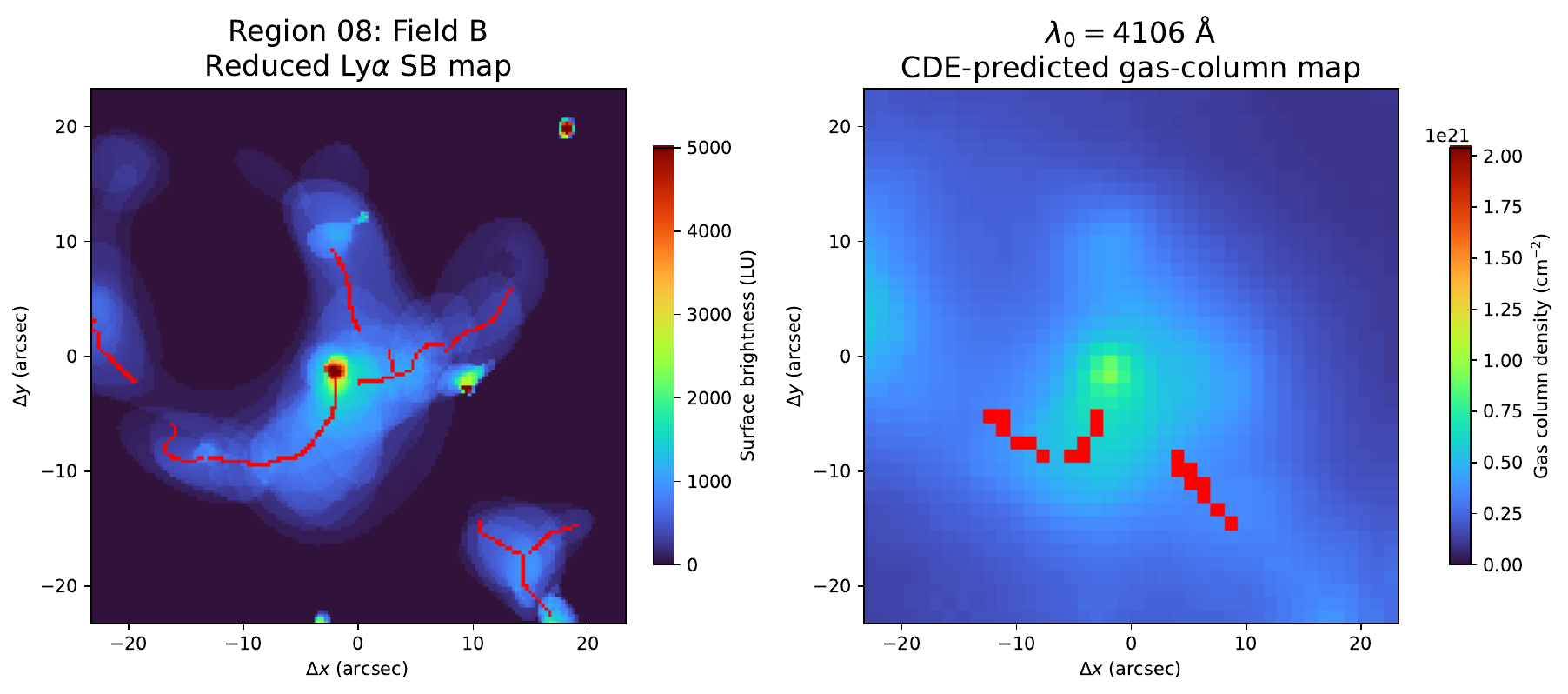}\vspace{0.6ex}
    \includegraphics[width=\linewidth]{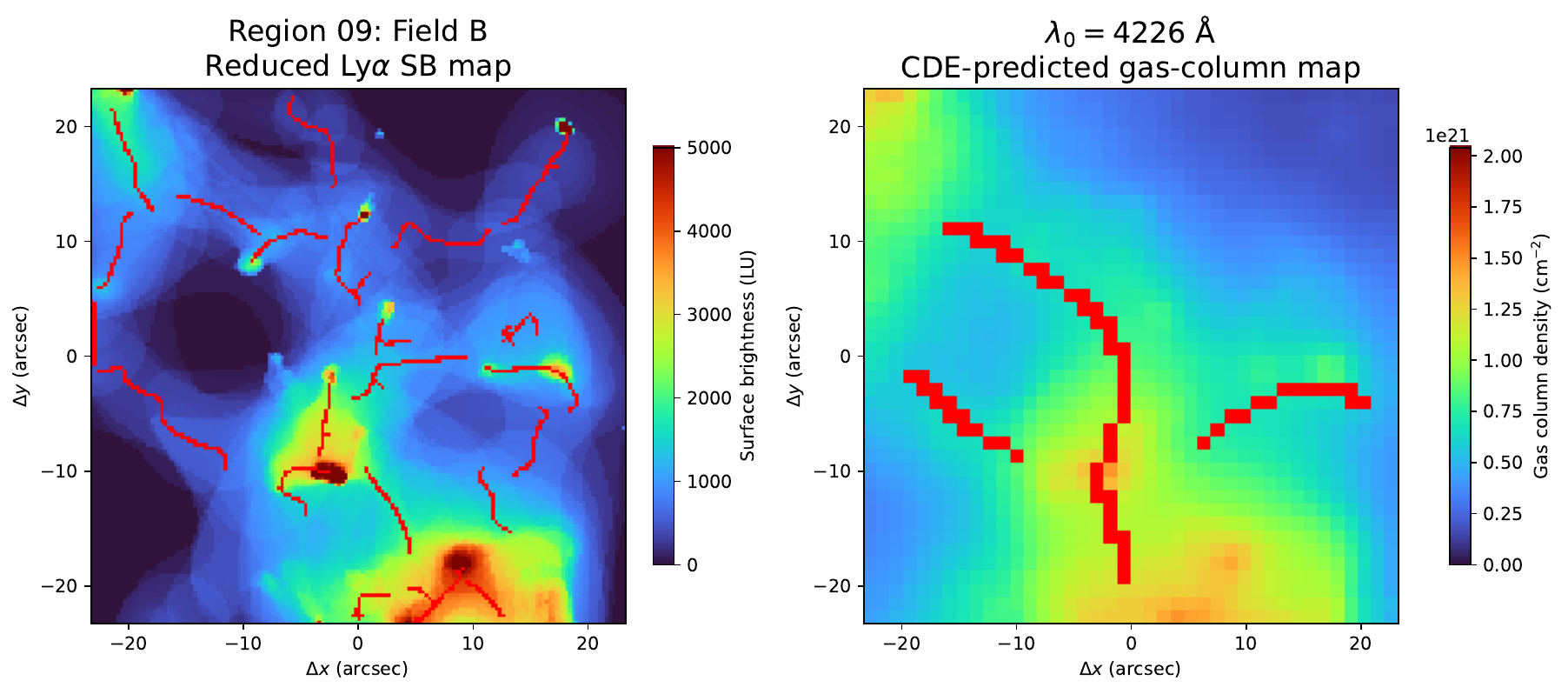}
  \end{minipage}

  \caption{Atlas of CDE reconstructions and filament detections for all ten KCWI regions (left column: regions 0--4; right column: regions 5--9). Each panel shows the adaptively smoothed Ly$\alpha$ surface-brightness map integrated over the retained 20 spectral channels and the corresponding CDE-predicted column density with SCMS filaments overlaid in red.}
  \label{fig:all_targets_atlas}
\end{figure*}

Figure~\ref{fig:all_targets_atlas} shows the CDE reconstructions and SCMS filament identifications for all ten observed emission regions in Field A and Field B. SCMS identifies large-scale filamentary structures in the CDE-predicted column density maps for all ten regions.

\clearpage
\bibliography{report_bib}{}
\bibliographystyle{aasjournal}

\end{document}